\documentclass[11pt,a4paper]{article}
\usepackage{epsfig}
\usepackage[T1]{fontenc}    
\usepackage{graphics}
\usepackage{graphicx}
\usepackage{pstricks,pst-coil,pst-fill,pst-plot}
\usepackage[fleqn]{amsmath}    
\usepackage{amssymb,amsthm}    
\usepackage{amsfonts}   
\usepackage{verbatim}   
\usepackage{mathrsfs}   
\usepackage{dsfont}
\usepackage{yfonts}
\usepackage{txfonts}
\usepackage{marvosym}
\usepackage{vmargin}        

\setmarginsrb{3.2cm}{2.5cm}{3.2cm}{2.5cm}{1cm}{1cm}{1cm}{1.6cm}

\makeatletter
\@addtoreset{equation}{section}
\makeatother

\def\pour#1{_{\,\vrule height 13pt depth 1pt\> {#1}\!}}

\def\sg{\sigma}

\def\veps{\varepsilon}

\renewcommand{\ln}{\log}
\def\Ad{\operatorname{Ad}}
\def\tr{\operatorname{tr}}

\def\id{\operatorname{id}}

\newcommand{\bra}[1]{\langle\,#1\,|}
\newcommand{\ket}[1]{|\,#1\,\rangle}

\let\tend=\rightarrow

\newtheorem{theorem}{Theorem}[section]
\newtheorem{prop}{Proposition}[section]
\newtheorem{cor}{Corollary}[section]
\newtheorem{defin}{Definition}[section]
\newtheorem{conj}{Conjecture}[section]
\newtheorem{lemme}{Lemma}[section]
{\theoremstyle{remark}
\newtheorem{rem}{Remark}[section]}

\def\Proof{\medskip\noindent {\it Proof --- \ }}

\let\qed=\cqfd

\newcommand\beq{\begin{equation}}
\newcommand\enq{\end{equation}}
\newcommand\bem{\begin{multline}}
\newcommand\enm{\end{multline}}

\def\beqa{\begin{eqnarray}}
\def\eeqa{\end{eqnarray}}
\def\ba{\begin{array}}
\def\ea{\end{array}}

\def\a{\alpha}

\def\det{\operatorname{det}}
\def\eps{\epsilon}
\def\ga{\gamma}
\def\la{\lambda}
\def\sg{\sigma}

\newcommand{\f}[2]{{\ensuremath{%
    \mathchoice%
    {\dfrac{#1}{#2}}
    {\dfrac{#1}{#2}}
    {\frac{#1}{#2}}
    {\frac{#1}{#2}}
}}}
\newcommand{\tf}[2]{\ensuremath{#1/#2}}
\newcommand{\pa}[1]{\ensuremath{\left(#1\right)}}
\newcommand{\paa}[1]{\ensuremath{\left\{#1\right\}}}
\newcommand{\pac}[1]{\ensuremath{\left[#1\right]}}
\newcommand{\paf}[2]{\ensuremath{\left(\f{#1}{#2}\right)}}
\newcommand{\pab}[2]{\ensuremath{\pa{\ba{c} #1 \\ #2 \ea }}}

\def\eps{\epsilon}
\def\ga{\gamma}
\def\Ga{\Gamma}
\def\de{\delta}
\def\dep{\deltaup}
\def\De{\Delta}
\def\la{\lambda}

\def\sg{\sigma}
\def\Sg{\Sigma}

\def\th{\theta}

\newcommand{\mc}[1]{\ensuremath{\mathcal{#1}}}
\newcommand{\mf}[1]{\ensuremath{\mathfrak{#1}}}
\newcommand{\msc}[1]{\ensuremath{\mathscr{#1}}}

\newcommand{\ov}[1]{\ensuremath{\overline{#1}}}
\newcommand{\wt}[1]{\ensuremath{\widetilde{#1}}}
\newcommand{\wh}[1]{\ensuremath{\widehat{#1}}}

\newcommand{\Int}[2]{\ensuremath{\int\limits_{#1}^{#2}}}
\newcommand{\Oint}[2]{\ensuremath{\oint\limits_{#1}^{#2}}}

\newcommand{\sul}[2]{\ensuremath{\sum\limits_{#1}^{#2}}}
\newcommand{\pl}[2]{\ensuremath{\prod\limits_{#1}^{#2}}}

\newcommand{\R}{\ensuremath{\mathbb{R}}}
\newcommand{\Cx}{\ensuremath{\mathbb{C}}}

\newcommand{\Dp}[1]{\ensuremath{\partial_{#1}}}
\newcommand{\D}[2]{\ensuremath{\f{\dd #1}{\dd #2}}}

\newcommand{\limit}[2]{\ensuremath{\underset{#1 \tend #2}{\longrightarrow} }}

\newcommand{\ex}[1]{\ensuremath{\e{e}^{#1}}}

\newcommand{\braket}[2]{\ensuremath{\langle #1 \mid  #2 \rangle }}

\newcommand{\ddet}[2]{\ensuremath{\det_{#1}\pac{#2}}}

\newcommand{\abs}[1]{\ensuremath{\left| #1 \right|}}
\newcommand{\norm}[1]{\ensuremath{\left\|#1\right\|}}

\newcommand{\Cont}[2]{\ensuremath {\msc{C}^{#1}\pa{#2} }}

\newcommand{\dd}{\mathrm{d}}
\newcommand{\e}[1]{\ensuremath{\mathrm{#1}}}

\newcommand{\intff}[2]{\ensuremath{\left [ \, #1 \,; #2 \, \right ] }}
\newcommand{\intfo}[2]{\ensuremath{\left [ \, #1 \,; #2 \, \right [ }}
\newcommand{\intof}[2]{\ensuremath{\left ] \, #1 \,; #2 \, \right ] }}
\newcommand{\intoo}[2]{\ensuremath{\left ] \, #1 \,; #2 \, \right [ }}

\begin{document}

\vspace*{1cm}
\begin{flushright}
LPENSL-TH-07/08\\
\end{flushright}
\par \vskip .1in \noindent

\vspace{24pt}

\begin{center}
\begin{LARGE}
{\bf  Riemann--Hilbert approach to a generalised sine kernel and applications}
\end{LARGE}

\vspace{50pt}

\begin{large}

{\bf N.~Kitanine}\footnote[1]{LPTM,  Universit\'e
de Cergy-Pontoise et CNRS, France, kitanine@u-cergy.fr},~~
{\bf K.~K.~Kozlowski}\footnote[2]{ Laboratoire de Physique, ENS Lyon et CNRS, France, karol.kozlowski@ens-lyon.fr},~~
{\bf J.~M.~Maillet}\footnote[3]{ Laboratoire de Physique, ENS Lyon et CNRS,  France, maillet@ens-lyon.fr},\\
{\bf N.~A.~Slavnov}\footnote[4]{ Steklov Mathematical Institute,
Moscow, Russia, nslavnov@mi.ras.ru},~~
{\bf V.~Terras}\footnote[5]{ Laboratoire de Physique, ENS Lyon et CNRS,  France, veronique.terras@ens-lyon.fr, on leave of
absence from LPTA, Universit\'e Montpellier II et CNRS, France}
\par

\end{large}

\vspace{80pt}

\centerline{\bf Abstract} \vspace{1cm}
\parbox{12cm}{\small%
 We investigate the asymptotic behaviour of a generalised sine kernel acting on a finite size interval $\intff{-q}{q}$.
 We determine its asymptotic resolvent as well as the first terms in the asymptotic expansion of its Fredholm determinant. Further, we apply our results to build the resolvent of truncated Wiener--Hopf operators generated by
 holomorphic symbols. Finally, the leading asymptotics of the Fredholm determinant allows us to establish the asymptotic
 estimates of certain oscillatory multidimensional coupled integrals that appear in the study of correlation functions of quantum integrable models.
 }
\end{center}


\newpage

\tableofcontents

\newpage

%
\section{Introduction}
%

The sine kernel
\begin{equation}\label{Sine_kernel}
S\pa{\la,\mu}= \f{\sin \frac{x}2\pa{\la-\mu}}{\pi\pa{\la-\mu}} ,
\end{equation}
is a very important object in mathematical physics. In particular,
the Fredholm determinant of the integral operator $I-S$ acting on
some interval $J \subset \R$ appears in random matrix theory
\cite{GaudinGUELevelSpacingasSineKernel}. In the bulk scaling limit,
$\ddet{J}{I-S}$  stands for the probability
\cite{GaudinMehtaDensityOfEigenvaluesRandomMatrices} that a matrix
belonging to the Gaussian unitary ensemble has no eigenvalues in $x
 J$. The kernel (\ref{Sine_kernel}) also appears in the theory of
 quantum integrable systems. In particular,
the determinant $\ddet{J}{I+\gamma S}$, $\ga$ being a
 parameter, describes various zero--temperature correlation functions  of
the impenetrable Bose gas 
\cite{Len64,JimMiwaMoriSatoSineKernelPVForBoseGAz}.

In all these interpretations  of the sine kernel, one is interested in the
  large $x$ behaviour of its Fredholm determinant.
The first attempt to analyze the $x\tend +\infty$ asymptotics of
$\ddet{J}{I-S}$ goes back to Gaudin and Mehta
\cite{GaudinGUELevelSpacingasSineKernel,GaudinMehtaDensityOfEigenvaluesRandomMatrices}. In 1973, Des Cloizeaux and
Mehta \cite{DescloizeauxMethaSineKernelFirstAsympotics} showed that
\begin{equation}
\ln\ddet{\intff{-1}{1}}{I-S}=-\f{x^2}{8}-\f{1}{4}\ln x +
\e{O}\pa{1}, \qquad x \tend + \infty \;.
\label{noyau sinus pur asymptotiques}
\end{equation}
\noindent Three years later, using Widom's formula \cite{WidomSzegoLimitonCircularArcs} for the asymptotics of Toeplitz
determinants supported on an arc,  Dyson \cite{DysonSineKernelInverseScatteringAsymptoticExpansions} gave a heuristic
derivation of the constant terms $c_0$ and proposed a recursive method to compute the subleading coefficients
$c_1,c_2,\dots $ in the asymptotic expansion:
\begin{equation}
\ln\ddet{\intff{-1}{1}}{I-S}=-\f{x^2}{8}-\f{1}{4}\ln x + c_0 +
\f{c_{1}}{x}+ \f{c_{2}}{x^2}+ \dots\;\;.
\label{Sine kernel full asymptotics}
\end{equation}
However, the forementioned results  were  heuristic. It was only in 1994 that Widom
 \cite{WidomSinekernelOnSingleIntervals} managed to prove rigorously the first term in the asymptotics
 \eqref{noyau sinus pur asymptotiques}:
\begin{equation}
\D{}{x}\ln\ddet{\intff{-1}{1}}{I-S}=-\f{x}{4}+\e{o}\pa{1} \; .
\end{equation}
One year later, this analysis was extended  to the multiple interval
case \cite{WidomSinekernelOnUnionOfIntervals}. While Widom studied
the asymptotic behaviour of the Fredholm determinant by operator techniques, Deift,
Its and Zhou applied the Riemann--Hilbert problem (RHP) formulation
for integrable integral operators
 \cite{ItsIzerginKorepinSlavnovDifferentialeqnsforCorrelationfunctions} to
the sine kernel acting on a union of intervals $\cup_{\ell} J_{\ell}$ and
proved the existence of the asymptotic expansion \eqref{Sine kernel full
asymptotics}.  However, their
method did not allow them to obtain an estimate for the constant
$c_0$, as they inferred the asymptotic expansion of
$\ln\ddet{J}{I-S}$ from that of
\begin{equation}
P_x=x\D{}{x} \ln \ddet{}{I-S} \; .
\end{equation}
The first proofs of Dyson's heuristic formula for $c_0$ appeared in the independent, and based on 
completely different methods, works of  Ehrhardt \cite{Ehrhardt2006} and Krasovsky
\cite{Krasovsky2004} and more recently in \cite{DeiftIKZ2007}.

We would like to point out that there is a very nice connection of the sine
kernel to the Painlev\'e V equation \cite{JimMiwaMoriSatoSineKernelPVForBoseGAz}, as $P_x$ solves this equation.
The link between Painlev\'e V and $P_x$ was also investigated in \cite{DieftItsZhouSineKernelOnUnionOfIntervals}
in the framework
 of RHP. It was shown that one can deduce this Painlev\'e equation directly from the RHP data.

\bigskip

This article is devoted to the study  a generalisation of the sine kernel \eqref{Sine_kernel}. This
kernel, that we will refer to as the generalised sine kernel (GSK),
is of the form
\begin{equation}
V\pa{\la,\mu}=\f{\ga \sqrt{F\pa{\la} F\pa{\mu}} }{2i\pi \pa{\la-\mu}}
  \big[ e_+\pa{\la}e_-\pa{\mu}-e_-\pa{\la}e_+\pa{\mu} \big], \quad
\label{generalised Sine}
\end{equation}
where
\begin{equation}\label{epm}
 e_\pm(\la)=\ex{\pm\tf{[ixp\pa{\la}+g\pa{\la}]}{2}}.
\end{equation}
We will be more specific about the functions  $F$, $p$ and $g$ later on.

Various particular cases of the  kernel (\ref{generalised Sine})
already appeared in the literature.
These particular kernels were mostly  used for the description of correlation functions of matrix models or quantum
integrable models equivalent to free fermions (see e.g.
\cite{McCoyPArkShrockSpinTimeAutoCorrAsModSineKernel,Len66,ItsIK90,
ItsIK90a,KorS90,ColIKT92,ColIKT93,ItsIzerginkorepinSlavnov1993,ItsIzerginkorepinVarguzin1991,CheianovZvonarevZeroTempforFreeFermAndPureSine}).
In the present paper we consider a rather general case, only based on the analytic properties of the functions $F$, $p$ and $g$. The GSK
(\ref{generalised Sine}) plays a crucial role in the study of
correlation functions of (non free-fermion) quantum integrable
systems \cite{KitKMST009}. It is also useful for the
asymptotic analysis of truncated Wiener--Hopf operators with Fischer--Hartwig singularities \cite{KozWienerHopfWithFischerHartwig}.

We investigate here the large $x$ asymptotic behaviour of the Fredholm determinant of the GSK in the
framework of RHP. Our work is a natural extension of an
unpublished analysis by Deift, Its and Zhou of the sine kernel $I+\ga S$
by RHP. This kernel was also analysed by RHP in \cite{CheianovZvonarevZeroTempforFreeFermAndPureSine}.

\bigskip

This article is organized as follows. In Section~\ref{sec-pb-results}, we  announce the main results of
the paper, namely,

\begin{itemize}
\item the large $x$ asymptotic behaviour of the Fredholm
determinant of the integral operator $I+V$, \textit{cf.} \eqref{generalised Sine};
\item the asymptotic resolvent of some Wiener--Hopf operators connected to \eqref{generalised Sine};
\item the asymptotic behaviour of coupled multiple integrals involving a cycle of kernels $V$ \eqref{generalised Sine} versus some holomorphic symmetric functions.
\end{itemize}

The proof of the asymptotic behaviour of $\ln\det[I+V]$ is given in
the core of the paper (Sections~\ref{sec-initialRHP}, \ref{Section
solution RHP}, \ref{Section Developpement asym N} and \ref{Section
asymptotique log det}). More precisely, in
Section~\ref{sec-initialRHP}, we recast the problem into a certain RHP. In Section~\ref{Section solution
RHP}, we transform this initial RHP into a RHP that can easily
be solved asymptotically. This asymptotic solution is
presented in Section~\ref{Section Developpement asym N} and
used in Section~\ref{Section asymptotique log det}  to obtain the leading and the first subleading
terms of $\ln\det[I+V]$ in the $x\tend + \infty$ limit.

In Section~\ref{section applications}, we apply these results to truncated Wiener--Hopf operators. We show how one can use the asymptotic resolvent
of the generalised sine kernel to construct asymptotic resolvents of
truncated Wiener--Hopf operators acting on $\intff{-x}{x}$, with $x$
large.  This asymptotic resolvent is used to reproduce the low magnetic field behaviour of the so-called
dressed charge arising in the theory of quantum integrable models solvable by the Bethe ansatz \cite{BogoliubiovIzerginKorepinBookCorrFctAndABA}.

Section~\ref{sec-integrales} is devoted to the study of the asymptotic behaviour of some particular type of coupled multiple integrals which can be obtained in terms of the GSK. This is in fact  our main motivation to study the GSK: indeed, from the knowledge of the asymptotic behaviour of this type of multiple integrals one can obtain the asymptotic behaviour of quantum integrable models correlation functions, as it is done in \cite{KitKMST009}.

Finally, in Section~\ref{sec-general-kernel}, we consider the case of further modifications of the GSK, in particular those useful for the correlation functions of the integrable Heisenberg spin chains \cite{KitKMST009}.

Some properties of confluent hypergeometric functions and proofs
of several lemmas are gathered in the appendices.

\section{Problem to solve and main results}\label{sec-pb-results}

\subsection{generalised sine kernel: assumptions and notations}
\label{subsec-hyp}

Let $I+V$ be the integral operator with kernel \eqref{generalised Sine} and acting on $L^2\pa{\intff{-q}{q}}$.

We assume that there exists some open relatively compact neighbourhood $U$ of
$\intff{-q}{q}$ such that  the functions $p$, $F$ and $g$, as well
as the parameter $\ga$, satisfy the following properties:
\begin{itemize}
\item $F$ and $g$ are holomorphic on $\ov{U}$, the closure of $U$;
\item $p$ is holomorphic and injective on $\ov{U}$, $p\pa{\intff{-q}{q}}\subset \R$,
and $p$ stabilizes the upper half plane ${\mathcal H}_+$ (resp. the
lower half plane ${\mathcal H}_-$),  i.e. $p\pa{U\cap {\mathcal
H}_{\pm}} \subset {\mathcal H}_{\pm}$;
\item $\ga \in D_{0,r}=\{\lambda\in\mathbb{C}: |\lambda|<r\}$, where $r$ is such that $\abs{r F}<1$ and
$\e{arg}\pa{1+\ga F} \in \intoo{-\pi}{\pi}$ on $\ov{U}$.
\end{itemize}

We study  the large $x$ expansion of the
Fredholm determinant of $I+V$ under these assumptions. This will be done by asymptotically solving
 a certain matrix RHP. It will become clear in the next section that
 the assumption $p\pa{\intff{-q}{q}}\subset \R$ is tantamount to
imposing the associated RHP to be of oscillatory nature. Moreover,
the case  $p\pa{U\cap {\mathcal H}_{\pm}} \subset {\mathcal
H}_{\mp}$ is obtained by the negation $\pa{\ga, g\pa{\la} } \mapsto
\pa{-\ga, -g\pa{\la}}$. 

Note that $\gamma$ plays here the role of a regularisation parameter; in particular it should be stressed that our method does not allow to reach the $|\gamma F| = 1$ case corresponding to \eqref{Sine kernel full asymptotics} which requires a different analysis \cite{Ehrhardt2006,Krasovsky2004,DeiftIKZ2007}.

\bigskip

Before presenting the main result of this article, let us introduce some convenient notations.

First, we define two auxiliary functions used in the
article:
\begin{align}
&\nu\pa{\la}= \f{-1}{2i\pi} \ln \pa{1+\ga F\pa{\la}}\; ,
        \label{def-nu}\\
&\kappa (\la;q) \equiv \kappa(\la) =
\exp\paa{ \Int{-q}{q} \f{\nu\pa{\la}-\nu\pa{\mu}}{\la-\mu} \dd \mu  } \; .
\label{nuit}
\end{align}
Note that $\kappa$ is a function of the two parameters $\la$ and
$q$, although we will sometimes omit the dependence on the second
parameter.

Finally, we will use the following simplified notations for the
values of the functions $p$ and $\nu$ and of their derivatives at
the points $\pm q$:
\begin{align}
 &p_\pm=p(\la)\pour{\la=\pm q}\;, \qquad p'_\pm=p'(\la)\pour{\la=\pm q}\;,
  \qquad etc.
       \label{notation-p-q}\\
 &\nu_\pm=\nu(\la)\pour{\la=\pm q}\;,  \qquad \nu'_\pm=\nu'(\la)\pour{\la=\pm
 q}\;, \qquad etc.
       \label{notation-nu-q}
\end{align}

\subsection{The main results}

We now give the asymptotic behaviour of the Fredholm determinant in
the $x\tend+\infty$ limit:
%
\begin{theorem}
\label{theorem asymptotiques order zero log det}
Let $V$ be the GSK \eqref{generalised Sine} with $p$, $g$, $F$ and
$\gamma$ satisfying the assumptions of Section~\ref{subsec-hyp}.
Then, in the $x\tend +\infty$ limit, $\ln\det[I+V]$ behaves as
\begin{equation}\label{asympt-V-ordre0}
 \ln\det[I+V]=\ln\det[I+V]^{(0)}+\e{o}(1),
\end{equation}
with
\begin{multline}
 \ln\det[I+V]^{(0)}
 = -i x \Int{-q}{q} \nu(\la)p'(\la)\ \dd\la
    -(\nu_+^2+\nu_-^2)\ln x - \Int{-q}{q} \nu(\la)g'(\la)\ \dd\la
  \\
    + \ln\pac{\f{G(1,\nu_+)\, G(1,\nu_-)\, \kappa^{\nu_+}(q;q)}
             {\pa{2q p'_+ }^{\nu_+^2}\pa{2q p'_- }^{\nu_-^2}
                 \kappa^{\nu_-}(-q;q)  }}
    +  \f{1}{2} \Int{-q}{q}\, \dd \la\, \dd \mu
       \f{\nu'(\la) \nu(\mu)- \nu(\la) \nu'(\mu) }{\la-\mu},
\label{asymptotique log det ordre zero}
\end{multline}
in which we have used the notations of Section~\ref{subsec-hyp}. The Barnes G-function \cite{BarnesDoubleGaFctn1,BarnesDoubleGaFctn2}
 admits the integral representation:
\begin{equation}
G\pa{z+1}= \pa{2\pi}^{\f{z}{2}}\exp\paa{ - \f{z\pa{z-1}}{2}+
\Int{0}{z} t \psi\pa{t} \dd t }, \quad \Re(z)>-1,\quad
\psi(z)=\frac{\Gamma'(z)}{\Gamma(z)}\;,
\label{fonction de Barnes}
\end{equation}
and we denote $G(1,z)\equiv G(1+z)G(1-z)$.
\end{theorem}

Using the perturbation theory for singular integral equations one
can  refine the theorem and obtain  sub-leading corrections.
Although, in principle, nothing opposes to derive the next
sub-leading corrections, the computations become more and more
involved. We have proved the structure of the first corrections to
the equation \eqref{asymptotique log det ordre zero}.

\begin{prop}
\label{proposition correstion en 1/x au resolvent}
Let V be the GSK \eqref{generalised Sine} with the conditions of
Section~\ref{subsec-hyp}. The leading asymptotics $\ln
\ddet{}{I+V}^{\pa{0}}$ of $\ln \ddet{}{I+V}$ as defined in
Theorem~\ref{theorem asymptotiques order zero log det} has
non-oscillating and oscillating corrections.

Let $0<\delta<q$ be such that the disks $D_{\pm q,\delta}$ of radius $\delta$ centered at $\pm q$ fulfill
$D_{\pm q,\delta}\subset \ov{U}$.
Let $\ov{\veps}=2 \sup_{\partial D_{q,\delta}\cup \partial
D_{-q,\delta}}\abs{\Re (\nu)}$. Then the first non-oscillating
corrections are of the form
\begin{equation}
 \frac{N_1}{x}+\e{O}\pa{\frac{1}{x^{2(1-\ov{\veps})}}},
\end{equation}
with
\begin{equation}
 N_1
 =i\sum_{\sg=\pm}\frac{\nu_\sg^2}{p'_\sg}\left\{
    2\sg \nu_\sg'\, \ln x
    + \sg \D{}{q}\ln u_\sg +  p'_\sg \D{}{q} \pa{\f{\nu_\sg}{p'_\sg}}
         - \f{\nu_{-\sg}}{q}
         \right\}.
\label{correction non oscillante log det}
\end{equation}
The first oscillating corrections are of the form
\begin{equation}
 \frac{O_1}{x^2}+\e{O}\pa{\frac{1}{x^{3(1-\ov{\veps})}}}.
\end{equation}
and the leading oscillating coefficient is given by
\begin{equation}
 O_1
 = \f{\nu_- \nu_+}{\pa{2q}^2 \, p_+'\,p_-'}
  \sum_{\sigma=\pm 1}
  \Big(\frac{u_+}{u_-}\Big)^\sigma x^{2\sigma(\nu_++\nu_-)}\,  \ex{i\sigma x \pa{p_+-p_-}
  }\;,
 \label{terme oscillant log det}
\end{equation}
where we have introduced
\begin{align}
 &u_+=e^{g(q)}\,\frac{\Ga(1-\nu_+)}{\Ga(1+\nu_+)}
       \left\{\frac{(2qp'_+)^{\nu_+}}{\kappa(q;q)}\right\}^2,\\
 &u_-=e^{g(-q)}\,\frac{\Ga(1+\nu_-)}{\Ga(1-\nu_-)}
       \left\{(2qp'_-)^{\nu_-}\kappa(-q;q)\right\}^{-2}.
\end{align}

\end{prop}

\begin{rem}
The GSK depends only on the combination
$ixp(\lambda)+g(\lambda)$ (see \eqref{epm}). Therefore the Fredholm
determinant and its asymptotics can only depend on this combination.
This observation allows us to obtain the complete asymptotic
expansion depending on the function $g(\lambda)$ from the asymptotic
expansion of the Fredholm determinant $I+V$ corresponding to $g=0$.
Namely, it is enough to replace in the obtained formulae
$p(\lambda)$ by $p(\lambda)-\frac ixg(\lambda)$ and then expand into
negative powers of $x$.
\end{rem}


It is quite interesting to apply the latter proposition in order to
obtain the first few terms of the asymptotic expansion of
$\ddet{}{I+V}$. The reason why we draw the reader's attention to
these asymptotics is because they present a very interesting
structure: the leading oscillating terms in the asymptotic expansion are
just given by the sum of the leading asymptotics evaluated at $\nu$ shifted by $1$ or $-1$.
This structure of the asymptotics seems to restore, at least partly,
the  original  periodicity $\nu\to\nu+n,~n\in\mathbb{Z}$, of the
Fredholm determinant of $I+V$.

\begin{cor}
\label{corollaire nu periodicite}
Let I+V be the GSK as above, $\ddet{}{I+V}^{\pa{0}}\pac{\nu}$ the
leading asymptotics of its Fredholm determinant just as in
Theorem~\ref{theorem asymptotiques order zero log det},  $N_1$ and
$O_1$ as in Proposition~\ref{proposition correstion en 1/x au
resolvent}. Note that we have emphasized the structure of
$\ddet{}{I+V}^{\pa{0}}\pac{\nu}$ as a functional of $\nu$. Then the
oscillating corrections $O_1$ can be reproduced from the
non-oscillating part via the shift of $\nu$ by $\pm1$:
\begin{equation}
\ddet{}{I+V}^{\pa{0}}\pac{\nu}  \frac{O_1}{x^2} = \ddet{}{I+V}^{\pa{0}}\pac{\nu+1}
+\ddet{}{I+V}^{\pa{0}}\pac{\nu-1}\;.
\label{terme oscillant et non oscillant du det}
\end{equation}
\end{cor}

This structure of the first  terms of the large $x$ asymptotic
expansion for $\ddet{}{I+V}$ leads us to raise the following
conjecture on the structure of the asymptotic series :
\begin{conj}
The asymptotic expansion of the Fredholm determinant $\ddet{}{I+V}$
of the GSK  restores the  periodicity $\nu\to\nu+n,~n\in\mathbb{Z}$,
of the determinant. In particular, this asymptotic expansion
contains all the $\mathbb{Z}$-periodized terms with respect to $\nu$
of the leading asymptotics $\ddet{}{I+V}^{\pa{0}}\pac{\nu}$. Thus, all the oscillating terms can be deduced from the
non-oscillating ones. More precisely, let
\begin{equation}
\mc{A}\pac{\nu}\pa{x} \sim  \ddet{}{I+V}^{\pa{0}}\pac{\nu} \pa{1+ \f{C_1\pa{\ln x}\pac{\nu}}{x}+\dots+ \f{C_M\pa{\ln x}\pac{\nu}}{x^M}+ \dots   }
\end{equation}
stand for the formal asymptotic series corresponding to the non-oscillating part of the asymptotic series for $\ln\ddet{}{I+V}$. There $C_k\pa{X}\pac{\nu}$ are polynomials of degree $k$ in $X$ whose coefficients are functionals in $\nu$. Moreover each of the $C_k$'s has no oscillating exponents of the type $\ex{\pm i xp_{\pm}}$. Then the formal asymptotic series for $\ddet{}{I+V}$ is given by
\begin{equation}
\ddet{}{I+V} \sim \sul{n \in \mathbb{Z}}{} \mc{A}\pac{\nu+n}\pa{x} \; .
\end{equation}
\end{conj}
This conjecture is supported by \eqref{terme oscillant et non oscillant du det} and also by the results of
\cite{McCoyTangSineKernelSubleadingFromPainleveV} where several sub-leading corrections to the
asymptotics of the  Fredholm determinant of the pure sine-kernel were computed.

The first application of the asymptotic behaviour of the GSK we consider in this article concerns the asymptotic inversion of
truncated Wiener--Hopf operators. We will prove in Section~\ref{section applications} the following
proposition:
\begin{prop}
Let $I+K$ be a truncated Wiener--Hopf operator on $\intoo{-x}{x}$, acting on functions $g\in L^2(\mathbb{R})$ as
\begin{equation}
\pac{\pa{I+K}.g} \pa{t}=g\pa{t} + \Int{-x}{x} K\pa{t-t'} g\pa{t'}
\dd t'\; .
\label{operateur de Wiener--Hopf a interval symetrique}
\end{equation}
The kernel $K$ is defined by its Fourier transform $F$:
\begin{equation}
 K\pa{t}=\mc{F}^{-1}\pac{F}\pa{t} \; ,
\end{equation}
and we suppose that there exists $\delta>0$ such that
\begin{itemize}

\item $F$ admits an analytic continuation to $\paa{z: \abs{\Im (z)} \leq \de}$;

\item  $\xi \mapsto F\pa{\xi \pm i\de}
\in L^{1}\pa{\R}$;

\item the analytic continuation of $1+F$ never vanishes for $\abs{\Im (z)} \leq \de$.
\end{itemize}
Then the resolvent $I-R$ of $I+K$ fulfills
%
%
\begin{equation}
R\pa{\la,\mu}= \Int{\R}{} \f{\dd \xi\, \dd \eta}{4\pi^2 i} \, F(\xi)\,
\paa{\f{\a_+(\eta)}{\a_-(\xi)} \ex{ix (\xi-\eta) }
-\f{\a_+(\xi)}{\a_-(\eta)} \ex{-ix (\xi-\eta)} } \f{\ex{i\pa{\mu\eta-\la\xi}}}{\xi-\eta}
 + \e{O}\pa{\ex{-2 \de x}},
\label{Resolvent Wiener--Hopf Asymptotique}
\end{equation}
%
where $\alpha(\lambda)$ is given by
\begin{equation}\label{Sol-alpha}
 \a\pa{\la}
 =\exp\paa{-\f{1}{2 i \pi }\Int{\R}{} \ln (1+ F(\mu))\, \f{\dd\mu}{\mu-\la}}.
\end{equation}
\end{prop}

Our main motivation to study the asymptotics of $\ln\ddet{}{I+V}$ comes
from the theory of one dimensional quantum integrable models.
Indeed, the generating function of the zero temperature
two-point correlation functions (at distance $x$) of different quantum integrable models \cite{KitKMST009}
has a series expansion in terms of cycle integrals of the type
\begin{equation}
\mc{I}_n\pac{\mc{F}_n}=\oint\limits_{\Gamma\pa{\intff{-q}{q}}}^{}
\hspace{-3mm}\f{\dd^n z}{\pa{2i\pi}^n}
\Int{-q}{q}\f{\dd^n \la}{\pa{2i\pi}^n}\;
  \mc{F}_n\pa{\ba{c} \paa{\la} \\ \paa{z} \ea  }
\pl{j=1}{n}\f{\ex{ix\pa{p\pa{z_j}-p\pa{\la_j}}  }}{\pa{z_j-\la_j}\pa{z_j-\la_{j+1}}}  \; .
\label{integrale multiple cycle definition}
\end{equation}
Here the function $\mc{F}_n$ is holomorphic in
some open neighbourhood of $\intff{-q}{q}^{2n}$ and symmetric in the
$n$ variables $\paa{\la}$ (we set $\lambda_{n+1}\equiv\lambda_1$)
and in the $n$ variables $\paa{z}$; $\Ga\pa{\intff{-q}{q}}$ is a counter clockwise closed contour around $\intff{-q}{q}$ inside this neighbourhood.

In Section \ref{sec-integrales}, using the above results for the GSK,
we prove the following asymptotic expansion of $\mc{I}_n\pac{\mc{F}_n}$ in the $x\tend +\infty$ limit :
\begin{prop}
Let $\mc{F}_n$ and $\mc{I}_n\pac{\mc{F}_n}$ be as above. Then for $x\tend +\infty$,
\begin{multline}
\mc{I}_{n}\pac{\mc{F}_n}
=\f{1}{2i\pi} \Int{-q}{q} \dd\la \,
\paa{ixp'(\la) + \partial_{\eps}}
\mc{F}_n\pa{\ba{c} \paa{\la}^{n} \\ \{\la+\eps\},\paa{\la}^{n-1} \ea}
\pour{\eps=0}
\\
+ \sul{\sg=\pm}{} \pa{b_n-c_n\ln\pa{2qp'_{\sg}x}}
\mc{F}_n\pa{\ba{c} \paa{\sg q}^{n} \\ \paa{\sg q}^{n} \ea}
 \\
 +\f{n}{\pa{2\pi}^2} \sul{\sg=\pm}{} \sul{p=1}{n-1}\Int{-q}{q} \dd \la
\f{\mc{F}_n\pa{\ba{c}\paa{\sg q}^{n} \\ \paa{\sg q}^{n} \ea }- \mc{F}_n\pa{\ba{c}\paa{\sg q}^{p} , \paa{\la}^{n-p}\\
\paa{\sg q}^{p} , \paa{\la}^{n-p}\ea } }{p\pa{n-p}\pa{q-\sg \la}}  \\
+ \f{n}{2\pa{2\pi}^2}\sul{p=1}{n-1}\Int{-q}{q} \f{\dd \la \dd \mu }{\pa{n-p}\pa{\la- \mu}}
\left\{
\partial_{\eps}\mc{F}_n\pa{ \ba{c}\paa{\la+\eps},\paa{\la}^{p-1},\paa{\mu}^{n-p}\\ \paa{\la+\eps} ,\paa{\la}^{p-1},\paa{\mu}^{n-p}\ea}\right.
                  \\
\left.
-\partial_{\eps}\mc{F}_n\pa{\ba{c} \paa{\mu+\eps} ,\paa{\mu}^{p-1},\paa{\la}^{n-p} \\ \paa{\mu+\eps},\paa{\mu}^{p-1},\paa{\la}^{n-p} \ea}
\right\}\pour{\eps=0}+\e{o}\pa{1},
\label{forme fonctionnelle de H}
\end{multline}
with
 \begin{equation}\label{5-bncn}
 c_n=\frac{(-1)^{n-1}}{(n-1)!}\left.\frac{\partial^n\nu_0^2}{\partial\gamma^n}
 \right|_{\gamma=0},\quad b_n=\frac{(-1)^{n-1}}{(n-1)!}
 \left.\frac{\partial^n\log G\left(1,\nu_0\right)}{\partial\gamma^n}
 \right|_{\gamma=0}, \quad \nu_0=\frac i{2\pi}\log(1+\gamma),
 \end{equation}
and where $\{\la\}^n$ denotes the set formed by $n$ copies of the same parameter $\la$.
\end{prop}

Moreover, in Section~\ref{sec-integrales} we will also describe the form of the sub-leading  corrections to this result.

\subsection{Comparison with known results}

There are several results in the literature concerning the
asymptotic behaviour of the Fredholm determinant $\ddet{}{I+\ga S}$.
This determinant corresponds to the GSK with $p=\e{id}$, $F=1$ and
$g=0$. It is clear that we reproduce the answer concerning the
leading asymptotics of
 $\ddet{}{I+\ga S}$ analyzed in \cite{BudynBuslaevPureGammaSineKernelAsympt} and
 \cite{BasorTracyProblemsWithTauFunctionAndSineKernel}.

As observed in \cite{JimMiwaMoriSatoSineKernelPVForBoseGAz}, $x
\D{}{x} \ln \ddet{}{I+\ga S}$ satisfies the fifth Painlev\'e
equation. The authors of
\cite{JimMiwaMoriSatoSineKernelPVForBoseGAz} used this property to
obtain an asymptotic expansion of $\ln \ddet{}{I+\ga S}$.  This fact was also exploited by the authors of
\cite{McCoyTangSineKernelSubleadingFromPainleveV} in order to
derive the first few terms in the sub-leading asymptotics of the
latter quantity. Their result reads
\begin{multline}
 x \D{}{x} \ln \det [I+\ga S]
 = -4i x \nu_0 -2\nu_0^2 - i\f{\nu_{{}_0}^3}{x}
     \\
 + i\f{\nu_{{}_0}^2}{4x}
\paa{ \paf{\Gamma\pa{-\nu_0}}{\Gamma\pa{\nu_0}}^2
\pa{4x}^{4\nu_{{}_0}} \ex{4ix} -
 \paf{\Gamma\pa{-\nu_0}}{\Gamma\pa{\nu_0}}^2  \f{\ex{-4ix}}{\pa{4x}^{4\nu_{{}_0}} }},
\label{reponse de Mc}
\end{multline}
with $\nu_0$ given in \eqref{5-bncn} and $q=2$. It is
straightforward to see that in such a limit $N_1 = i \nu_0^3$ and
\begin{equation}
O_1  \tend  \f{\nu_{{}_0}^2}{\pa{2q}^2} \paa{ \ex{2iqx}
\pa{2qx}^{4\nu_{{}_0}} \paf{\Gamma\pa{-\nu_0}}{\Gamma\pa{\nu_0}}^2 +
\f{\ex{-2iqx}}{\pa{2qx}^{4\nu_{{}_0}} }
\paf{\Gamma\pa{\nu_0}}{\Gamma\pa{-\nu_0}}^2} \; ,
\end{equation}
which reproduces the oscillating terms \eqref{reponse de Mc} after setting $q=2$ and taking the $q$ derivative.
%

%
\section{The initial Riemann--Hilbert problem}
\label{sec-initialRHP}
%
%
The GSK \eqref{generalised Sine} belongs to a special algebra of
integral operators, the so-called integrable integral operators.
This algebra was first singled out in
\cite{ItsIzerginKorepinSlavnovDifferentialeqnsforCorrelationfunctions}
and then studied more thoroughly in
\cite{DieftItsZhouSineKernelOnUnionOfIntervals}. It is well known
that many properties of these integrable operators can be obtained
from the solution of a certain RHP.

In this section, we formulate our problem in terms of a RHP
that we then asymptotically solve.

\subsection{Notations}
An important property of completely integrable integral operators is
that their resolvent still lies in the same algebra. However, before
presenting the formula for the resolvent we introduce some quite
useful vector notations. Namely, let
\begin{equation}
\ket{E^R\pa{\la}}=\f{\gamma \sqrt{F\pa{\la}}}{2i\pi}
                            \begin{pmatrix}
                                e_+\pa{\la}\\
                                e_-\pa{\la}
                            \end{pmatrix}, \quad
\bra{E^L\pa{\la}}=\sqrt{F\pa{\la}}
                           \begin{pmatrix}
                               -e_-\pa{\la}\, , \,
                                e_+\pa{\la}
                            \end{pmatrix},
\end{equation}
\noindent so that the kernel $V$ has a simple expression in terms of $\ket{E^{R}\pa{\la}}$ and $\bra{E^L\pa{\la}\!}$:
\begin{equation}
V\pa{\la,\mu}=\f{\braket{E^L\pa{\la}}{E^R\pa{\mu}}}{\la-\mu} \;.
\end{equation}
Observe that
 \begin{equation}\label{Orthog}
  \braket{E^L\pa{\la}}{E^R\pa{\la}}=0,
  \end{equation}
and, hence, the kernel $V$ is not singular at $\lambda=\mu\,$.

Let $\ket{F^R\pa{\la}}$ be the solution to the integral equation:
\begin{equation}
\ket{F^R\pa{\mu}}+\Int{-q}{q}  V\pa{\la,\mu} \ket{F^{R}\pa{\la}} \dd
\la= \ket{E^{R}\pa{\mu}}\;,
\end{equation}
\noindent and $\bra{F^L\pa{\la}}$ be the solution to the corresponding dual equation.
It is convenient to write $\ket{F^R\pa{\la}}$ as well as its dual $\bra{F^L\pa{\la}}$ in a form similar to
$\ket{E^R\pa{\la}}$ and $\bra{E^L\pa{\la}}$:
\begin{equation}
\ket{F^R\pa{\la}}=\f{\gamma \sqrt{F\pa{\la}}}{2i\pi}
 \begin{pmatrix}
 f_+\pa{\la}\\
 f_-\pa{\la}
 \end{pmatrix} , \quad
\bra{F^L\pa{\la}}= \sqrt{F\pa{\la}}
 \begin{pmatrix}
 -f_-\pa{\la} \, , \,
  f_+\pa{\la}
 \end{pmatrix}.
\end{equation}
Then the resolvent of the kernel $V$ defined by $I - R = (I + V)^{-1}$ reads:
\begin{equation}
R(\la,\mu)= \f{\braket{F^L\pa{\la}}{F^R\pa{\mu}}}{\la-\mu}= \f{\gamma\sqrt{F(\la)F(\mu)}}{2i\pi \pa{\la-\mu}}
\pac{f_{+}(\la)f_{-}(\mu)-f_{+}(\mu)f_{-}(\la)}.
\label{Reconstruction du Resolvent}
\end{equation}
%
%

\subsection{The Riemann--Hilbert problem associated to the GSK}

\begin{figure}[h]

\begin{pspicture}(3,2)
\end{pspicture}
\begin{pspicture}(8,2)
\psline{*-*}(2,1)(6,1) \psline[linewidth=2pt]{->}(4,1)(4.1,1)
\rput(1.8,1.3){$-q$}

\rput(6.2,0.7){$q$}

\rput(4,1.3){$+$}

\rput(4,0.7){$-$}

\end{pspicture}

\caption{Original contour for the RHP.\label{Contour du RHP pour chi}}
\end{figure}
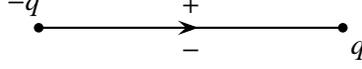

\begin{prop}
\label{definition RHP chi}
Let $V$ be the GSK \eqref{generalised Sine} understood as acting on
$L^2\pa{\intff{-q}{q}}$, and such that $\ddet{}{I+V}\not=0$. Then,
there exists a $2\times 2$ matrix $\chi\pa{\la}$ such that
\begin{equation}
\ket{F^{R}\pa{\la}}= \chi\pa{\la}\ket{E^{R}\pa{\la}}, \qquad  \bra{F^{L}\pa{\la}}=
\bra{E^{L}\pa{\la}} \chi^{-1}\pa{\la}.
\label{reconstruction F en fonction E}
\end{equation}
The matrix $\chi\pa{\la}$ is the unique solution of the RHP:
\begin{itemize}
\item $\chi$ is analytic on  $\Cx \setminus\intff{-q}{q}$  ;
\item $\chi(\la) = \e{O}\left(\ba{cc} 1 & 1 \\
                                 1 & 1 \ea   \right)
                                 \ln\abs{\la^2-q^2}
                                 \quad \text{for}\ \la \tend \pm q$;
\item $\chi(\la) \underset{\la \tend \infty}{\tend} I_2=\pa{ \ba{c c} 1& 0 \\ 0 & 1 \ea }$ ;
\item $\chi_{+}\pa{\la} G_{\chi}\pa{\la}=\chi_-\pa{\la}  \quad \text{for} \  \la \in \intoo{-q}{q} \;\; .$
\end{itemize}
The jump matrix $G_\chi$ for this RHP reads
\begin{equation}
 G_\chi\pa{\la} = \left(\ba{cc}
                   1-\gamma F\pa{\la} & \gamma F\pa{\la} e_+^{2}\pa{\la} \\
                   -\gamma F\pa{\la} e_-^{2}\pa{\la} & 1+\gamma F\pa{\la}
                   \ea \right)
 = I+2i\pi \ket{E^R\pa{\la}}\bra{E^L\pa{\la}} \; .
\end{equation}
Finally, $\chi$ and its inverse can be expressed in terms of $\ket{F^R\pa{\la}}$ and of its dual $\bra{F^L\pa{\la}\!}$:
\begin{equation}
\chi(\la)=I_2-\Int{-q}{q}
\f{\ket{F^R(\mu)}\bra{E^L(\mu)}}{\mu-\la}\, \dd \mu , \quad
\chi^{-1}(\la)=I_2+\Int{-q}{q}
\f{\ket{E^R(\mu)}\bra{F^L(\mu)}}{\mu-\la}\, \dd \mu .
\label{forme explicite de chi en terms f plus moins}
\end{equation}
\end{prop}
We emphasize that the big $\e{O}$ symbol, $\e{O}\left(\ba{cc} 1 & 1 \\ 1 &
1 \ea   \right)$, is to be understood entrywise. Moreover,
$\chi_{\pm}\pa{\mu}$ stands for the non-tangential limit of
$\chi\pa{\la}$ when $\la$ approaches a point $\mu$ belonging to the
jump curve from the left, resp. right, side of the contour (see
Fig.~\ref{Contour du RHP pour chi}).

\Proof The unicity of the solution to this RHP is proved along the same line as in
\cite{KuilajaarsMVVUniformAsymptoticsForModifiedJacobiOrthogonalPolynomials}.
The proof of existence of the solution is based on the
equivalence between RHP and  singular integral equations which, in
the case of the above RHP, implies
\begin{equation}
\chi\pa{\la}=I_2+ \Int{-q}{q} \f{\dd \mu}{\la-\mu} \chi_+\pa{\mu} \ket{E^R\pa{\mu}}\bra{E^L\pa{\mu}\!} \quad, \;\;
 \la \in \mathbb{C}\setminus\intff{-q}{q} \; .
\end{equation}
The solution to this equation can be expressed in terms
of the resolvent kernel $I-R$ of $I+V$
\begin{equation}
\chi\pa{\la}=I_2+\Int{-q}{q} \f{\dd \mu}{\la-\mu}
\paa{\ket{E^R}.\pa{I-R}}\pa{\mu}\bra{E^L\pa{\mu}}\;.
\label{forme explicite de chi}
\end{equation}

In its turn, the resolvent kernel exists as $\ddet{}{I+V}\not=0$. Moreover, the explicit construction of the resolvent
through a Fredholm series shows that $\pa{\la,\mu}\mapsto R\pa{\la,\mu}$ is analytic in $\ov{U}\times \ov{U}$. Hence,
so is $\ket{F^R\pa{\mu}}=\pac{\ket{E^R}.\pa{I-R}}\pa{\mu}$.
The estimate $\abs{\chi}=O\pa{\ln\abs{\la^2-q^2}}\; , \; \la \tend \pm q$ follows from the integral
representation \eqref{forme explicite de chi} supplemented with the fact that both $\bra{E^L}$ and $\ket{F^R}$ are
smooth on $\intff{-q}{q}$.

Applying (\ref{forme explicite de chi en terms f plus moins}) to
$|E^R(\lambda)\rangle$ and $\langle E^L(\lambda)|$ we obtain the
equations (\ref{reconstruction F en fonction E}). Hereby one can
easily check that due to the orthogonality condition (\ref{Orthog})
the transform (\ref{reconstruction F en fonction E}) is continuous
across $[-q,q]$.  \qed

It is also possible to express logarithmic derivatives of $\ddet{}{I+V}$ either in terms of the resolvent $R$ of $I+V$
or in terms of $\chi$. Indeed, we have the
\begin{lemme}
The derivative of $\ln \ddet{}{I+V}$ with respect to $x$ is related to the following trace involving the matrix $\chi$
\begin{equation}
\Dp{x} \ln \ddet{}{I+V}= \oint\limits_{\Ga\pa{\intff{-q}{q}}}{}
\hspace{-3mm}\f{\dd \la}{4\pi} p(\la)\,
\e{tr}\pac{\Dp{\la}\chi\pa{\la}\sg_3\chi^{-1}\pa{\la}},
\label{the x derivative path}
\end{equation}
with $\sg_3=\begin{pmatrix}
 1&0\\0&-1
\end{pmatrix}$ and $\Ga\pa{\intff{-q}{q}}$ defined as in (\ref{integrale multiple cycle definition}),
whereas its derivatives with respect to  $\gamma$ and $q$ are
expressed in terms of the resolvent as
\begin{equation}
\Dp{\ga} \ln \ddet{}{I+V}= \Int{-q}{q} \f{\dd \la}{\ga}
R\pa{\la,\la},\quad 
 \Dp{q} \ln \ddet{}{I+V}=R\pa{q,q} + R\pa{-q,-q}.
\label{the q and gamma derivative path}
\end{equation}
\end{lemme}

\Proof
The last two equations are easily proved by the multiple integral series expansion of $\ln\ddet{}{I+V}$.
We shall only focus on the equation relating the $x$ derivative of $\ln \ddet{}{I+V}$ to $\chi$. Clearly
\begin{equation}
\Dp{x}\ln \ddet{}{I+V}
  =  \Int{-q}{q} \pac{\Dp{x}V.\pa{I-R}}\pa{\la,\la} \dd\la \; ,
\end{equation}
with
\begin{equation}
\Dp{x}V\pa{\la,\mu}=
-\oint\limits_{\Ga\pa{\intff{-q}{q}}}{} \f{\dd z}{4\pi} \f{p\pa{z}}{\pa{z-\la}\pa{z-\mu}} \bra{E^{L}\pa{\la}}\sg_3\ket{E^R\pa{\mu}}.
\end{equation}
So that, using the representation \eqref{Reconstruction du
Resolvent} of the resolvent $R$ in terms  of $\bra{F^L}$ and
$\ket{F^R}$ and the fact that
$\bra{F^L(\mu)}\,\sg_3\,\ket{F^R(\la)}=\tr \left[ \sg_3
\ket{F^R(\la)}\bra{F^L(\mu)}\right]$, we get
\begin{multline}
\Dp{x}\ln \ddet{}{I+V}
= -\oint\limits_{\Ga\pa{\intff{-q}{q}}}{}\hspace{-4mm} \f{\dd z}{4\pi}\, p\pa{z}
\Int{-q}{q} \dd \la \, \f{\bra{E^{L}\pa{\la}}\sg_3\ket{E^R\pa{\la}}}{\pa{z-\la}^2} \\
 \hspace{1.5cm}+\e{tr}\paa{\;\oint\limits_{\Ga\pa{\intff{-q}{q}}}{} \hspace{-4mm}\f{\dd z}{4\pi}\, p\pa{z} \Int{-q}{q} \dd \la \dd \mu\,
\ket{F^R\pa{\la}}\bra{E^L\pa{\la}} \right.
 \\
\times \left. \pa{ \f{1}{\la-z}-\f{1}{\la-\mu} } \sg_3\f{\ket{E^R\pa{\mu}}\bra{F^L\pa{\mu}} }{\pa{\mu-z}^2}
 \vphantom{\oint\limits_{\Ga\pa{\intff{-q}{q}}}{}} }.
\end{multline}
Using the integral expressions \eqref{forme explicite de chi en
terms f plus moins} for $\chi$ and $\chi^{-1}$, we obtain
\begin{align}
\Dp{x}\ln \ddet{}{I+V}
&= - \hspace{-3mm} \oint\limits_{\Ga\pa{\intff{-q}{q}}}{}\hspace{-4mm} \f{\dd z}{4\pi}\, p\pa{z} \Int{-q}{q} \dd \la \, \f{\bra{E^{L}\pa{\la}}\sg_3\ket{E^R\pa{\la}}}{\pa{z-\la}^2}
            \nonumber\\
&\qquad
+\e{tr}\paa{\;\oint\limits_{\Ga\pa{\intff{-q}{q}}}{} \hspace{-4mm}\f{\dd z}{4\pi}\, p(z) \Int{-q}{q} \dd \mu
\pa{\chi(\mu)-\chi(z)} \sg_3\f{\ket{E^R(\mu)}\bra{F^L(\mu)} }{\pa{\mu-z}^2} }
            \nonumber\\
 &=\oint\limits_{\Ga\pa{\intff{-q}{q}}}{} \hspace{-4mm}\f{\dd
z}{4\pi} \, p\pa{z} \e{tr}\paa{ \Dp{z}\chi\pa{z} \sg_3
\chi^{-1}\pa{z}}, \hspace{3cm}
\end{align}
where we used \eqref{reconstruction F en fonction E}.
\qed

It is worth noticing that formula \eqref{the x derivative path} is
particularly effective when $p$ is a rational function as then the
contour of integration can be deformed to the poles of $p$  (including the pole at $\infty$). The integrals can be then easily calculated. In particular, in the
case $p\pa{\la}=\la$, we have the following result:
\begin{cor}
Let $\chi_{{}_1}$ be the first non-trivial coefficient of the
expansion of $\chi$ around $\infty$, i.e.
\begin{equation}
\chi\pa{\la}=I_{2}+\f{\chi_{{}_1}}{\lambda}+\e{o}\paf{1}{\lambda}.
\end{equation}
Then
\begin{equation}
\Dp{x}\ln \ddet{}{I+V}\mid_{p=\e{id}}=-\f{i}{2} \e{tr}\paa{\chi_1 \sg_3}.
\label{the x derivative for MSK}
\end{equation}
\end{cor}

\Proof $-\e{tr}\pa{\sg_3 \chi_1}$ is the residue of the pole at infinity  of $\lambda
\mapsto \lambda\, \e{tr}\paa{ \Dp{\la}\chi(\lambda)\, \sg_3
\chi^{-1}(\lambda)}$. \qed

In this way, we recover one of the formulae derived for the 
sine kernel \cite{DieftItsZhouSineKernelOnUnionOfIntervals}, but
also for more general kernels as in
\cite{ItsIK90,ItsIK90a,ItsIzerginkorepinVarguzin1991}. We emphasize  that \eqref{the x derivative
for MSK} is valid not only for the  sine kernel as it was
originally derived, but also for the generalised sine kernel with
$p=\e{id}$.

\section{Transformations of the original RHP}
\label{Section solution RHP}

In this section we perform several transformations on the RHP for
$\chi$ so as to implement Deift--Zhou's steepest descent method
\cite{DieftZhouSteepestDescentForOscillatoryRHP}. The first
substitution maps the RHP for the matrix $\chi$ into a RHP for a
matrix $\Xi$ whose jump matrix has 1 on its lower diagonal entry. This
jump matrix is then easily factorized into  upper/lower triangular matrices. This
factorization allows us to define another RHP for an unknown matrix
$\Upsilon$ whose jump matrices are already exponentially close to
identity uniformly away from the endpoints $\pm q$. It  remains
to construct the parametrices at $q$ and $-q$. These parametrices enable us to define a matrix $\Pi$ satisfying a RHP with jump matrices
uniformly $I_2+\e{o}\pa{1}$ when  $x\tend +\infty$.

\subsection{The first step $\chi \tend \Xi$}

Let
\begin{equation}
\a\pa{\la}=\exp\paa{ \Int{-q}{q}\f{\nu\pa{\mu}}{\mu-\la}\; \dd \mu} = \kappa\pa{\la} \paf{\la-q}{\la+q}^{\nu\pa{\la}} \; .
\end{equation}
Then clearly, $\a\pa{\la}$  solves the scalar RHP
\begin{equation}
 \a_-\pa{\la}=\a_+\pa{\la}(1+\gamma F\pa{\la}),\qquad
 \la\in \intff{-q}{q},\quad \alpha(\lambda)\to
 1\quad\mbox{at}\quad\lambda\to\infty\;.
\label{RHP for Alpha}
\end{equation}
The functions $\kappa\pa{\la}$ and $\nu\pa{\la}$ were already introduced in \eqref{nuit} and \eqref{def-nu}. In the following, we shall also use another representation for the
function $\alpha(\lambda)$ :
\begin{equation}\label{Sol-alpha-p}
 \a\pa{\la}
 =\kappa_p(\la)
\bigg[\f{p(\la)-p_{+}}{p(\la)-p_{-}}\bigg]^{\nu(\la)}\;,
\end{equation}
where $\kappa_p$ is defined as
\begin{equation}
 \ln \kappa_p(\la;q)\equiv\ln\kappa_p(\la)
 = \Int{-q}{q}  \left(\nu\pa{\la} \f{p'\pa{\mu} }{p\pa{\la}-p\pa{\mu}}
  -\f{\nu\pa{\mu}}{\la-\mu}\right) \dd \mu\;. \label{fonction kappa}
\end{equation}
We specify that we chose the principal branch of the logarithm, i.e.
$\e{arg} \in \intoo{-\pi}{\pi}$. Due to our assumptions on $\ga$,
$F$ and $p$, Morera's theorem implies that the functions $\nu$,
$\ln\kappa$ and $\ln\kappa_p$ are holomorphic on $\ov{U}$ . Moreover
we have $|\Re \bigl(\nu \pa{\la}\bigr)|< \tf{1}{2}$, $\forall \la \in
\ov{U}$. Indeed
\begin{equation}
\nu\pa{\la}=\f{i}{2\pi} \ln \abs{1+\ga F\pa{\la}}-\f{1}{2\pi} \e{arg}\pa{1+\ga F\pa{\la}} \;,
\end{equation}
and we have assumed that $\e{arg}\pa{1+\ga F}\in\intoo{-\pi}{\pi}$.

We use the function $\a$ to transform the RHP for $\chi$. Let us
define the matrix $\Xi\pa{\la}$ according to
\begin{equation}
\Xi\pa{\la} = \chi\pa{\la}
            \begin{pmatrix}
                    \a\pa{\la}& 0\\
                     0&  \a^{-1}\pa{\la}
            \end{pmatrix} .
\label{substitution 1 }
\end{equation}
This new matrix $\Xi\pa{\la}$ satisfies the following RHP:
\begin{itemize}
\item $\Xi$ is analytic on $\mathbb{C}\setminus\intff{-q}{q}$;
\item $\abs{\Xi(\la)} =  \e{O}\left(\ba{cc}
                                1 & 1 \\
                                1 & 1 \ea\right) \abs{\la^2-q^2}^{\pm \sg_3 \Re(\nu_{\pm})}
\ln\abs{\la^2-q^2} \quad$  for $\la \tend \pm q\; $;
\item $\Xi (\la)\underset{\la \tend \infty}{\tend} I_2 \; $;
\item $\Xi_{+}(\la)\, G_{\Xi}(\la)=\Xi_-(\la)   \quad\text{for}\ \la \in \intff{-q}{q} \; .$
\end{itemize}
Here the new jump matrix $G_\Xi$ reads
 \begin{equation}
 G_\Xi\pa{\la}=\left( \ba{cc}
                     1+P(\la)Q(\la) & P(\la) \ex{ix p(\la) }\\
                     Q(\la) \ex{-ix p(\la) }& 1
                    \ea\right),
\label{jump matrix for Phi}
\end{equation}
and
 \begin{align}
 &P(\la)
  =\f{\gamma F(\la)}{1+\gamma F(\la)}\,\a_+^{-2}(\la)\,\ex{g\pa{\la}}
  = -2i \ex{i\pi \nu\pa{\la}} \,
   \f{\sin \pi \nu(\la)}{\a_+^{2}\pa{\la}}\, \ex{g\pa{\la}} ,
      \\
 &Q(\la)
  =-\f{\gamma F(\la)}{1+\gamma F(\la)}\,\a_-^2(\la)\,\ex{-g\pa{\la}}
  = 2i \ex{i\pi \nu\pa{\la}} \,
  \f{\sin \pi \nu (\la)}{\ex{g\pa{\la}}}\, \a_-^{2}\pa{\la} .
\label{Definition de P et Q}
\end{align}

The solution of this RHP for $\Xi$ exists  as it can be
constructed from $\chi$. Moreover it is unique as seen by arguments
similar to those providing  uniqueness  of the solution to the RHP
for $\chi$.
%

\subsection{The second step $\Xi \tend \Upsilon $}

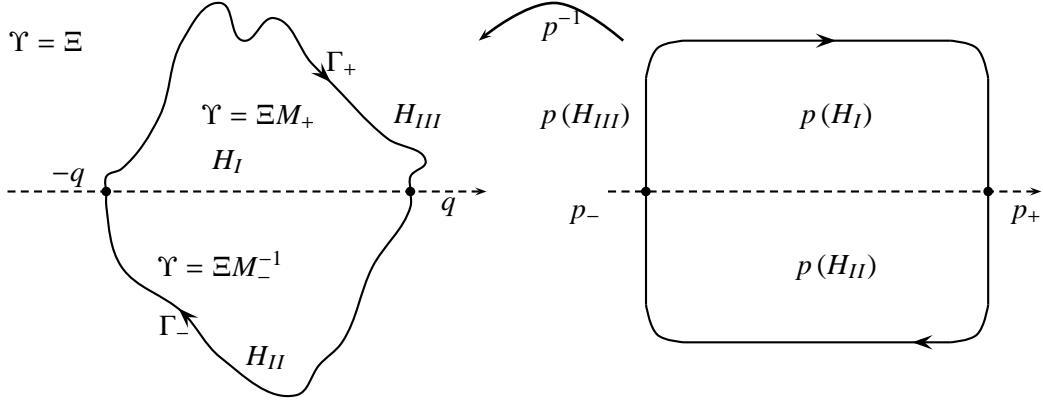
\begin{figure}[h]
\begin{center}

\begin{pspicture}(14.5,7)

\psline[linestyle=dashed, dash=3pt 2pt]{->}(0.2,4)(6.5,4)

\rput(1,4.2){$-q$}%
\psdots(1.5,4) \rput(6,3.8){$q$} \psdots(5.5,4)
\pscurve(1.5,4)(1.5,3.8)(1.7,3)(2.5,2.4)(3,1.8)(4,1.3)(4.3,1.7)(4.6,2)(5,3)(5.5,3.8)(5.5,4)
(5.5,4.2)(5.7,4.4)(5.2,4.7)(4.4,5.5)(4.2,5.7)(3.7,6.3)(3.3,6)(3,6.5)(2.5,6)(2.2,5)(1.7,4.3)(1.5,4.2)(1.5,4)

\psline[linewidth=2pt]{<-}(2.45,2.45)(2.55,2.35)
\rput(2.4,2.2){$\Gamma_-$}
\psline[linewidth=2pt]{<-}(4.45,5.45)(4.35,5.55)
\rput(4.6,5.7){$\Gamma_+$}

\rput(0.7,6){$\Upsilon=\Xi$} \rput(3.5,5){$\Upsilon=\Xi M_+$}
\rput(3,3){$\Upsilon=\Xi M_-^{-1}$}

\pscurve[linewidth=1pt.]{->}(8.2,6)(7.3,6.5)(6.3,6)
\rput(7.4,6.2){$p^{-1}$}

\psline[linestyle=dashed, dash=3pt 2pt]{->}(8,4)(13.7,4)
\psdots(8.5,4)(13,4) \rput(7.7,3.7){$p_{-}$}
\rput(13.5,3.7){$p_{+}$}

\psline{-}(8.5,2.5)(8.5,5.5)
\pscurve{-}(8.5,5.5)(8.65,5.9)(9,6)

\psline{-}(9,6)(12.5,6)
\pscurve{-}(12.5,6)(12.85,5.9)(13,5.5)

\psline{-}(13,5.5)(13,2.5)
\pscurve{-}(13,2.5)(12.85,2.1)(12.5,2)

\psline{-}(12.5,2)(9,2)
\pscurve{-}(9,2)(8.65,2.1)(8.5,2.5)


\rput(11,5){$p\pa{H_I}$}

\rput(11,3){$p\pa{H_{II}}$}
\rput(7.7,5){$p\pa{H_{III}}$}


\rput(3,4.4){$H_I$}

\rput(3.5,1.8){$H_{II}$}
\rput(5.5,5){$H_{III}$}

\psline[linewidth=2pt]{<-}(12,2)(12.1,2)
\psline[linewidth=2pt]{<-}(11,6)(10.9,6)

\end{pspicture}

\caption{Contours $\Gamma_+$ and
$\Gamma_-$ associated with the RHP for $\Upsilon$.\label{contour pour le RHP de Y}}
\end{center}
\end{figure}

As already mentioned, the jump matrix $G_{\Xi}$ admits an explicit
factorization into a product of  upper and  lower triangular
matrices:
\begin{equation}
G_{\Xi}=M_+\, M_-  \;.
\label{factorization of GPhi}
\end{equation}
The matrices $M_\pm$ are given by
\begin{equation}
 M_+\pa{\la}= \left(\ba{cc}
                             1& P\pa{\la}\ex{ix p\pa{\la}}\\
                             0& 1 \ea\right),
\qquad   M_-\pa{\la}= \left(\ba{cc}
                                    1& 0\\
                                     Q\pa{\la} \ex{-ixp\pa{\la}}& 1 \ea \right),
\label{matrixM_+M-}
\end{equation}
and can be continued to $U\cap {\mathcal H}_+$, resp. $U\cap {\mathcal
H}_-$, where we recall that ${\mathcal H}_{\pm}$ is the upper/lower
half plane and $\ov{U}$ is the domain of holomorphy of all the
functions appearing in the RHP. Then we draw two new contours
$\Gamma_\pm$ in $p\pa{U}$ and define a new matrix $\Upsilon(\la)$
according to Fig.~\ref{contour pour le RHP de Y}.

As readily checked, $\Upsilon\pa{\la}$ is continuous across $\intoo{-q}{q}$ and thus holomorphic in
the interior of $\Gamma_+\cup \Gamma_-$. We have thus removed the cut along $\intff{-q}{q}$ and
replaced it with cuts along $\Gamma_+ \cup \Gamma_-$.

The matrix $\Upsilon$ solves the following RHP:
\begin{itemize}
\item $\Upsilon$ is analytic in $\mathbb{C}\setminus \Gamma_+\cup\Gamma_-$ ;
\item $\Upsilon(\la)
= \e{O} \begin{pmatrix}
  1&1 \\
  1&1
     \end{pmatrix}
     \begin{pmatrix}
  \abs{\la \mp q}^{\pm \Re(\nu_{\pm}) }& \abs{\la\mp q}^{\mp \Re(\nu_{\pm}) }\\
   0    &  \abs{\la \mp q}^{\mp \Re(\nu_{\pm})}
     \end{pmatrix}
\ln\abs{\la\mp q}  , \quad  \la \underset{ \la \in H_I }{\longrightarrow} \pm q  $;
\item $\Upsilon(\la)
 = \e{O} \begin{pmatrix}
  1&1 \\
  1&1
     \end{pmatrix}
     \begin{pmatrix}
           \abs{\la \mp q}^{\pm \Re(\nu_{\pm})} & 0  \\
           \abs{\la\mp q}^{\pm \Re(\nu_{\pm}) }    &  \abs{\la \mp q}^{\mp \Re(\nu_{\pm})}
     \end{pmatrix}
\ln\abs{\la\mp q}  , \quad   \la \underset{ \la \in H_{II} }{\longrightarrow} \pm q  $;
\item $\Upsilon(\la)
  = \e{O} \begin{pmatrix}
  1&1 \\
  1&1
     \end{pmatrix}
 \abs{\la \mp q}^{\pm \sg_3\Re(\nu_{\pm})}
\ln\abs{\la\mp q}  , \quad  \la \underset{ \la \in H_{III} }{\longrightarrow} \pm q  $;
\item $\Upsilon(\la) \underset{\la \tend \infty}{\tend} I_2$;
\item $\left\{ \ba{ll} \Upsilon_{+}\pa{\la} M_+\pa{\la}=\Upsilon_-\pa{\la}   \quad
  &\text{for}\ \la \in \Gamma_+ \, , \\
                      \Upsilon_{+}\pa{\la} M_-^{-1}\pa{\la}=\Upsilon_-\pa{\la}   \quad
  &\text{for}\ \la \in \Gamma_-\, ,
                \ea \right. $
\end{itemize}
where the domains $H_{I}$, $H_{II}$, $H_{III}$ are shown on the
Figure \ref{contour pour le RHP de Y}.

Clearly, the solution of the RHP for $\Upsilon$ exists and is unique.
 Hence, the matrices $\Upsilon$ and $\chi$ are in a one to one
correspondence.

Note that, except in some vicinities of $q$ and $-q$,
the jump matrices $M_+$ and $M_-^{-1}$ for $\Upsilon$ are exponentially close to the identity matrix. Therefore, to study the asymptotic solution of the RHP, it is enough to study the local problems in the vicinities of $q$ and $-q$.

\subsection{Parametrix around $-q$}

\begin{figure}[h]
\begin{center}

\begin{pspicture}(8.2,5)
\pscurve{-}(3,0.8)(3.5,1)(4,2.35)(4,2.4)(4,2.5)(4,2.6)(4,2.65)(4.6,4.4)
\pscircle(4,2.5){2}
\psline[linewidth=2pt]{->}(6,2.5)(6,2.6)

\rput(4.3,2.5){$-q$}
\psdots(4,2.5)

\rput(3.7,3.5){$\Gamma_+$}
\psline[linewidth=2pt]{->}(4.15,3.2)(4.2,3.32)

\rput(4.4,1.5){$\Gamma_-$}
\psline[linewidth=2pt]{->}(3.89,1.7)(3.94,1.85)

\psline{->}(7.1,2.5)(8,2.5) \rput(8,2.2){$\Re (\la)$}

\psline{->}(7.1,2.5)(7.1,3.4) \rput(6.7,3.2){$\Im (\la)$}

\end{pspicture}

\caption{Contours in the RHP for $\mc{P}$.\label{contours for the RHP for P}}
\end{center}
\end{figure}
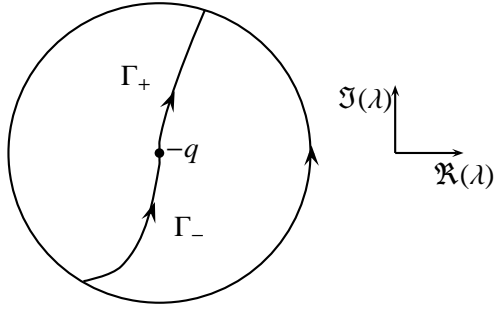

We first present the parametrix $\mc{P}$ on a small disk
$D_{-q,\delta}\subset U$ of radius $\delta$ and centered at $-q$,
that is an exact solution of the RHP:
\begin{itemize}
\item $\mc{P}$ is analytic on $D_{-q,\delta}\setminus \paa{\Gamma_+\cup\Gamma_-} \; ;$

\item $\mc{P}(\la) = \e{O} \left(\ba{cc}
                                1&1 \\
                                1&1 \ea\right)
\pa{\ba{cc} \abs{\la + q}^{-\Re(\nu_{-})} & \abs{\la + q}^{ \Re(\nu_{-}) }\\
            0    &  \abs{\la + q}^{ \Re(\nu_{-})} \ea  }
\ln\abs{\la+q}  , \;  \la \underset{ \la \in H_I }{\longrightarrow}
-q  \; ;$
\item $\mc{P}(\la) = \e{O} \left(\ba{cc}
                                1&1 \\
                                1&1 \ea\right)
\pa{\ba{cc} \abs{\la + q}^{- \Re(\nu_{-})} & 0  \\
            \abs{\la + q}^{- \Re(\nu_{-}) }    &  \abs{\la + q}^{ \Re(\nu_{-})} \ea  }
\ln\abs{\la+ q}  , \;  \la \underset{ \la \in H_{II}
}{\longrightarrow}\! - q \; ;$
\item $\mc{P}(\la) = \e{O} \left(\ba{cc}
                                1&1 \\
                                1&1 \ea\right) \abs{\la + q}^{-\sg_3\Re(\nu_{-})}
\ln\abs{\la+ q}  , \quad  \la \underset{ \la \in H_{III} }{\longrightarrow} - q  \; ; $
\item $\mc{P}(\la)= I_2+ \e{O}\pa{\f{1}{x^{1-\veps}}}, \quad \text{uniformly for } \la \in \partial D_{-q,\delta} $,
\item $\left\{ \ba{ll}
   \mc{P}_{+}\pa{\la} M_+\pa{\la}=\mc{P}_-\pa{\la} \quad
   &\text{for }\la \in \Gamma_+ \cap D_{-q,\delta}, \\
   \mc{P}_{+}\pa{\la} M_-^{-1}\pa{\la}=\mc{P}_-\pa{\la} \quad
   &\text{for }\la \in \Gamma_- \cap D_{-q,\delta}.
                \ea \right.$
\end{itemize}
Here $\veps= 2 \!\!\underset{\partial
D_{-q,\delta}}{\sup}\!\!\abs{\Re (\nu)}< 1$. The canonically
oriented contour $\partial D_{-q,\delta}$ is depicted in
Fig.~\ref{contours for the RHP for P}.
The RHP for $\mc{P}$ admits a class of solutions.  Each element of
this class is related to another one through a left multiplication
by a holomorphic matrix that is uniformly
$I_2+O\pa{\tf{1}{x^{1-\veps}}}$ on $\partial D_{-q,\delta}$. In
order to construct the solution $\mc{P}$ to this problem, we first
focus on the simpler case where the functions $F$, $g$ and
$\kappa_p$ are constant. Then the solution to the RHP for
$\mc{P}_{const}$ can be obtained by the differential equation method
\cite{Its81,DeiIZ93,DeiZ94}. This leads to the solution
\begin{equation}
\mc{P}_{const}\pa{\la}=\Psi\pa{\la  } L\pa{\la} \pac{\zeta_{-q}}^{-\nu\sg_3} \ex{\f{i\pi\nu}{2}}\,\; . \;\;
\label{parametrice en -q}
\end{equation}
Here $\zeta_{-q}=x\pa{p\pa{\la}-p_{-}}$,~ $\nu=\tf{
i\log\pa{1+\ga F }}{2\pi}$,
\begin{equation}
\Psi\pa{\la}= 
              \begin{pmatrix}
                        \Psi\pa{-\nu,1;-i\zeta_{-q}}   &ib_{12}\, \Psi\pa{1+\nu,1;i\zeta_{-q}} \\
                        -ib_{21}\, \Psi\pa{1-\nu,1;-i\zeta_{-q}}  & \Psi\pa{\nu,1;i\zeta_{-q}}
              \end{pmatrix}
%
,
\end{equation}
and finally
\begin{align}
b_{12}\pa{\la}&=-i \f{\sin \pac{\pi \nu}\Gamma^{2}\pa{1+\nu} }{\pi \kappa_p^2 \pac{x\pa{p_{+}-p\pa{\la} }}^{2\nu}}
\ex{ixp_{-}+g }   ,\\
b_{21}\pa{\la}&= -i\f{\pi \kappa_p^{2} \pac{x\pa{p_{+}-p\pa{\la}}}^{2\nu}}
                        {\sin\pac{\pi \nu} \Gamma^{2}\pa{\nu}}
            \ex{-ixp_{-}-g} \;\; .
\end{align}
$\Psi\pa{a,c;z}$ denotes Tricomi confluent hypergeometric function
(CHF) of the second kind (see Appendix~\ref{App-A}). It solves the differential equation
\begin{equation}\label{ODE}
zy'' + \pa{c-z} y' -a y=0 \;.
\end{equation}
Recall that $\Psi$ has a cut along $\R^-$. Note that this choice for the cut of $\Psi$ implies the use of the principal branch
of the logarithm: $-\pi < \e{arg}\pa{z} < \pi$.
The expression for the piecewise constant matrix $L$ depends on the
region of the complex plane. Namely,
\begin{equation}
L\pa{\la}= \left\{\ba{cc}  I_2                            & -\tf{\pi}{2}<\e{arg}\pac{p\pa{\la}-p_{-}}< \tf{\pi}{2} \vspace{4mm},\\
                  \left( \ba{cc} 1  &0 \\
                                0& \ex{-2i\pi \nu}
                                    \ea \right)  & \tf{\pi}{2}<\e{arg}\pac{p\pa{\la}-p_{-}}< \pi \vspace{4mm},\\
                  \left( \ba{cc} \ex{-2i\pi \nu}  &0 \\
                                     0& 1
                                    \ea \right)  & -\pi<\e{arg}\pac{p\pa{\la}-p_{-}}< -\tf{\pi}{2}.
        \ea \right.
\label{matrice L constante}
\end{equation}

The reader can check  using the monodromy properties of Tricomi CHF
\eqref{cut-Psi-1} and \eqref{cut-Psi-2} that the jump condition for
constant functions $F$ and $g$ are satisfied by the matrix
$\mc{P}_{const}$. Moreover the asymptotic expansion for
$\Psi\pa{a,c;z}$ at $z \tend \infty$ allows one to check that
$\mc{P}_{const}$ has the correct behaviour at infinity. We remind that this parametrix also appeared recently in the work \cite{ItsK2008}.

In order to extend this result to the case of  arbitrary holomorphic functions
$F\pa{\la}$, $g\pa{\la}$ and $\kappa_p\pa{\la}$, it is enough to add
the $\la$ dependency in all places where these functions appear. One
ends up with the following solution to the RHP for $\mc{P}$ :
\begin{equation}
\mc{P}\pa{\la}=\Psi\pa{\la  } L\pa{\la} \pac{\zeta_{-q}}^{-\nu\pa{\la}\sg_3} \ex{\f{i\pi\nu\pa{\la}}{2}}\,\; . \;\;
\label{parametrice en -q1}
\end{equation}
Here $\zeta_{-q}=x\pa{p\pa{\la}-p_{-}}$,
\begin{equation}
\Psi\pa{\la}= 
              \begin{pmatrix}
                        \Psi\pa{-\nu\pa{\la},1;-i\zeta_{-q}}   &ib_{12}(\la)\, \Psi\pa{1+\nu\pa{\la},1;i\zeta_{-q}} \\
                        -ib_{21}(\la)\, \Psi\pa{1-\nu\pa{\la},1;-i\zeta_{-q}}  & \Psi\pa{\nu\pa{\la},1;i\zeta_{-q}}
              \end{pmatrix}
%
,
\end{equation}
with
\begin{align}
b_{12}\pa{\la}&=-i \f{\sin \pac{\pi \nu\pa{\la}}\Gamma^{2}\pa{1+\nu\pa{\la}} }{\pi \kappa_p^2\pa{\la} \pac{x\pa{p_{+}-p\pa{\la} }}^{2\nu\pa{\la}}}
\ex{g\pa{\la}+ixp_{-} }  = -i \nu\pa{\la} u\pa{\la;x} ,\\
b_{21}\pa{\la}&= -i\f{\pi \kappa_p^{2}\pa{\la} \pac{x\pa{p_{+}-p\pa{\la}}}^{2\nu\pa{\la}}}
                        {\sin\pac{\pi \nu\pa{\la}} \Gamma^{2}\pa{\nu\pa{\la}}}
            \ex{-ixp_{-}-g\pa{\la}}= -i \f{\nu\pa{\la}}{u\pa{\la;x}},
\end{align}
and finally
\begin{equation}
u\pa{\la;x}=\f{\Gamma\pa{1+\nu\pa{\la}}}{\Gamma\pa{1-\nu\pa{\la}}}
    \left\{\kappa_p(\la)\,x^{\nu(\la)} \pac{p_{+}-p\pa{\la}}^{\nu\pa{\la}}
  \right\}^{-2}e^{ix p_-+g(\la)}.
\label{definition de u}
\end{equation}

In the above formulae we have explicitly stressed the dependence of
the functions $b_{12}$, $b_{21}$ and $\nu$ on $\la$. Finally, the matrix $L\pa{\la}$ is given by \eqref{matrice L constante} with $\nu$ replaced by the function $\nu\pa{\la}$.

This construction originates from the observation that the
replacements $F \mapsto F\pa{\la}$, $g\mapsto g\pa{\la}$ and
$\kappa_p \mapsto \kappa_p\pa{\la}$ preserve the jump conditions as
the latter hold pointwise. Of course, once the parametrix $\mc{P}$
is guessed it is not a problem to check directly that it solves the
RHP in question. The asymptotic behaviour is inferred from
\eqref{asy-Psi}, whereas the jump conditions can be verified thanks
to \eqref{cut-Psi-1} and \eqref{cut-Psi-2}. Furthermore, due to the
definition of the matrix $L$, the solution is continuous across the
line $\e{arg}\pac{p\pa{\la}-p_{-}}=\pi$ and thus analytic in the
whole domain $\paa{\la\in\mathbb{C};
\Re\pac{p\pa{\la}-p_{-}}<0}$.

\subsection{Parametrix around $q$}

The RHP for the parametrix $\wt{\mc{P}}$ around $q$ reads
\begin{itemize}
\item $\wt{\mc{P}}$ is analytic on $D_{q,\de}\setminus \paa{\Gamma_+\cup\Gamma_-} \; ;$

\item $\wt{\mc{P}}(\la) = O \left(\ba{cc}
                                1&1 \\
                                1&1 \ea\right)
\pa{\ba{cc} \abs{\la - q}^{+\Re(\nu_{+})} & \abs{\la - q}^{- \Re(\nu_{+}) }\\
            0    &  \abs{\la - q}^{ - \Re(\nu_{+})} \ea  }
\ln\abs{\la-q}  , \quad  \la \underset{ \la \in H_I }{\longrightarrow} q  \; ;$
\item $\wt{\mc{P}}(\la) = O \left(\ba{cc}
                                1&1 \\
                                1&1 \ea\right)
\pa{\ba{cc} \abs{\la - q}^{+\Re(\nu_{+})} & 0  \\
            \abs{\la - q}^{+ \Re(\nu_{+}) }    &  \abs{\la - q}^{ -\Re(\nu_{+})} \ea  }
\ln\abs{\la- q}  , \quad  \la \underset{ \la \in H_{II} }{\longrightarrow}  q  \; ;$
\item $\wt{\mc{P}}(\la) = O \left(\ba{cc}
                                1&1 \\
                                1&1 \ea\right) \abs{\la - q}^{\sg_3\Re(\nu_{+})}
\ln\abs{\la- q}  , \quad  \la \underset{ \la \in H_{III} }{\longrightarrow}  q  \; ; $
\item $\wt{\mc{P}}(\la)= I_2+ O\pa{\f{1}{x^{1-\wt{\veps}}}}\quad \text{uniformly for } \la \in \partial D_{q,\delta} $;
\item $\left\{ \ba{ll}
 \wt{\mc{P}}_{+}\pa{\la} M_+\pa{\la}=\wt{\mc{P}}_-\pa{\la}  \
 &\text{for }\la \in \Gamma_+ \cap D_{q,\de}, \\
  \wt{\mc{P}}_{+}\pa{\la} M_-^{-1}\pa{\la}=\wt{\mc{P}}_-\pa{\la} \
 &\text{for }\la \in \Gamma_- \cap D_{q,\de}.
                \ea \right. $
\end{itemize}
and $\wt{\veps}=2\sup_{\partial D_{q,\delta}}\abs{\Re(\nu)}<1$ .
\begin{figure}[h]
\begin{center}

\begin{pspicture}(8.2,5)
\pscurve{-}(3,0.8)(3.5,1)(4,2.35)(4,2.4)(4,2.5)(4,2.6)(4,2.65)(4.6,4.4)
\pscircle(4,2.5){2}
\psline[linewidth=2pt]{->}(6,2.5)(6,2.6)

\rput(4.3,2.5){$q$}
\psdots(4,2.5)

\rput(3.7,3.5){$\Gamma_+$}
\psline[linewidth=2pt]{<-}(4.15,3.2)(4.2,3.32)

\rput(4.4,1.5){$\Gamma_-$}
\psline[linewidth=2pt]{<-}(3.89,1.7)(3.94,1.85)

\psline{->}(7.1,2.5)(8,2.5) \rput(8,2.2){$\Re (\la)$}

\psline{->}(7.1,2.5)(7.1,3.4) \rput(6.7,3.2){$\Im (\la)$}

\end{pspicture}

\caption{Contours in the RHP for $\wt{\mc{P}}$.\label{Contours for RHP for P tilde}}
\end{center}
\end{figure}
The solution of the RHP for the parametrix $\wt{\mc{P}}$ around $q$ can be formally obtained from the one at
$-q$ through the transformation $q \rightarrow -q$ and $\nu \tend -\nu$ on the solution to the RHP for $\mc{P}$.
 Indeed, the two RHP are identical modulo this negation.

Just as for the parametrix around $-q$, we focus on the  solution
\begin{equation}
\wt{\mc{P}}\pa{\la}=\wt{\Psi}\pa{\la} \wt{L}\pa{\la} \zeta_q^{\nu\pa{\la}\sg_3}
\ex{-\f{i\pi\nu\pa{\la}}{2}},
\end{equation}
where $\zeta_q=x\pac{p\pa{\la}-p_{+}}$, and
\begin{equation}
\widetilde{\Psi}\pa{\la}= \left( \ba{cc}
                        \Psi\pa{\nu\pa{\la},1;-i\zeta_q}   &i\tilde{b}_{12}\pa{\la} \Psi\pa{1-\nu\pa{\la},1;i \zeta_q} \\
                        -i\tilde{b}_{21}\pa{\la} \Psi\pa{1+\nu\pa{\la},1;-i\zeta_q}  & \Psi\pa{-\nu\pa{\la},1;i \zeta_q} \ea \right)
.
\end{equation}
Here
\begin{align}
&\tilde{b}_{12}(\la)
 =i \f{\sin[\pi\nu(\la)]\,\Gamma^{2}(1-\nu(\la)) }
                         {\pi \kappa_p^2(\la)}
            [x(p(\la)-p_{-})]^{2\nu(\la)} \ex{g(\la)+ixp_{+} }
= i \nu(\la) \tilde{u}(\la;x), \nonumber \\
&\tilde{b}_{21}(\la)
 = i\f{\pi \kappa_p^{2}(\la)\, \ex{-g\pa{\la}-ixp_{+}}}
      {\sin[\pi \nu(\la)]\, \Gamma^{2}(-\nu(\la))\, [x(p(\la)-p_{-})]^{2\nu(\la)} }
 =i \f{\nu(\la)}{ \tilde{u}(\la;x) }, \nonumber
\end{align}
and
\begin{equation}
\tilde{u}\pa{\la;x}
   = \f{\Gamma\pa{1-\nu\pa{\la}}}{\Gamma\pa{1+\nu\pa{\la}}}\,
  \paa{ \f{x^{\nu(\la)}\pac{p\pa{\la}-p_{-}}^{\nu\pa{\la}} }{\kappa_p(\la)}}^{2}e^{ixp_++g(\la)}.
\label{definition de u tilde}
\end{equation}
Just as for the parametrix around $-q$, the matrix $\tilde
L\pa{\la}$ depends on the quadrant of the complex plane:
\begin{equation}
\tilde
L\pa{\la}=\left\{\ba{cc}  I_2                            & -\tf{\pi}{2}<\e{arg}\pac{p\pa{\la}-p_{+}}< \tf{\pi}{2} \vspace{4mm},\\
                  \left( \ba{cc} 1  &0 \\
                                0& \ex{2i\pi \nu\pa{\la}}
                                    \ea \right)  & \tf{\pi}{2}<\e{arg}\pac{p\pa{\la}-p_{+}}< \pi \vspace{4mm},\\
                  \left( \ba{cc} \ex{2i\pi \nu\pa{\la}}  &0 \\
                                     0& 1
                                    \ea \right)  & -\pi<\e{arg}\pac{p\pa{\la}-p_{+}}< -\tf{\pi}{2}.
        \ea \right.
\end{equation}
%

\subsection{The last transformation $\Upsilon\tend \Pi$}
\label{Section RHP Pour PI}
Let
\begin{equation}
\Pi\pa{\la}=\left\{  \ba{ll}
                    \Upsilon(\la) \wt{\mc{P}}^{-1}(\la) &\text{for } \la\in D_{q,\delta},\\
                    \Upsilon(\la) \mc{P}^{-1}(\la) &\text{for } \la \in D_{-q,\delta}, \\
                    \Upsilon(\la) &\text{for } \la \in \mathbb{C}\setminus\paa{\ov{D}_{q,\delta}\cup \ov{D}_{-q,\delta}}.
            \ea \right.
\end{equation}

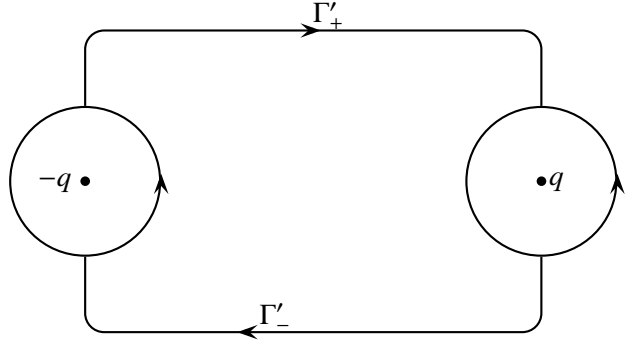
\begin{figure}[h]
\begin{center}

\begin{pspicture}(10,6.5)

\pscircle(2,4){1}
\psdots(2,4)
\rput(1.6,4){$-q$}
\psline[linewidth=2pt]{->}(3,4)(3,4.1)

\pscircle(8,4){1}
\psdots(8,4)
\rput(8.2,4){$q$}
\psline[linewidth=2pt]{->}(9,4)(9,4.1)

\psline[linearc=.25]{-}(2,5)(2,6)(8,6)(8,5)
\psline[linewidth=2pt]{->}(5,6)(5.1,6)
\rput(5.2,6.2){$\Gamma_+'$}

\psline[linearc=.25]{-}(2,3)(2,2)(8,2)(8,3)
\psline[linewidth=2pt]{<-}(4,2)(4.1,2)
\rput(4.5,2.2){$\Gamma_-'$}

\rput(3.5,1){$\Sg_\Pi=\Ga_-'\cup\Ga_+'\cup\Dp{}D_{-q,\de}\cup\Dp{}D_{q,\de}$}
\end{pspicture}

\caption{Contour $\Sg_\Pi$ appearing in the RHP for $\Pi$.\label{Contour for RHP for R}}
\end{center}
\end{figure}

Introduce the curve $\msc{C}=\paa{\Gamma_+\cup \Gamma_-} \cap
\paa{\ov{D}_{q,\delta}\cup \ov{D}_{-q,\delta}}$. Then $\Pi$ is
continuous across $\msc{C} \setminus\paa{q,-q}$. Since $\Pi$ is
holomorphic in a vicinity of $\msc{C}$, we have that $\Pi$ is
holomorphic in $D_{q,\delta}\cup D_{-q,\delta} \setminus
\paa{q,-q}$. This, in turn, due to the estimates for $\mc{P}$,
$\wt{\mc{P}}$ and $\Upsilon$ around the points $\pm q$, ensures that
the singularities at these points are of a removable type. Hence
$\Pi$ is  holomorphic on the disks $D_{q,\delta}\cup D_{-q,\delta}$.
Finally, we see that $\Pi$ satisfies the following RHP:
\begin{itemize}
\item $\Pi$ is analytic in $\Cx\setminus \Sg_\Pi$ (\textit{cf} Fig.~\ref{Contour for RHP for R}) ;
\item $\Pi\pa{\la} = I_2+ O\pa{\tf{1}{\la}} \quad \text{for }  \la \tend \infty $;
\item $\left\{ \ba{l l}
 \Pi_{+}\pa{\la} M_+\pa{\la}=\Pi_-\pa{\la}
        &\text{for }\la \in \Gamma'_+  ,\\
 \Pi_{+}\pa{\la} M_-^{-1}\pa{\la}=\Pi_-\pa{\la}
        &\text{for }\la \in \Gamma'_- ,\\
 \Pi_{+}\pa{\la} \mc{P}\pa{\la}=\Pi_-\pa{\la}
        &\text{for }\la \in \partial D_{-q,\delta}  ,\\
 \Pi_{+}\pa{\la} \wt{\mc{P}}\pa{\la}=\Pi_-\pa{\la}
        &\text{for }\la \in \partial D_{q,\delta} . \\
                \ea \right.$
\end{itemize}
The solution to the RHP for $\Pi$, exits and is unique as seen by standard arguments.

The jump matrices for $\Pi$ are uniformly exponentially close to
$I_2$ in $x$ on $\Gamma'_-\cup\Gamma'_+$ and uniformly
$I_2+\e{O}\pa{x^{\ov{\veps}-1}}$ on $\partial D_{q,\delta}\cup
\,\partial D_{-q,\delta}$, with $\ov{\veps}=2\sup_{\partial
D_{q,\delta}\cup \,\partial D_{-q,\delta}}\abs{\Re(\nu)}$. As a
consequence, $I_2$ is the unique solution of the RHP, up to
uniformly $\e{O}\pa{x^{\ov{\veps}-1}}$ corrections. In addition,
using the equivalence between singular integral equations and RHP,
the asymptotic expansion of $\Pi$ can be obtained by a Neumann
series. This will be done in the upcoming section.


\section{Asymptotic solution of the RHP}
\label{Section Developpement asym N}

In this Section we asymptotically solve the above RHP for $\Pi$.

We derive an asymptotic expansion into negative powers of $x$ for
the jump matrices for $\Pi$, and use it to prove the existence
of an asymptotic series for $\Pi$. The corresponding asymptotic series for $\chi$
follows readily. One can finally infer the asymptotic behaviour of
the resolvent of the GSK up to any order in
$1/x$.

\subsection{Asymptotics of the jump matrices}

Denote the jump matrices for $\Pi$ by $I_2+\Delta\pa{\la}$. Then the
matrix $\Delta\pa{\la}$ has the asymptotic expansion in the limit
$x\tend+\infty$:
\begin{equation}\label{expansion-Delta}
\De\pa{\la}=\sul{m=1}{M}\f{\De^{(m)}\pa{\la;x}}{x^m} \;\;+\;\;
\e{o}\pa{x^{-M-1+\ov{\veps}}},
\end{equation}
with $\ov{\veps}=2\sup_{\partial D_{q,\delta}\cup \,\partial
D_{-q,\delta}}\abs{\Re(\nu)}$.

The explicit form of the matrices $\De^{(n)}\pa{\la;x}$ depends on
the position of $\la$ in the contour $\Sg_\Pi$: they vanish to any
order in $\tf{1}{x}$ on $\Ga_+'\cup\Ga_-'$, whereas the asymptotic
expansion for $\De$ on $\Dp{}D_{q,\de}\cup D_{-q,\de}$ follows
promptly from the asymptotic expansion of Tricomi CHF
\eqref{asy-Psi}. More explicitly, for any $n \in \mathbb{N}^*$,

\begin{equation}
\Delta^{(n)}\pa{\la;x}=\left\{ \ba{c l}
                   \f{\De_{_{(-)}}^{(n)}\pa{\la;x}}{\pac{p\pa{\la}-p_{-}}^{n}}\,
         &\text{for } \la \in \partial D_{-q,\delta},\vspace{2mm}\\
                     \f{{\De}_{_{(+)}}^{(n)}\pa{\la;x}}{\pac{p\pa{\la}-p_{+}}^{n}}\,  &\text{for }
                    \la \in \partial D_{q,\delta}, \vspace{4mm}\\
                    0 &\text{for } \la \in \Ga_+' \cup \Ga_-'.
                    \ea \right.
\end{equation}
We have separated the jump matrices into their pole parts
$\pac{p\pa{\la}-p_\pm}^{-n}$ and
 regular parts $\De_{_{(-)}}^{(n)}$ and $\De_{_{(+)}}^{(n)}$, with
\begin{equation}
\De_{_{(-)}}^{(n)}(\la;x)=
                \f{i^n}{n! }
  \begin{pmatrix}
     1 & -\f{n\, u(\la;x)}{\nu(\la)} \\
     \f{n}{\nu(\la) u(\la;x)}    & 1
  \end{pmatrix}
  \begin{pmatrix}
    \pa{-1}^n \pa{-\nu(\la)}^2_n & 0 \\
    0 & \pa{\nu(\la)}^2_n
  \end{pmatrix}
\end{equation}
for $\la\in \partial D_{-q,\delta}$, and
\begin{equation}
\De_{_{(+)}}^{(n)}(\la;x)=
                \f{i^n}{n! }
  \begin{pmatrix}
     1 & \f{n\, \tilde{u}(\la;x)}{\nu(\la)} \\
     -\f{n}{\nu(\la) \tilde{u}(\la;x)}    & 1
  \end{pmatrix}
  \begin{pmatrix}
    \pa{-1}^n \pa{\nu(\la)}^2_n & 0 \\
    0 & \pa{-\nu(\la)}^2_n
  \end{pmatrix}
\end{equation}
for $\la\in \partial D_{q,\delta}$ . Here we use the standard
notation $\pa{\nu}_n=\tf{\Ga\pa{\nu+n}}{\Ga\pa{\nu}}$ and
$u\pa{\la;x}$, resp. $\tilde{u}\pa{\la;x}$, have been defined in
\eqref{definition de u}, resp. \eqref{definition de u tilde}. Thus,
the matrices $\De^{(n)}$ depend on $x$, but their entries
are a $\e{O}(x^{\ov{\veps}})$.

\subsection{Asymptotic expansion for $\Pi$}

Using the equivalence between RHP and singular integral equations we
can  express $\Pi$  in terms of its boundary value from the ``$+$''
side of the contour  $\Sigma_\Pi$
\begin{equation}\label{int-Pi}
\Pi\pa{\la}=I_2+\f{1}{2i\pi} \Int{\Sigma_\Pi}{} \Pi_{+}(s)\,
\Delta(s)\, \f{\dd s}{\la-s} \; .
\end{equation}
In its turn $\Pi_{+}(\la)$ belongs to  $L^{2}\pa{\Sg_\Pi}$ and
fulfills the linear singular integral equation of Cauchy type
\begin{equation}\label{Cauchy-IE}
\Pi_+(z)=I_2+ C^{+}_{\Sg_\Pi}\pac{\Pi_+ \Delta }(z)\;.
\end{equation}
Recall that the Cauchy operator on $L^{2}\pa{\Sg_\Pi}$ is defined as
\begin{equation}
C^{+}_{\Sg_\Pi}[g](z) = \lim_{t \tend z^+ }C_{\Sg_\Pi}[g](t) \qquad \text{and}\qquad
C_{\Sg_\Pi}[g](t)=\f{1}{2i\pi}\Int{\Sg_\Pi}{} \f{g(s)\, \dd s}{t-s}
,\;\ t\not\in \Sg_\Pi \;.
\end{equation}
The notation $t\tend z^+$ stands for the non-tangential limit of $t$
approaching $z$ from the $"+"$ side of the contour $\Sg_\Pi$. Recall
that the Cauchy operator is bounded: i.e. there exists a constant
$c_2$ such that, for any function $g\in L^{2}\pa{\Sg_\Pi}$, one has
$\norm{C_{\Sg_\Pi}^+\pac{g}}\leq c_2 \norm{g}$, where $\norm{.}$ is
the canonical $L^2\pa{\Sg_\Pi}$ norm.

The matrix $\Pi_+$ can be asymptotically approximated by the following series:

\begin{prop}
Let $\Pi_{+}^{(k)}$ be defined recursively according to
\begin{equation}
\Pi_{+}^{(k)}=\sul{p=1}{k} C^+_{\Sg_\Pi}\pac{\Pi_{+}^{(k-p)}\, \De^{(p)} } \quad\text{with}\quad \Pi_{+}^{(0)}=I_2.
\end{equation}
Then, for any integer $M>0$, there exists a constant $C(M)>0$ such that
\begin{equation}\label{asympt-Pi+}
\norm{\Pi_+-\sul{p=0}{M-1} x^{-p}\, \Pi_{+}^{(p)}} \leq
\f{C(M)}{x^{M\pa{1-\ov{\veps} }}}.
\end{equation}
\end{prop}

\Proof
Let us prove this statement by induction on $M$.
For $M=1$ we have that
\begin{equation}
\norm{\Pi_+ - I_2}=\norm{C_{\Sg_\Pi}^+\pac{\pa{\Pi_+-I_2}\De} +
C_{\Sg_\Pi}^+\pac{\De}  }
\leq c_2 \norm{\Pi_+-I_2}\norm{\De} + c_2 \norm{\De}.
\end{equation}
Therefore, for  $x$ large,
\begin{equation}
\norm{\Pi_+-I_2} \leq \f{c_2 \norm{\De}}{1-c_2 \norm{\De}} \leq
\f{C(1)}{x^{1-\ov{\veps}}}\;.
\label{norme N+ cas M=0}
\end{equation}

Let us now suppose that the result holds up to $M$. Then,
\begin{align}
\norm{\Pi_+ - \sul{k=0}{M}x^{-k}\, \Pi_{_{+}}^{(k)} }
&=\norm{C_{\Sg_\Pi}^+\pac{\Pi_+\De
     -\sul{k=1}{M}x^{-k}\sul{p=1}{k}\Pi_{_{+}}^{(k-p)} \De^{(p)} } }
                \nonumber\\
&\hspace{-2.6cm}
 =\norm{C_{\Sg_\Pi}^+\pac{\Pi_+\pa{\De-\sul{k=1}{M} x^{-k} \De^{(k)} }
        + \sul{p=1}{M}\pa{\Pi_+ -\sul{k=0}{M-p}\Pi_{_{+}}^{(k)}
                           x^{-k}} \De^{(p)} x^{-p} } }
                 \nonumber\\
&\hspace{-2.6cm} \leq c_2 \norm{\Pi_+} C_{\De_M}\,
x^{-M-1+\ov{\veps}}
    +\sul{p=1}{M} c_2\norm{\De^{(p)}}\, x^{-p}\, C(M-p)\, x^{\pa{p-M-1}\pa{1-\ov{\veps}}}
                 \nonumber\\
&\hspace{-2.6cm} \leq \f{C(M+1)}{x^{\pa{M+1}\pa{1-\ov{\veps}}}} \;,
\hspace{3cm}
\label{norme N+ cas M qcq}
\end{align}
for some constants $C_{\De_M}$ and $C(M+1)$. We used the fact that all $\De^{(p)}$ are in $L^2\pa{\Sg_\Pi}$ and that $\norm{\Pi_+}$ is bounded in virtue of \eqref{norme N+ cas M=0}.
\qed

Let us now extend this result for points $\lambda$ being uniformly
away from the contour $\Sg_\Pi$. Define the matrices
\begin{align}
 &\Pi^{(0)}(z)=I_2, \qquad
 \Pi^{(p)}(z)=\sul{k=1}{p}C_{\Sg_\Pi}\pac{\Pi_{+}^{(p-k)} \De^{(k)} }(z), \ \; p>0,
         \label{def-Pip}\\
 &\Pi(z;M)=\sum_{p=0}^M x^{-p}\,\Pi^{(p)}(z),
         \label{def-PiM}
\end{align}
which are analytic away from  $\Sg_\Pi$. Then we have the following
result:

\begin{prop}\label{prop-Pi}
Let $K$ be any compact subset of  $\; \Cx\setminus \Sg_\Pi$.
Then, $\forall k \in \mathbb{N}$, $\forall M \in \mathbb{N}^\star$,
\begin{equation}\label{asympt-Pi}
\abs{\Dp{\la}^k \Pi\pa{\la}
-\Dp{\la}^k \Pi(\la;M-1)
 }
\leq  \f{k!\; C(M)\ \e{lgth}(\Sg_\Pi) }{d(K,\Sg_\Pi)^{k+1}
x^{M\pa{1-\ov{\veps}}}}, \quad \la \in K .
\end{equation}
Here $\abs{.}$ denotes the usual max norm
$\abs{\Pi}\equiv{\max}_{i,j}\abs{\Pi_{i,j}}$, $d(.,.)$ is any distance on $\Cx$ and $\e{lgth}(\Sg_\Pi)$ is the length of the curve $\Sg_\Pi$.
\end{prop}

\Proof
Let $k\in \mathbb{N}$, $M\in \mathbb{N}^\star$, then
\begin{align*}
&\abs{\Dp{\la}^k \Pi(\la)- \sul{p=0}{M-1}x^{-p}\,\Dp{\la}^{k} \Pi^{(p)}(\la) }
  \nonumber\\
&\hspace{1.8cm} \leq \abs{  \f{k!}{2i\pi} \Int{\Sg_\Pi}{} \f{ \dd
s}{\pa{\la-s}^{k+1}}
\paa{ \Pi_+(s)\, \De(s) - \sul{p=1}{M-1}x^{-p} \sul{l=1}{p} \Pi^{(p-l)}_{+}(s)\, \De^{(l)}(s) }}
   \nonumber\\
&\hspace{1.8cm} \leq \f{k!\,C(M)\,
\e{lgth}(\Sg_\Pi)}{x^{M\pa{1-\ov{\veps}}}\, d^{k+1}\pa{K,\Sg_\Pi} }
\end{align*}
due to \eqref{asympt-Pi+}.
\qed

\subsection{The functions  $f_{\pm}$ to the leading order}
\label{sec-fpm-ordre0}

We now perform the transformations from $\Pi$ back to $\chi$.

The solution to the RHP of Proposition~\ref{definition RHP
chi} reads
\begin{equation}
 \chi(\la)=\Pi(\la)\; \chi^{(0)}(\la)\;.
\end{equation}
We call $\chi^{(0)}$  the zero$^{\e{th}}$ order solution (i.e.
obtained for $\Pi=I_2$). In the vicinities of the endpoints of
$\intff{-q}{q}$, $\chi^{(0)}$ is given as
\begin{equation*}
\chi^{(0)}(\la)
  =\left\{ \ba{ll}
   \mc{P}(\la)\, M_+^{-1}(\la)\,\a(\la)^{-\sg_3} ,
        &\ \la \in D_{-q,\delta}\cap \paa{0<\e{arg}[p(\la)-p_-]<\tf{\pi}{2}},
                \vspace{1mm} \\
   \wt{\mc{P}}(\la)\, M_{+}^{-1}(\la)\, \a(\la)^{-\sg_3},
        &\ \la \in D_{q,\delta}\cap \paa{\tf{\pi}{2}<\e{arg}[p(\la)-p_+]<\pi}. \ea \right.
\end{equation*}
Similarly, on $\intff{-q}{q}$, and uniformly away from the endpoints,
\begin{equation*}
\chi^{(0)}(\la)=M_+^{-1}(\la)\,\a_+(\la)^{-\sg_3},\qquad
             \la \in \intoo{-q+\de}{q-\de}\;.
\end{equation*}

In the  $\Im\pa{\la}=0^+$ limit and for
$\Re\pa{\la}\in\intff{-q}{q}$,
\begin{equation}
M_+^{-1}\a_{+}^{-\sg_3} \left(\ba{c} e_{+}\pa{\la} \\
                                 e_{-}\pa{\la} \ea\right) =
       \pa{\a_+e_{-}}^{-\sg_3} \left(\ba{c}  \ex{2i\pi\nu} \\
                                 1 \ea\right) ,
\end{equation}
so that, for $\la \in \intoo{-q+\de}{q+\de}\,$,
\begin{equation}
\left(\ba{c} f^{\pa{0}}_{+}\pa{\la} \\
            f^{\pa{0}}_{-}\pa{\la} \ea\right)=\ex{i\pi \nu\pa{\la}} \pac{\a_+^{-1}\pa{\la} e_+\pa{\la} \ex{i\pi\nu\pa{\la}} }^{\sg_3}
                        \left(\ba{c}  1 \\
                                 1 \ea\right) ,
\label{fplus moins bulk}
\end{equation}
where we have explicitly written all the dependence on $\la$.

 When  $\la \in \intff{-q}{-q+\delta}$,
we should multiply the latter expression by $\mc{P}$. Using the
decomposition \eqref{Phi s'ecrit comme Psi} of Humbert CHF into a
sum of two Tricomi CHF we get
\begin{equation}
\left(\ba{c} f^{\pa{0}}_{+}\pa{\la} \\
            f^{\pa{0}}_{-}\pa{\la} \ea\right)= \ex{\f{i\pi \nu}{2}}
            \paf{e_{+}{\pa{\la}}}{\kappa_p\pa{\la} \zeta_q ^{\nu}}^{\sg_3}
                    \left(\ba{c} \Gamma\pa{1+\nu} \Phi\pa{-\nu,1;-i\zeta_{-q}} \\
                                \Gamma\pa{1-\nu} \Phi\pa{\nu,1;i\zeta_{-q}}\ea\right) ,
\label{fplusmoins en -q}
\end{equation}
with $\zeta_q=x\pac{p_{+}-p\pa{\la}}$ and $\zeta_{-q}=x\pac{p\pa{\la}-p_{-}}$.

Analogously, for $\la\in \intff{q-\delta}{q}$,
\begin{equation}
\left(\ba{c} f^{\pa{0}}_{+}\pa{\la} \\
            f^{\pa{0}}_{-}\pa{\la} \ea\right)= \ex{\f{i\pi \nu}{2}}
            \paf{e_{+}{\pa{\la}}\zeta_{-q}^{\nu}}{\kappa_p\pa{\la} }^{\sg_3}
                    \left(\ba{c} \Gamma\pa{1-\nu} \Phi\pa{\nu,1;i\zeta_q} \\
                                \Gamma\pa{1+\nu} \Phi\pa{-\nu,1;-i\zeta_q}\ea\right).
\label{fplusmoins en q}
\end{equation}

Note that the piecewise expressions for the functions
$f^{\pa{0}}_{\pm}\pa{\la}$ are in fact analytic in a vicinity of
their respective domain of validity, although  they have been
obtained by taking  the limit   of $\la$ approaching  a point in
$\intff{-q}{q}$ from the upper half plane. More precisely, the
formula \eqref{fplusmoins en -q} holds on $D_{-q,\de}$,
\eqref{fplusmoins en q} on $D_{q,\de}$, and \eqref{fplus moins bulk}
on the connected component of the interior of $\Sg_{\pi}$ containing $\intff{\de-q}{q-\de}$. This observation follows from (\ref{Orthog}),
but  of course it can be checked by a direct computation
 based on the expression for the matrix $\chi$ in the lower half plane.
%
%

\subsection{Integral bounds for the resolvent}
We now introduce a function $R^{\pa{0}}\pa{\la,\mu}$  and show that it is a good approximation  of the resolvent in the sense that
\begin{equation}
\e{tr}\pa{R-R^{\pa{0}}}=\e{O}\pa{x^{\ov{\veps}-1}}\; .
\end{equation}
Such estimates are necessary for the integration of the $\ga$-derivative of $\log\ddet{}{I+V}$.
\begin{defin}
Let $\Pi_{\tau}\pa{\la}$ denote the solution of the RHP given in  Subsection~\ref{Section RHP Pour PI}
whose jumps are on circles of radius $\tau$ and on the corresponding curves $\Ga_+'$ and $\Ga_-'$.
\end{defin}
We can then write the solution of the RHP for $\chi$ as $\chi\pa{\la}=\Pi_{\tau}\chi^{\pa{0}}_{\tau}$. There
$\chi^{\pa{0}}_{\tau}$ do not depend explicitly on $\tau$. The radius $\tau$ only determines which patch we should use for the definition of the matrix $\chi_{\tau}^{\pa{0}}$. Moreover the whole combination $\Pi_{\tau}\chi^{\pa{0}}_{\tau}$ does not depend on the radius $\tau$ at all. Hence, we can represent the exact resolvent as
\begin{equation}
R\pa{\la,\mu}= \bra{E^L\pa{\la}} \pac{\chi^{\pa{0}}_{\tau}\pa{\la}}^{-1} \f{\Pi_{\tau}^{-1}\pa{\la} \Pi_{\tau}\pa{\mu}}{\la-\mu} \chi^{\pa{0}}_{\tau}\pa{\mu} \ket{E^R\pa{\mu}} \; .
\end{equation}
There, without altering the value of $R\pa{\la,\mu}$, we can chose different values of $\tau$ depending on the point we consider. This is quite useful as we can take one value of the radius $\tau$ in order to have estimates around
$\pm q$ and another one  to perform estimates in the bulk $\intff{\de-q}{q-\de}$. This will become clearer during the proof of the proposition below.

\begin{defin}
Let us fix $\de$, $q>\de>0$ and define what we call the diagonal zero$^{\e{th}}$ order resolvent
\begin{equation}
R^{(0)}\pa{\la,\la}=
\f{\ga F\pa{\la}}{2i\pi} \pa{\Dp{\la }f^{\pa{0}}_+ \pa{\la} f_-^{\pa{0}}\pa{\la} -\Dp{\la }f^{\pa{0}}_- \pa{\la} f_+^{\pa{0}}\pa{\la}},
\end{equation}
where the functions $f_{\pm}^{\pa{0}}\pa{\la}$ are given by
\eqref{fplus moins bulk}  for $\la \in \intff{\de-q}{q-\de}$,
\eqref{fplusmoins en -q} for $\la \in \intfo{-q}{\de-q}$ and
\eqref{fplusmoins en q} for $\la \in \intof{q-\de}{q}$. Similarly,
$\ket{F^{R;\pa{0}}\pa{\la}}$ and $\bra{F^{L;\pa{0}}\pa{\la}}$ are
defined in terms of the same functions $f_{\pm}^{\pa{0}}$.
\end{defin}
We stress that the radius $\tau$ previously introduced to build the
exact solution
$\Pi_{\tau}\hspace{-0.2mm}(\la)\chi^{\pa{0}}_{\tau}\hspace{-0.2mm}\pa{\la}$
and $\de$ appearing in the definition are, a priori, unrelated.
\begin{prop}\label{est-trace}
Let $R\pa{\la,\mu}$ be the exact resolvent of the generalised sine kernel. Then
\begin{equation}
\e{tr}\pa{R-R^{(0)}}=\e{O}\pa{x^{\ov{\veps}-1}},
\end{equation}
where the $\e{O}$ is uniform in $\ga \in D_{0,r}$.
\end{prop}

\Proof According to the preceding observations we have, for $\la \in \intfo{-q}{-q+\de} \cup \intof{q-\de}{q}$,
\begin{equation}
R\pa{\la,\la}=
        R^{(0)}\pa{\la,\la}+ \bra{F^{L;\pa{0}}\pa{\la}}
        \Pi_{2\de}\pa{\la}\Dp{\la}\Pi_{2\de}\pa{\la}\ket{F^{R;\pa{0}}\pa{\la}},
\end{equation}
and
\begin{equation}
  R\pa{\la,\la}=
   R^{(0)}\pa{\la,\la}+ \bra{F^{L;\pa{0}}\pa{\la}}
   \Pi_{\tf{\de}{2}}\pa{\la}\Dp{\la}\Pi_{\tf{\de}{2}}\pa{\la}\ket{F^{R;\pa{0}}\pa{\la}},
\end{equation}
for $\la \in \intff{\de-q}{q-\de} $.
The advantage of using two different matrices $\Pi$ for the
corrections of $R\pa{\la,\la}$ with respect to the zero$^{\e{th}}$
order resolvent $R^{(0)}\pa{\la,\la}$ is that the corrections are
always analytic on the whole domain where they are considered. One
does not need to take into account that $\Pi_{\de}\pa{\la}$ has a
jump across $\la=\pm\pa{q-\de}$. This might be problematic as, for
instance, the integral of $\Dp{\la}\Pi\pa{\la}$ on
$\intff{-q}{\de-q}$ might be ill-defined. Moreover the uniform
estimates that we have derived for the matrix $\Pi\pa{\la}$ only
hold uniformly away from the jump contour. As we will only
integrate the terms containing $\Pi_{2\de}$ on
$\intfo{-q}{\de-q}\cup\intof{q-\de}{q}$, we will be in this
situation. The same holds for the terms involving
$\Pi_{\tf{\de}{2}}$. However, we would not be able to use the
uniform estimates \eqref{prop-Pi} for $\Dp{\la}\Pi_{\de}$ when integrating it on
$\intff{-q}{\de-q}$, as we would not always be uniformly away from the boundary
of the jump contour for $\Pi_{\de}$.

With this way of understanding the corrections we have
\begin{multline}
\e{tr}\pa{R-R^{(0)}}=\Bigg(\Int{-q}{\de-q}+\Int{q-\de}{q}\Bigg) \dd \la\;
\bra{F^{L;\pa{0}}\pa{\la}}
\Pi_{2\de}\pa{\la}\Dp{\la}\Pi_{2\de}\pa{\la}\ket{F^{R;\pa{0}}\pa{\la}}\\
+\Int{\de-q}{q-\de} \dd \la \; \bra{F^{L;\pa{0}}\pa{\la}} \Pi_{\tf{\de}{2}}\pa{\la}\Dp{\la}\Pi_{\tf{\de}{2}}\pa{\la}\ket{F^{R;\pa{0}}\pa{\la}}.
\end{multline}

Let us start by the bulk part of integral, i.e. the part on
$\intff{\de-q}{q-\de}$. From the explicit form for
$f_{\pm}^{\pa{0}}$  on
$\intff{\de-q}{q-\de}$ given in \eqref{fplus moins bulk} we see that these functions are uniformly $\e{O}\pa{1}$.
Moreover, the uniform estimates for the matrices
$\Pi_{\tf{\de}{2}}\pa{\la}$ for $\la$ uniformly away from the jump contour
guarantee that
\begin{equation}
\bra{F^{L;\pa{0}}\pa{\la}}\Pi_{\tf{\de}{2}}\pa{\la}\Dp{\la}\Pi_{\tf{\de}{2}}\pa{\la}\ket{F^{R;\pa{0}}\pa{\la}}
=\e{O}\pa{x^{\ov{\veps}-1}},
\end{equation}
the $\e{O}\pa{x^{\ov{\veps}-1}}$ being uniform in $\ga$, at least
for $\ga$ small enough.

The situation at the boundaries is a little more complex. We only
consider the right boundary as the other case is treated similarly.
We still have that
$\Pi_{2\de}\pa{\la}=I_{2}+\e{O}\pa{x^{\ov{\veps}-1}}$ and
$\Dp{\la}\Pi_{2\de}\pa{\la}=\e{O}\pa{ x^{\ov{\veps}-1} }$ uniformly
on $\intff{q-\de}{q}$. However the functions
$f_{\pm}^{\pa{0}}\pa{\la}$ are no longer uniformly a $\e{O}\pa{1}$
on this interval. We should thus estimate the following integral
\begin{equation}
\sul{\sg,\sg'=\pm}{} \Int{q-\de}{q}
f^{\pa{0}}_{\sg}\pa{\la}f^{\pa{0}}_{\sg'}\pa{\la}
G_{\sg,\sg'}\pa{\la} \dd \la
\label{estimation correction resolvent en q}
\end{equation}
with $G_{\sg,\sg'}\pa{\la}=\e{O}\pa{x^{\ov{\veps}-1}}$ being related
to the entries of
$\Pi_{\tf{\de}{2}}\pa{\la}\Dp{\la}\Pi_{\tf{\de}{2}}\pa{\la}$. The
situation being similar for all the possible choices of $\sg$ and
$\sg'$, we explain the mechanism for $\pa{\sg,\sg'}=\pa{+,+}$. The
asymptotics of Humbert CHF guarantees that
\begin{equation}
\Phi\pa{a,1;\pm i t} = \f{c_{\pm}}{\abs{t}^{a}}\pa{1+\e{o}\pa{1}} \;\; t \tend +\infty
\end{equation}
for some computable constants $c_{\pm}$ depending on $a$. These constants are continuous with respect to $a$ belonging to an open neighbourhood of $\nu\pa{\intff{q-\de}{q}}$, and so is the $\e{o}\pa{1}$ term. Hence, there exist an $a$ independent constant $C$ such that
\begin{equation}
\abs{\pa{1+\abs{t}}^{\Re\pa{a}} \Phi\pa{a,1;\pm i t}  } \leq C \;\; .
\end{equation}
Indeed  the latter function is continuous on $\R$ and has a finite
limit at $\infty$. Moreover the constant $C$ can be chosen in such a
way that the estimate holds for $a$ belonging to some small vicinity
of $\nu\pa{\intff{q-\de}{q}}$. Hence, by explicitly extracting the
$x^{\ov{\veps}-1}$ factor coming from $G_{+,+}\pa{\la}$ we get that, 
for some constant $C'$,
\begin{equation}
\abs{f_{+}^{\pa{0}}\pa{\la}f_{+}^{\pa{0}}\pa{\la}G_{+,+}\pa{\la}}
\leq C'x^{\ov{\veps}-1} \varphi_{x}\pa{p\pa{\la}-p_{+}}\; ,
\end{equation}
with $\varphi_x\pa{t}=x^{2\Re\pa{\nu\pa{\la}}} \pa{  1+ x\abs{t}  }^{-2\Re\pa{\nu\pa{\la}}}$. The function $\varphi_x\pa{t}$ fulfills
\begin{equation}
\abs{\varphi_x\pa{p\pa{\la}-p_{+}}} \leq \tilde{C}\abs{p\pa{\la}-p_{+}}^{-2\Re\pa{\nu\pa{\la}}}
\end{equation}
as, for any $\a \in \R$, $t\mapsto \tf{t^{\a}}{\pa{1+t^{\a}}}$ is bounded. The latter function is integrable on $\intff{q-\de}{q}$ (we consider the case $\abs{\Re\pa{\nu}}<\tf{1}{2}$).
 Thus the integrals in
\eqref{estimation correction resolvent en q} do eventually yield
$\e{O}\pa{x^{\ov{\veps}-1}}$ contributions. \qed

One can prove, in a very similar way, the estimates for the
Hilbert--Schmidt norm of the resolvent. Namely,
\begin{prop}
Under the assumptions of the previous proposition,
\begin{equation}
\norm{ R-R^{(0)}}_2=\e{O}\pa{x^{\ov{\veps}-1}}
\end{equation}
with $\norm{.}_2$ being the Hilbert--Schmidt norm.
\end{prop}

\subsection{Asymptotic expansion of the resolvent}

We now prove that the asymptotic expansion for $\Pi$ can be used to
obtain an asymptotics expansion for the diagonal of the resolvent
$R(\la,\la)$. We derive point-wise bounds for the latter as this quantity appears in the
$q$-derivative of the Fredholm determinant:
\begin{equation}
\Dp{q}\ln\ddet{}{I+V}=R\pa{q,q}+R\pa{-q,-q} \;.
\end{equation}
We need to estimate the error when we replace the exact resolvent $R$ by  the approximate one $R^{\pa{0}}$. The magnitude for the error term follows from the following result:
\begin{prop}
\label{proposition DA du resolvent}
Let $\chi^{\pa{0}}$ be the solution of the RHP for $\chi$ up to the
leading order in $x$, that is to say the one obtained from $\Pi=I_2$ and corresponding to the contour $\Sg_{\Pi}$
with disks $D_{\pm q, \de}$ having radius $\de$.
Define the leading  vectors $\bra{F^{L;\pa{0}}}$  and
$\ket{F^{R;\pa{0}}}$ as
\begin{equation}
\bra{F^{L;\pa{0}}\pa{\la}}= \bra{E^{L}\pa{\la}}\,\chi^{\pa{0}}(\la)^{-1}, \qquad
\ket{F^{R;\pa{0}}\pa{\la}}= \chi^{\pa{0}}(\la)\,
\ket{E^{R}\pa{\la}}\;,
\end{equation}
and the leading order of the resolvent by
\begin{equation}\label{LO-resolvent}
 R^{(0)}\pa{\la,\mu}=\frac{\langle F^{L;\pa{0}}\pa{\la}
\ket{F^{R;\pa{0}}\pa{\mu}} }{\lambda-\mu}\;.
\end{equation}
Then
\begin{equation}\label{developpement Asymptotique resolvent}
R\pa{\la,\la}= R^{(0)}\pa{\la,\la}
+\sul{p=1}{k} \f{R^{(p)}\pa{\la,\la}}{x^p}
+\e{O}\paf{x^{\ov{\veps}}}{x^{\pa{k+1}\pa{1-\ov{\veps} }}}\; ,
\end{equation}
for $\la$ uniformly away from $\Sg_\Pi$ and belonging to $\intff{-q}{q}$. Here,
\begin{align}
&R^{(0)}\pa{\la,\la}
  = - \braket{F^{L;\pa{0}}\pa{\la}}{ \Dp{\la} F^{R;\pa{0}}\pa{\la}} ,
        \label{R0}\\
&R^{(p)}\pa{\la,\la}
  = - \bra{F^{L;\pa{0}}\pa{\la}}\, \hat{\Pi}^{(p)}(\la) \,
      \ket{F^{R;\pa{0}}\pa{\la}},
 \quad p>0,
         \label{Rp}
\end{align}
in which
\begin{equation}\label{dev-Pik}
\Pi^{-1}\pa{\la;k}  \Dp{\la} \Pi\pa{\la;k}
 =\sul{p=1}{k} \hat{\Pi}^{(p)}(\la)\, x^{-p}
 + \e{O}\paf{1}{x^{\pa{k+1}\pa{1-\ov{\veps}}}}  .
\end{equation}
\end{prop}

\Proof
Clearly,
\begin{align*}
R(\la,\mu)
 &= \f{\braket{F^{L;(0)}(\la)}{F^{R;(0)}(\mu)}}{\la-\mu}
 +\bra{F^{L;(0)}(\la)} \f{\Pi^{-1}(\la)\,\Pi(\mu)-I_2 }{\la-\mu}\ket{F^{R;(0)}(\mu)},\\
 &\underset{\la\tend\mu}{\tend}
 - \braket{F^{L;(0)}(\la)}{\Dp{\la} F^{R;(0)}(\la)}
 - \bra{F^{L;(0)}(\la)} \Pi^{-1}(\la)\,\Dp{\la}\Pi(\la)\ket{F^{R;(0)}(\mu)}.
\end{align*}
The corrections to the leading order for the resolvent are
given here by the second term.

The inversion operator on $\mc{M}_2\pa{\Cx}$: $u \mapsto u^{-1}$ is continuously
differentiable around the identity $I_2$. Thus there exists an open
neighbourhood $W$ of the identity matrix $I_2$ and a constant $C>0$
such that, $\forall \; A, \, B \, \in  W$, one has $\norm{A^{-1}-B^{-1}}
\leq C \norm{A-B}$. Here $\norm{.}$ denotes any matrix norm.
The matrices $\Pi\pa{\la}$ and $\Pi\pa{\la;k}\,$ belong to  $W$ for $x$ sufficiently large, as they both go to $I_2$
in the $x\tend +\infty$ limit for $\la$ uniformly
away from $\Sg_\Pi$, and we get, from Proposition~\ref{prop-Pi},
\begin{align*}
&\norm{\Pi^{-1}\pa{\la} \Dp{\la}\Pi\pa{\la}-\Pi^{-1}\pa{\la;k} \Dp{\la} \Pi\pa{\la;k} } \\
&\hspace{2cm}\leq
C\norm{\Pi\pa{\la}-\Pi\pa{\la;k}} \norm{\Dp{\la} \Pi\pa{\la}}  \\
&\hspace{3cm} + C \pa{\norm{\Pi\pa{\la}}+ \norm{\Pi\pa{\la}-\Pi\pa{\la;k}}}
\norm{\Dp{\la}\Pi\pa{\la}-\Dp{\la}\Pi\pa{\la;k}} \\
&\hspace{2cm}\leq \f{\wt{C}\pa{k}}{x^{\pa{k+1}\pa{1-\ov{\veps}}}},
\end{align*}
for some constant $\wt{C}(k)$. Thus, uniformly away from $\Sg_\Pi$ and on the real axis, one has
%
\begin{equation}
\abs{    \bra{F^{L;\pa{0}}(\la)} \Pi^{-1}(\la)\, \Dp{\la}\Pi(\la) - \Pi^{-1}(\la;k)\, \Dp{\la} \Pi(\la;k) \ket{F^{R;\pa{0}}(\la)}    }
 = \e{O}\paf{x^{\ov{\veps}}}{x^{\pa{k+1}\pa{1-\ov{\veps}}}}.
\end{equation}
%
In the last equality, we used the fact that $f_{\pm}^{(0)}$ are at
most of order $\e{O}(x^{\ov{\veps}})$ on the real axis, as follows
from their behaviour around
 $\pm q$.
\qed
%
%

%
\section{Leading asymptotic behaviour of $\ln\ddet{}{I+V}$}
\label{Section asymptotique log det}
%
%

In this Section, we prove the result of Theorem~\ref{theorem
asymptotiques order zero log det};  that is to say, we compute the leading
asymptotic behaviour  $\det[I+V]^{(0)}$  of $\det[I+V]$ up to
$\e{o}(1)$ corrections in the $x\tend +\infty$ limit. More
precisely, we show that
\begin{multline}
\ln\ddet{}{I+V}^{(0)}
  =2 \Int{-q}{q}\dd\la\, \nu(\la) \ln'[e_-(\la)]
  +\sum_{\sigma=\pm}\ln\pac{\f{G(1,\nu_\sigma)\; \kappa^{\sigma \nu_\sigma}(\sigma q;q)}
          {\pa{2q p'_\sigma x}^{\nu_\sigma^2} }}  \\
  +  \f{1}{2} \Int{-q}{q} \dd \la \,\dd \mu \,
         \f{\nu'(\la)\, \nu(\mu)- \nu(\la)\, \nu'(\mu)}{\la-\mu}\;.
\label{asymptotique-ordre_zero}
\end{multline}

This result will be obtained  by two different methods based on the
integration of equations $\eqref{the q and gamma derivative
path}$. The first one, which uses the derivative of the Fredholm
determinant over $\gamma$, is based on the uniformness of the
asymptotic expansion for the resolvent for $\gamma$ small enough. It
is worth mentioning that this way is technically quite involved.
The second method deals with the derivative of the Fredholm
determinant over $q$. Although we have not been able to provide a full rigorous proof for it, we would like to draw the reader's attention to this method as it is much more direct and simple.

%
\subsection{The leading asymptotics from the $\ga$-derivative method}

Due to Proposition \ref{est-trace}, the proof of the leading
asymptotics  of the Fredholm determinant from the first equation
\eqref{the q and gamma derivative path},
\begin{equation}
\Dp{\ga} \ln \ddet{}{I+V}= \Int{-q}{q} \f{\dd \la}{\ga} R\pa{\la,\la},
\end{equation}
only necessitates the use of $R^{(0)}\pa{\la,\la}$ defined in
\eqref{LO-resolvent}. Recall that $R^{(0)}\pa{\la,\la}$ has
different leading asymptotics in the bulk $\intoo{-q}{q}$ and near
the boundary. Let $\de>0$ be sufficiently small. Then
\begin{equation}
\f{R^{(0)}(\la,\la)}{\ga}
 =\left\{ \ba{l l }
  R^{(0)}_{q}\pa{\la,\la}      & \la \in \intff{q-\de}{q},
               \vspace{1mm}\\
  R^{(0)}_{\e{bk}}\pa{\la,\la}
                       & \la \in \intff{-q+\de}{q-\de},
               \vspace{1mm}\\
  R^{(0)}_{-q}\pa{\la,\la}  & \la \in \intff{-q}{-q+\de}, \ea \right.
\label{resolvent R0 par partie}
\end{equation}
where
\begin{equation*}
R^{(0)}_{\e{bk}}(\la,\la)=\f{F(\la)}{2i\pi \pa{1+\ga F(\la)}}
     \paa{ 2 \Dp{\la}\ln e_+(\la)
          -2 \Dp{\la} \ln \pac{\kappa_p(\la) \paf{p_+-p(\la)}{p(\la)-p_-}^{\nu(\la)} }  } ,
\end{equation*}
\begin{multline*}
R^{(0)}_{-q}\pa{\la,\la}=- \nu  \varphi\pa{\nu;x\pac{p-p_-}}
\Big\{ 2  {\nu}' \ln  x   -2 \Dp{\la}\pac{ \nu \ln \pa{p_+-p} -2\Dp{\la} \ln \kappa_p }   \\
 + {\nu}'  \pac{\psi\pa{1+\nu}+\psi\pa{1-\nu}  }   +  g'\Big\}
+i x \nu p' \tau\pa{\nu; x \pac{p-p_-}} 
 + \nu{\nu}'\rho\pa{\nu;x\pac{p-p_-}},
\end{multline*}
\begin{multline*}
R^{(0)}_q\pa{\la,\la}=- \nu \varphi\pa{\nu;x\pac{p_+-p}}
\Big\{ 2 {\nu}' \ln  x   -2 \Dp{\la}\pac{ \nu \ln \pa{p-p_-} -2\Dp{\la}\ln \kappa_p}   \\
 +  \Dp{\la}\nu  \pac{\psi\pa{1+\nu}+\psi\pa{1-\nu}  }
   +  g' \Big\}
+i x \nu p' \tau\pa{\nu; x \pac{p_+-p}} 
- \nu {\nu}' \rho\pa{\nu; x \pac{p_+-p} }.
\end{multline*}
Here $\psi(z)=\frac d{dz}\log\Gamma(z)$ and we have introduced the
shorthand notations
\begin{align*}
&\varphi\pa{\nu;t}= \Phi\pa{-\nu,1;-i t}\Phi\pa{\nu,1;i t} , \\
&\rho\pa{\nu;t} = \pa{\Dp{1}\Phi}\pa{\nu,1;it} \Phi\pa{-\nu,1;-it}+
\pa{\Dp{1}\Phi}\pa{-\nu,1;-it} \Phi\pa{\nu,1;it} ,\\
&\tau\pa{\nu;t}=-\Phi\pa{-\nu,1;-i t}\Phi\pa{\nu,1;i t} +  \pa{\Dp{z}\Phi}\pa{-\nu,1;-i t}\Phi\pa{\nu,1;i t}\\
&\hspace{5.85cm}+
\Phi\pa{-\nu,1;-i t}\pa{\Dp{z}\Phi}\pa{\nu,1;i t}.
\end{align*}
Moreover, in order to lighten the above expressions and similar ones in the following, we omit the explicit dependence on the argument $\la$ of the different functions involved (like $\nu$, $p$, and their derivatives ${\nu}'$, $p'$, etc.).

We can now split the integration contour into three parts
\begin{equation}
\Int{-q}{q} R^{(0)}\pa{\la,\la} \f{\dd \la}{\ga} = \Int{-q}{-q+\de} \f{\dd \la}{\ga} R^{(0)}_{-q}\pa{\la,\la}
+\Int{-q+\de}{q-\de} \f{\dd \la}{\ga} R^{(0)}_{\e{bk}}\pa{\la,\la}+
\Int{q-\de}{q} \f{\dd \la}{\ga} R^{(0)}_q\pa{\la,\la}.
\end{equation}
The bulk integral is carried  out straightforwardly. The integrals
over the vicinities of the endpoints are more involved. Consider,
for instance, the integration over $\intff{-q}{-q+\de}$.

Using the asymptotic series for Humbert CHF $\Phi$
\eqref{asymptotiques phi} and the equations \eqref{integrales tau et
phi0}, \eqref{integrales tau et phi} we get that
\begin{align*}
&\varphi\pa{a;t}-\ex{i\pi a} \; ,\\
&\rho\pa{a;t}+\f{\ex{i\pi a}}{\Ga\pa{1-a}\Ga\pa{1+a}} \paa{2\ln t -\psi\pa{1-a}-\psi\pa{1+a} - \f{4 i a}{1+t}} \;, \\
&\tau\pa{a;t} + \ex{i\pi a}\pa{1- \f{2i a}{1+t}  } \;,
\end{align*}
are uniformly Riemann integrable on $\R^+$ in the sense of the
definition of Lemma \ref{lemme preparatoire} (See Appendix B). Using the integration
Lemma \ref{lemme preparatoire} as well as the estimates for the
integrals of $\tau$ and $\varphi$ \eqref{integrales tau et phi0},
\eqref{integrales tau et phi}, we find
\begin{align}
&\Int{-q}{-q+\de} R^{(0)}\pa{\la,\la} \dd \la
= -\Int{-q}{-q+\de} \f{
\dd \la \;\ex{i\pi \nu}}{\Ga\pa{\nu}\Ga\pa{1-\nu}}
\left\{ 2 \nu' \ln x  -2 \Dp{\la}\pac{\nu \ln \pa{p_+-p}}  \right.
          \nonumber\\
&\hspace{7cm}
\left.+ \nu' \pa{\psi\pa{\nu}+\psi\pa{-\nu}}+g'    \right\}  \nonumber\\
&+\Int{-q}{-q+\de} \! \f{ \dd \la \;\ex{i\pi \nu}
\nu'}{\Ga\pa{\nu}\Ga\pa{1-\nu}}\paa{
2\ln \pac{x\pa{p-p_-}}-\psi\pa{-\nu}-\psi\pa{\nu}+\f{4i\nu}{1+x\pa{p-p_-}} }  \nonumber\\
&+ ix \Int{-q}{-q+\de} \dd \la \;p' \nu
\paa{\f{2i\nu}{x\pa{p-p_-}+1}-1 } \nonumber\\
&\hspace{3.5cm}
+\f{\nu_-\ex{i\pi
\nu_-}}{\Ga\pa{1-\nu_-}\Ga\pa{\nu_-}}
\paa{2 -\psi\pa{\nu_-} - \psi\pa{-\nu_-}} +\e{o}\pa{1} \;
.\label{4-lines}
\end{align}
Here the $\e{o}\pa{1}$ is with respect to the successive limits
$x\de \tend +\infty$ and $\de \tend 0$. The two terms proportional to
$\ln x$  compensate  each other. The remaining part of the first
three lines of \eqref{4-lines} is an $\e{O}\pa{\de}$ and can thus be
dropped. The integral in the last two lines of \eqref{4-lines} is
evaluated thanks to the second integration Lemma \ref{lemme
preparatopire 2} (see Appendix \ref{Two lemme preparatopire}). We
get,
\begin{multline}
\Int{-q}{-q+\de} R^{(0)}\pa{\la,\la} \dd \la = -i\Int{-q}{-q+\de} \f{\ex{i\pi \nu} p'}{\Ga\pa{\nu} \Ga\pa{1-\nu}} \dd \la \\
+\f{ \ex{i\pi\nu_-} \nu_- }{\Ga\pa{\nu_-} \Ga\pa{1-\nu_-}} \paa{-2\ln \pac{x \pa{p\pa{\de-q}-p_-}}
+ 2-\psi\pa{\nu_-}-\psi\pa{-\nu_-}} +\e{o}\pa{1} \; .
\end{multline}

The integration over $\intff{q-\delta}{q}$ can be
treated similarly. The result reads
\begin{multline}
\Int{q-\de}{q} R^{(0)}\pa{\la,\la} \dd \la= -i\Int{q-\de}{q} \f{\ex{i\pi \nu} p'}{\Ga\pa{\nu} \Ga\pa{1-\nu}} \\
+\f{\nu_+ \ex{i\pi\nu_+}}{\Ga\pa{\nu_+} \Ga\pa{1-\nu_+}} \paa{-2\ln \pac{ x \pa{p_+-p\pa{q-\de}}}
+\pac{2-\psi\pa{\nu_+}-\psi\pa{-\nu_+}} }+\e{o}\pa{1} \; .
\end{multline}
So that,
\begin{multline}\label{int-res}
 \Int{-q}{q} R^{(0)}\pa{\la,\la}\,\dd \la = \Int{-q}{q} \f{\ga
 F}{2i\pi\pa{1+\ga F}} \left[i xp'+g' -2\Dp{\la} \ln \kappa_p \right]\,\dd \la\\
+ \Int{\de-q}{q-\de}  \nu \ln \paf{p_+-p}{p-p_-}
\Dp{\la}\paa{\f{\ga F}{1+\ga F}}\f{\dd \la }{i\pi}
-\pac{\f{\ga \nu  F}{i\pi\pa{1+\ga F}} \ln \paf{p_+-p}{p-p_-}  }_{\de-q}^{q-\de} \\
+\f{ \ga \nu_- F_-}{2i\pi\pa{1+\ga F_-}}\paa{2 \ln \pac{x\pa{p\pa{\de-q}-p_-}} -2 +\psi\pa{\nu_-}+\psi\pa{-\nu_-}} \\
+\f{\ga \nu_+ F_+}{2i\pi\pa{1+\ga F_+}} \paa{2\ln \pac{
x\pa{p\pa{q-\de}-p_+}} - 2 +\psi\pa{\nu_+}+\psi\pa{-\nu_+}}
+\e{o}\pa{1},
\end{multline}
where we used $\f{\ex{i\pi \nu }}{\Ga\pa{\nu}\Ga\pa{1-\nu}}=-\f{\ga
F}{2i\pi \pa{1+\ga F}}$. Using the integral representation for the  Barnes
$G$-function \eqref{fonction de Barnes}, it is not a problem  to see that
\begin{multline}
\Int{-q}{q} R^{(0)}\pa{\la,\la} \f{\dd \la}{\ga} = \Dp{\ga} \paa{
\Int{-q}{q} \nu \, \Dp{\la} \ln e_- \,\dd \la+
\ln \paf{ G\pa{1,\nu_+} G\pa{1,\nu_-} }{ \pac{ x\pa{p_+-p_-} }^{\nu_+^2+\nu_-^2}  } } \\
+2 \Int{-q}{q} \Dp{\ga}\nu \, \Dp{\la}\ln\kappa_p\,\dd \la - 2
\Int{-q}{q}  \pac{\Dp{\la}\Dp{\ga}\nu} \nu \ln
\paf{p_+-p}{p-p_-}\,\dd \la\;.
\end{multline}
Now we should recast the last line as a  derivative with respect to
$\ga$. We have
\begin{multline}
2 \Int{-q}{q} \left[\Dp{\ga} \nu \Dp{\la} \ln \kappa_p
-\pac{\Dp{\la}\Dp{\gamma}\nu} \nu \ln\paf{p_+-p}{p-p_-}
\right]\,\dd \la \\
=\Dp{\ga} \ln \paa{ \f{\kappa^{\nu_+}\pa{q;q}  }{
\kappa^{\nu_-}\pa{-q;q} } \paf{p_+-p_-}{2qp'_+}^{2\nu_+^2}
\paf{p_+-p_-}{2qp'_-}^{2\nu_-^2} } \\
+\sul{\sigma=\pm}{}\Bigl[ \sigma \Dp{\ga}\nu_{\sigma} \ln
\kappa\pa{\sigma q ; q}-
\sigma \nu_{\sigma} \Dp{\ga} \ln\kappa\pa{\sigma q; q}\Bigr]\\
-2\Int{-q}{q}\left[ \partial_\gamma\partial_\lambda \nu\vphantom{\ln \paf{q-\la}{q+\la}  }\right] \left[\ln \kappa+ \nu
\ln \paf{q-\la}{q+\la}  \right] \dd \la\;.
\end{multline}
It remains to apply the identity
\begin{multline}
\Dp{\ga}\Int{-q}{q}
\f{\nu'\pa{\la}\nu\pa{\mu}-\nu'\pa{\mu}\nu\pa{\la}
}{2\pa{\la-\mu}}\,\dd \la\, \dd \mu = -2\Int{-q}{q}
\left[ \partial_\gamma\partial_\lambda \nu\vphantom{\ln \paf{q-\la}{q+\la}  }\right] \left[\ln \kappa+ \nu
\ln \paf{q-\la}{q+\la}  \right] \dd \la \\
+\sul{\sigma=\pm}{} \Bigl[\sigma \Dp{\ga}\nu_{\sigma} \ln
\kappa\pa{\sigma q ; q}- \sigma \nu_{\sigma} \Dp{\ga}
\ln\kappa\pa{\sigma q; q}\Bigr]
\,.
\label{egalite derivees gamma}
\end{multline}
Indeed,  we have for the r.h.s. of \eqref{egalite derivees gamma}
\begin{align}
%
RHS &= \Int{-q}{q}\left( \f{\nu_+ \Dp{\ga} \nu- \nu \,
\Dp{\ga} \nu_+ }{q- \la} + \f{\nu_-\Dp{\ga}\nu-\nu \Dp{\ga}
\nu_-}{q+\la}\right)\,\dd \la
               \nonumber\\
&\hspace{2cm}
+  \Int{-q}{q}  \nu\pa{\mu} \partial_\gamma\partial_\lambda\nu \pa{\la}\paa{\f{1}{\la-\mu+i0}+\f{1}{\la-\mu-i0}  }\,\dd \la\, \dd \mu
              \nonumber\\
&= \frac12\Int{-q}{q}  \sul{\sigma=\pm}{}
\pa{\nu_{\sigma}\Dp{\ga}\nu + \nu \Dp{\ga} \nu_{\sigma} }
\paa{\f{1}{q-\sigma\la + i0}+ \f{1}{q-\sigma\la - i0} }\,\dd \la
                \nonumber\\
&\hspace{2cm}
+ \Int{-q}{q} \nu\pa{\mu} \Dp{\ga}\nu\pa{\la}
\paa{\f{1}{\pa{\la-\mu+i0}^2}+\f{1}{\pa{\la-\mu-i0}^2}  }\,\dd \la
\,\dd \mu  \, .\label{shutka}
\end{align}
There we have regularized all the integrands and then performed an
integration by parts. On the other hand, one has for the l.h.s. of
\eqref{egalite derivees gamma}
\begin{multline}
%
LHS =
\Int{-q}{q}  \pac{\Dp{\ga} \nu\pa{\mu}  \Dp{\la} \nu\pa{\la} +
\nu\pa{\mu}
\partial_\gamma\partial_\lambda\nu\pa{\la}}
\paa{ \f{1}{\la-\mu+i0} + \f{1}{\la-\mu-i0} } \,\dd \la\, \dd \mu\,
.
\end{multline}
Taking the last integral by parts we arrive at \eqref{shutka}.

Thus, the l.h.s. of \eqref{int-res} is presented as a derivative
with respect to $\gamma$. Since the asymptotic  expansion is uniform
in $\ga$ we can integrate this result from $0$ to $\ga$. As $\ln
\ddet{}{I+V}\mid_{\ga=0}=0$ we get the desired result. 

%
\subsection{The leading asymptotics from the $q$-derivative method}
\label{sec-leading-q}

The method we use here is based on the second equation in \eqref{the
q and gamma derivative path},
\begin{equation}\label{derivative-q}
\partial_q \ln \ddet{}{I+V}=R\pa{q,q}+R\pa{-q,-q} .
\end{equation}
For the purpose of this sub-section, we assume that $\abs{\Re\pa{\nu\pa{\la}}}<\tf{1}{4}$.
Indeed we are then able to use the pointwise estimates for the resolvent established in
Proposition \ref{proposition DA du resolvent}. Such a restriction on \abs{\Re\pa{\nu\pa{\la}}} could be relaxed by much more refined estimates. Recall that one has for $\la$ uniformly away
from the boundary $\Sg_{\Pi}$ corresponding to disks of radius $\de$, 
\begin{equation}
\abs{R\pa{\la,\la}-R^{\pa{0}}\pa{\la,\la}}\leq
\f{C\pa{q}}{x^{1-2\ov{\veps}}}\leq \f{C\pa{q}}{x^{1-2\mf{e}}} \;
\;\;, \;  \e{with}\;  \mf{e}=2\underset{\ov{U}}{\sup}\abs{\Re \nu}
\;\; ,
\end{equation}
so that the $q$ anti-derivative of $R\pa{q,q}+R\pa{-q,-q}-R^{\pa{0}}\pa{q,q}-R^{\pa{0}}\pa{-q,-q}$ will be a
$\e{o}\pa{1}$ in the $x\tend +\infty$ limit.

Equation \eqref{resolvent R0 par partie} allows us to determine the value of
\begin{align}
R^{(0)}\pa{\la,\la}
  &= -\braket{F^{L;\pa{0}}\pa{\la}}{\Dp{\la}F^{R;\pa{0}}\pa{\la}}
               \nonumber\\
  &= \frac{\gamma F(\la)}{2i\pi}
     f_+^{\pa{0}}(\la)\, f_-^{\pa{0}}(\la)\,
            \paa{\Dp{\la} \ln f_+ - \Dp{\la} \ln f_-   }
\end{align}
at both endpoints $q$ and $-q$.

Consider, for instance, $R^{(0)}\pa{-q,-q}$. We have, for $\la \in
D_{-q, \de}$,
\begin{align}
R^{(0)}\pa{\la,\la}
 &=- \nu(\la)\, \Phi\pa{-\nu,1;-i\zeta_{-q}} \Phi\pa{\nu,1;i\zeta_{-q}}
    \bigg\{2 \Dp{\la} \ln e_+ (\la)
             \nonumber\\
 &\hspace{1.3cm}
    - 2 \Dp{\la} \big[\ln \kappa_p(\la) +\nu(\la)\,\ln\zeta_q\big]
    +{\nu}'(\la)  \big[\psi\pa{1+\nu}+\psi\pa{1-\nu}  \big]
             \nonumber\\
 &\hspace{1.3cm}
 -ix \, p'(\la) \pac{  \f{\pa{\Dp{z}\Phi}\pa{\nu,1;i\zeta_{-q}}}{ \Phi\pa{\nu,1;i\zeta_{-q} }}
 +\f{\pa{\Dp{z}\Phi}\pa{-\nu,1;-i\zeta_{-q}}}{ \Phi\pa{-\nu,1;-i\zeta_{-q} } }  }
            \nonumber\\
 &\hspace{1.3cm}
\left.      -{\nu}'(\la) \pac{  \f{\pa{\Dp{1}\Phi}\pa{\nu,1;i\zeta_{-q}}}{ \Phi\pa{\nu,1;i\zeta_{-q} }}
 +\f{\pa{\Dp{1}\Phi}\pa{-\nu,1;-i\zeta_{-q}}}{ \Phi\pa{-\nu,1;-i\zeta_{-q} } }  }                      \right\},
\end{align}
where, so as to lighten the formula, we have omitted the argument $\la$
of $\nu(\la)$  when $\nu$ appears as an argument of another
function (here $\psi$ or $\Phi$). The symbol $\Dp{z}$ stands for the
derivative of a CHF with respect to its variable, whereas $\Dp{1}$
stands for the derivative with respect to its first argument. Recall
also that $\zeta_{-q}=x\pac{p\pa{\la}-p_{-}}$ and
$\zeta_{q}=x\pac{p_{+}-p\pa{\la}}$.

It is remarkable that the last two terms involving  derivatives of
CHF vanish in the $\la \tend -q$ limit. The resulting expression can
be further simplified thanks to the identities:
\begin{align}
 &\ln\kappa_p(\la)
   =\ln\kappa(\la)+\nu(\la)\left\{\ln\bigg(\frac{q-\la}{p_+-p(\la)}\bigg)
   -\ln\bigg(\frac{\la+q}{p(\la)-p_-}\bigg)\right\},
         \\
 &\nu(\la)\,\nu'(\la)\,\big[\psi(1+\nu)+\psi(1-\nu)\big]
  = \Dp{\la}\ln G(1,\nu) +2\nu(\la)\,\nu'(\la),
\end{align}
Thus, we obtain
\begin{multline}
 R^{(0)}\pa{-q,-q}
  =-2 \nu_- \big[ \partial_\la \ln e_+(\la)\big]\pour{\la=-q} + 2\nu_-'\nu_-\ln x  \\
   + 2\nu'_- \nu_- \ln (2q p_-')
   -\frac{\nu_-^2}{q}+\nu_-^2\frac{p_-''}{p_-'}-2\nu_- \nu_-'
   + 2\nu_- \big[\Dp{\la} \ln \kappa(\la)\big]\pour{\la=-q},
\end{multline}
where we have used the notations \eqref{notation-p-q},
\eqref{notation-nu-q}.

The final aim is to integrate \eqref{derivative-q} over the variable
$q$. One should keep in mind that the function $\kappa(\la)\equiv
\kappa(\la;q)$ is actually a function of the two parameters $\la$
and $q$. Therefore, one should replace partial $\la$
derivatives at $\la=\pm q$ by total $q$ derivatives thanks to
\begin{equation}
\f{\dd}{\dd q} [ \ln \kappa (-q;q) ]
 = -\Dp{\la}\ln \kappa(\la;q)\pour{\la=-q}
   + \Dp{q}\ln \kappa(\la;q)\pour{\la=-q}.
\end{equation}
Then $R^{(0)}\pa{-q,-q}$ is almost a total $q$ derivative:
\begin{multline}
R^{(0)}\pa{-q,-q}
  =-2 \nu_- \big[ \partial_\la \ln e_+(\la)\big]\pour{\la=-q}
   + \D{}{q} \ln \pac{\f{ G\pa{1,\nu_-} }{\pa{2q p'_- x}^{\nu_-^2}} }
\\
   -2 \nu_- \D{}{q} \pac{\ln \kappa\pa{-q;q}}
   +\nu_- \f{\pa{\nu_+-\nu_-}}{q}  .
\label{R moins qu moins q}
\end{multline}
Similar calculations based on the expressions \eqref{fplusmoins en q} for $f_{\pm}^{\pa{0}}$ around $q$ lead to
\begin{multline}
R^{(0)}\pa{q,q}
  =-2 \nu_+ \big[\partial_\la \ln e_+(\la)\big]\pour{\la=q}
   + \D{}{q} \ln \pac{\f{ G\pa{1,\nu_+} }{\pa{2q p'_+ x}^{\nu_+^2}} }
\\
   +2 \nu_+ \D{}{q} \pac{\ln \kappa\pa{q;q}}
   -\nu_+ \f{\pa{\nu_+-\nu_-}}{q} \,  .
\label{R qu qu}
\end{multline}
Hence, we have
\begin{multline}\label{Deriv-q}
\partial_q \ln \ddet{}{I+V}
  = 2 \sum_{\sigma=\pm}\nu_\sigma \big[\partial_\la \ln e_-(\la)\big]\pour{\la=\sigma q}
   +\D{}{q}\ln\pac{\f{G\pa{1,\nu_+}G\pa{1,\nu_-}}
                  {\pa{2q p'_- x}^{\nu_-^2}\pa{2q p'_+ x}^{\nu_+^2}}}
  \\
 + 2\sum_{\sigma=\pm} \sigma \nu_\sigma \D{\ln\kappa (\sigma q;q)}{q}
 -\f{\pa{\nu_+-\nu_-}^2}{q} +\e{o}\pa{1} .
\end{multline}

It remains to express the last line as a total $q$-derivative thanks to Lemma \ref{lemme h et derivee kappa NLS}. After an integration with respect to $q$ of \eqref{Deriv-q} we arrive to
\begin{multline}\label{leading-up-to-constant}
\ln\ddet{}{I+V}
  =2 \Int{-q}{q} \!\dd\la\, \nu(\la) \ln'[e_-(\la)]
   + \ln\pac{\f{G(1,\nu_+)\, G(1,\nu_-)\, \kappa^{\nu_+}(q;q)}
            {\pa{2q p'_+ x}^{\nu_+^2}\pa{2q p'_- x}^{\nu_-^2}
                                  \kappa^{\nu_-}(-q;q)  }} \\
   +  \f{1}{2} \Int{-q}{q}\! \dd \la \dd \mu\,
             \f{\nu'(\la)\, \nu(\mu) - \nu(\la)\, \nu'(\mu) }{\la-\mu}
  +C + \e{o}(1)\, ,
\end{multline}
where $C$ is a $q$-independent integration constant still to be
determined.

One can give arguments that this constant should be also
$\gamma$-independent. Indeed, the asymptotic expansion of the
Fredholm determinant, being a functional of the holomorphic function $\gamma F(\lambda)$ in $\ov{U}$, can depend on this function either in
the integral form with integration over $[-q;q]$, or through the
values of $\gamma F$ and of its derivatives at the ends of the integration
contour $-q$ and $q$. In both cases the result should depend on $q$. Hence, the
$q$-independent constant $C$ can not depend on $\gamma F(\lambda)$
and, thus, it is $\gamma$-independent. We can then fix the constant $C$ by setting $\ga=0$ in the asymptotic formula. This yields $C=0$. A rigourous proof of this equality within this $q$-derivative method is however still missing. Indeed, although the above statement (about the functional form of the asymptotic expansion of the Fredholm determinant) is clear  in the case of one-dimensional oscillatory integrals without saddle point, its generalisation to the needed series of multiple oscillatory integrals would require additional work.

%
\subsection{The first corrections to the leading asymptotics of 
the Fredholm determinant}\label{sec-sub-asympt}
%

We close this section  by deriving the sub-leading corrections from the
$x$-derivative  \eqref{the x
derivative for MSK} of $\ln\ddet{}{I+V}$. This will constitute the proof of
Proposition~\ref{proposition correstion en 1/x au resolvent}.

In order to prove the claim of the Proposition~\ref{proposition
correstion en 1/x au resolvent}, one has to derive the first two
sub-leading corrections for the matrix $\Pi$. As one might expect
the computations are, by far, simpler that those necessary to fix the constant. We also would like
to point out that one can obtain the sub-leading asymptotic of
$\ddet{}{I+V}$ by the $q$-derivative method. However, the
computations are quite involved, so we omit the presentation of this method.

We derive the first term in the $\tf{1}{x}$ expansion of $\log \ddet{}{I+V}$ thanks to \eqref{the x derivative path}:
\begin{equation}
\Dp{x}\log \ddet{}{I+V}= \oint\limits_{\Ga\pa{\intff{-q}{q}}}{}
\hspace{-3mm}\f{\dd \la}{4\pi}\, p(\la)\,
\e{tr}\pac{\Dp{\la}\chi\pa{\la}\sg_3\chi^{-1}\pa{\la}}\;,
\end{equation}
where we chose the contour $\Ga\pa{\intff{-q}{q}}$ to lie outside of the contour $\Sigma_\Pi$ but still in the region of holomorphy for $p$.
There the solution for the RHP for $\chi$ has a simple form:
\begin{equation}
 \chi\pa{\la}=\Pi\pa{\la}\, \a^{-\sg_3}\pa{\la}.
\end{equation}
In order
to derive the first correction to the leading asymptotics, it is enough to consider the first two terms in the
asymptotic expansion for $\Pi(\la)$:
\begin{equation}
\Pi\pa{\la}=I_2+\f{ \Pi^{\pa{1}}\pa{\la} }{x}+\f{
\Pi^{\pa{2}}\pa{\la} }{x^2}+\e{O}\pa{x^{3\pa{\ov{\veps}-1}}}.
\end{equation}
There,
as follows from $\eqref{asympt-Pi}$, the $\e{O}$ is uniform on the
whole contour $\Gamma\pa{\intff{-q}{q}}$. Thus
\begin{multline}
\Dp{x}\log
\ddet{}{I+V}=-\hspace{-4mm}\oint\limits_{\Gamma\pa{\intff{-q}{q}}}{}
\hspace{-4mm}\f{\dd \la}{2\pi}\, p\pa{\la} \f{
\Dp{\la}\a\pa{\la}}{\a\pa{\la}}
+ \f{1}{x}\hspace{-4mm} \oint\limits_{\Gamma\pa{\intff{-q}{q}}}{} \hspace{-4mm}\f{\dd \la}{4\pi }\, p\pa{\la}  \e{tr}\pac{\sg_3\,\Dp{\la}\Pi^{\pa{1}}\pa{\la}}\\
+ \f{1}{x^2} \hspace{-4mm} \oint\limits_{\Gamma\pa{\intff{-q}{q}}}{}
\hspace{-4mm} \f{\dd \la}{4\pi  }\, p(\la) \,
 \e{tr}\paa{\sg_3 \pac{
\Dp{\la}\Pi^{\pa{2}}(\la)
-\Pi^{\pa{1}}(\la)\Dp{\la}\Pi^{\pa{1}}(\la) }}
+\e{O}\pa{x^{3\pa{\ov{\veps}-1}}}\;.
\label{the x derivative in action}
\end{multline}

The first term in this expansion will yield the leading correction. Indeed,
\begin{equation}
-\hspace{-4mm}\oint\limits_{\Gamma\pa{\intff{-q}{q}}}{}
\hspace{-4mm}\f{\dd \la}{2\pi}\, p\pa{\la}\f{
\Dp{\la}\a\pa{\la}}{\a\pa{\la}}
=\hspace{-2mm}\oint\limits_{\Gamma\pa{\intff{-q}{q}}}{}
\hspace{-4mm}\f{\dd \la}{2\pi}\, p'\pa{\la}\log
\a\pa{\la}=-i\Int{-q}{q} \dd \la \, p'\pa{\la} \nu\pa{\la}\;.
\end{equation}
Here we shrunk the contour to $\intff{-q}{q}$ and used the jump
condition for $\a$.

In order to evaluate the higher order
corrections in \eqref{the x derivative in action} we need to derive
the expressions for the matrices $\Pi^{\pa{1}}=C_{\Sigma_\Pi}\big[\Delta^{(1)}\big]$ and
$\Pi^{\pa{2}}=C_{\Sigma_\Pi}\big[\Pi_+^{(1)}\Delta^{(1)}+\Delta^{(2)}\big]$ outside of $\Sigma_\Pi$. An elementary
computation of residues yields:
\begin{equation}
\Dp{\la}\Pi^{\pa{1}}\pa{\la}=-\f{\De_{\pa{+}}^{\pa{1}}\pa{q;x} }{\pa{\la-q}^2p'_+}
-\f{\De_{\pa{-}}^{\pa{1}}\pa{-q;x} }{\pa{\la+q}^2p'_-},
\end{equation}
as well as
\begin{multline}
\Dp{\la}\Pi^{\pa{2}}\pa{\la}-\Pi^{\pa{1}}\pa{\la}\Dp{\la}\Pi^{\pa{1}}\pa{\la}\\
\hspace{5mm}
=-\sul{\sg=\pm}{} \f{\pa{\Dp{\la}\De_{\pa{\sg}}^{\pa{2}}}\pa{\sg q;x}+
\De_{\pa{\sg}}^{\pa{1}}\pa{\sg q;x}\pa{\Dp{\la}\De_{\pa{\sg
}}^{\pa{1}}}\pa{\sg q;x}}
{\pac{\pa{\la-\sg q}p_{\sg}'}^2} \\
\hspace{5mm}
+ \sul{\sg=\pm }{} \f{2 \De_{\pa{\sg}}^{\pa{2}}\pa{\sg q;x}+\pac{\De_{\pa{\sg }}^{\pa{1}}\pa{\sg q;x}}^2}{2 \pa{p'_{\sg}}^2}
\paa{\f{p_{\sg}''}{p_{\sg}'\pa{\la- \sg q}^2}-\f{2}{\pa{\la- \sg q}^3}  } \\
+\f{ \pac{\De_{\pa{+}}^{\pa{1}}\pa{q;x}, \De_{\pa{-}}^{\pa{1}}\pa{-q;x}} }{2q \, p'_+p'_- (\la-q)(\la+q)}
.
\end{multline}
Thus the  $\tf{1}{x}$ term in \eqref{the x derivative in action} gives the coefficient of $\ln x$ in \eqref{asymptotique log det ordre zero}. Indeed,
\begin{align}
\oint\limits_{\Gamma\pa{\intff{-q}{q}}}{} \hspace{-4mm}
\f{\dd
\la}{4\pi }\, p\pa{\la}  \e{tr}\pac{\sg_3\,\Dp{\la}\Pi^{\pa{1}}\pa{\la}}
&=\f{1}{2i} \e{tr}\pac{\sg_3 \De_{\pa{-}}^{(1)}\pa{-q;x}+\sg_3 \De_{\pa{+}}^{(1)}\pa{q;x}  }\nonumber\\
&=-\pa{\nu_+^2+\nu_-^2}  .\label{terme-log-x}
\end{align}

We now focus on the last term in \eqref{the x derivative in action}. It yields, after an $x$ integration,
the first correction to \eqref{asymptotique log det ordre zero}. A straightforward computation leads to
\begin{multline}
\oint\limits_{\Gamma\pa{\intff{-q}{q}}}{} \hspace{-4mm}
\f{\dd \la}{2i\pi  }\; p\pa{\la}  \e{tr}\paa{\sg_3\pac{
\Dp{\la}\Pi^{\pa{2}}\pa{\la}-\Pi^{\pa{1}}\pa{\la}\Dp{\la}\Pi^{\pa{1}}\pa{\la}    } } \hspace{3cm}\\
\hspace{7mm}=\f{p_+-p_-}{4q^2 p_-'p_+'} \e{tr}\paa{\sg_3\pac{ \De_{\pa{+}}^{(1)}\pa{q;x}\, ,\,\De_{\pa{-}}^{(1)}\pa{-q;x}   }  }\\
%
%
-\sum_{\sg=\pm}\f{1}{p_\sg'} \e{tr}\paa{\sg_3\pac{\pa{\Dp{\la}\De_{\pa{\sg}}^{\pa{2}}}\pa{\sg q;x}+ \De_{\pa{\sg}}^{\pa{1}}\pa{\sg q;x}\pa{\Dp{\la}\De_{\pa{\sg}}^{\pa{1}}}\pa{\sg q;x}}}  .
\end{multline}
The first term corresponds to the oscillating correction:
\begin{equation}
\e{tr}\paa{\sg_3\pac{ \De_{\pa{+}}^{(1)}\pa{q;x}\, ,\,\De_{\pa{-}}^{(1)}\pa{-q;x}   }  }=
2\nu_+\nu_-\paa{\f{\tilde{u}\pa{q;x }  }{u\pa{-q;x}}-\f{u\pa{-q;x }  }{\tilde{u}\pa{q;x}}  }  .
\end{equation}
The last term gives the non-oscillating one:
\begin{align}
&\e{tr}\paa{\sg_3\pa{\Dp{\la}\De_{\pa{\sg}}^{\pa{2}}}\pa{\sg q;x}   } =-2 \f{\dd \nu^3_{\sg} }{ \dd q}, \displaybreak[0]\\
\intertext{and}
&\sul{\sg=\pm}{} \f{1}{p'_{\sg}} \e{tr}\paa{\sg_3\; \De_{\pa{\sg}}^{\pa{1}}\pa{\sg q;x}\pa{\Dp{\la}\De_{\pa{\sg}}^{\pa{1}}}\pa{\sg q;x}   }   \nonumber
\\
&\hspace{2cm} =\f{2}{p'_+} \nu_+^2 \Dp{\la}\pa{\ln
\tilde{u}\pa{\la;x}}\pour{\la=q}
 +\f{2}{p'_-} \nu_-^2 \Dp{\la}\pa{\ln
u\pa{\la;x}}\pour{\la=-q} \nonumber\\
&\hspace{2cm} =
\sum_{\sg=\pm}\frac{2\nu_\sg^2}{p'_\sg}
\paa{2\sg\nu'_\sg\ln x
+\sg \D{}{q} \log u_{\sg}
+\frac{p'_\sg}{\nu_\sg}\D{}{q}\pa{\frac{\nu_\sg^2}{p'_\sg}}
-\frac{\nu_{-\sg}}{p'_-} }\; ,
\end{align}
with
\begin{align}
 &u_+=e^{g(q)}\,\frac{\Ga(1-\nu_+)}{\Ga(1+\nu_+)}
       \left\{\frac{(2qp'_+)^{\nu_+}}{\kappa(q;q)}\right\}^2,\\
 &u_-=e^{g(-q)}\,\frac{\Ga(1+\nu_-)}{\Ga(1-\nu_-)}
       \left\{(2qp'_-)^{\nu_-}\kappa(-q;q)\right\}^{-2}.
\end{align}
Putting all this together we obtain
\begin{multline}
%
 \Dp{x}\log \ddet{}{I+V}=-i \Int{-q}{q} \dd \la
\;\nu\pa{\la} p'\pa{\la} - \f{\nu_+^2+\nu_-^2}{x}
- 2i \f{\log x -1}{x^2}
\sul{\sg=\pm}{}\f{\nu_{\sg}^2}{p'_{\sg}}\D{\nu_\sg}{q}
\\
\hspace{3mm}-\f{i}{x^2} \sul{\sg=\pm}{} \f{\nu_{\sg}^2}{p'_{\sg}}
 \paa{
\sg\D{}{q} \log u_{\sg} + p'_\sg\D{}{q}\paf{\nu_{\sg}}{p_{\sg}'}
-\f{\nu_{-\sg}}{q}   } \\
\hspace{3mm}+ \f{i\pa{p_+-p_-}\nu_-\nu_+}{\pa{2q}^2 p'_+ p'_- x^2} \paa{\f{u_+}{u_-}x^{2\pa{\nu_+ + \nu_-}} \ex{ix\pa{p_+ - p_-}}
 -\f{u_-}{u_+}x^{-2\pa{\nu_+ + \nu_-}} \ex{ix\pa{p_- - p_+}} }   \\
+ \e{O}\pa{\f{\ex{ix\pa{p_+-p_-}}}{x^{3\pa{1-\ov{\veps}}}},
\f{1}{x^{3\pa{1-\ov{\veps}}}}  }.
\end{multline}
The first two terms reproduce the already known answer  for the
leading asymptotics. The remaining ones reproduce the first
oscillating and non-oscillating corrections as given in
Proposition~\ref{proposition correstion en 1/x au resolvent}. Note
that for the oscillating terms, one only should integrate the
exponent with respect to $x$ as all the other terms will give
subdominant contributions.

%
\section{Applications to truncated Wiener--Hopf operators}
\label{section applications}

Truncated Wiener--Hopf operators appear in many domains of
mathematical physics such as scattering or
diffusion processes. Moreover, many observables (dressed energy,
momentum or dressed charge) related to quantum integrable
models are solutions of integral equations of
truncated Wiener--Hopf type  \eqref{operateur de Wiener--Hopf a
interval symetrique}.

Let us recall that a truncated Wiener--Hopf operator can be
interpreted as an integral operator $I+K$ on $L^{2}\pa{\R}$ such
that it acts on $L^2\pa{\R}$ functions according to
\begin{equation}
\pa{I+K}.\varphi\pa{t}= \varphi\pa{t}+ \Int{0}{x} \dd t' K\pa{t-t'}
\varphi\pa{t'} \dd t'.
\label{operateur de Wiener1--Hopf}
\end{equation}
The kernel $K$ is characterized in terms of its Fourier transform $F$:
\begin{equation}
K(t)=\mc{F}^{-1}\pac{F}\pa{t}, \quad\text{with}\quad
\mc{F}^{-1}[F](t)= \f{1}{2\pi} \Int{\R}{} \dd \xi\, F(\xi)\, \ex{-it
\xi},\quad  \forall F \in L^{1}(\R) .
\end{equation}

The study of truncated Wiener--Hopf
operators is equivalent to a $2\times2$ matrix RHP. Another facet of
this equivalence is the correspondence between a truncated
Wiener--Hopf operator and the GSK acting on $\R$
in which $p=\id$ and $g=0$.
Indeed, it is easy to see that
\begin{equation}
K.\varphi=\mc{F}^{-1} \circ \wt{V} \circ \mc{F}\pac{\varphi},
\label{Wiener--Hopf comme Fourier du GSK}
\end{equation}
\noindent where $\wt{V}$ acts in $L^{2}\pa{\R}$ with a kernel
\begin{equation}
\widetilde{V}(\xi,\eta)=F\pa{\xi} \f{\ex{ix\pa{\xi-\eta}}-1}{2i\pi \pa{\xi-\eta}}.
\end{equation}
The operator identity
\begin{equation}
I+K= \mc{F}^{-1} .\pa{I+\wt{V}} \mc{F},
\end{equation}
together with the facts that $\wt{V}$ is trace-class  and
$\mc{F}^{\pm1}$ are continuous, ensures the equality between the
Fredholm determinants:
\begin{equation}
\ddet{}{I+K} = \ddet{}{I+\wt{V}}.
\end{equation}
\noindent The kernel $\wt{V}$ is related to
\begin{equation}
V\pa{\xi,\eta}= \sqrt{F\pa{\xi} F\pa{\eta}}\, \f{\sin\pa{x \tf{\pa{\xi-\eta}}{2}}}{\pi \pa{\xi-\eta}}
\end{equation}
\noindent by a similarity transformation. Hence,
\begin{equation}
\ddet{}{I+K}_{L^{2}\pa{0,x}}=  \ddet{}{I+V}_{L^{2}\pa{\R}}.
\label{noyau sinus et Wiener--Hopf}
\end{equation}
%

\subsection{The Akhiezer--Kac formula}

Our study of the generalised sine kernel allows us to recover the
Akhiezer--Kac formula describing the large $x$ behaviour of Fredholm
determinants of truncated Wiener--Hopf operators:
\begin{theorem}[Akhiezer--Kac \cite{AchiezerKacFormulaforTruncatedWienerHopf,KacAcheizerTruncatedWienerHopf}]
\label{theor-Akhiezer-Kac}
Let $I+K$ be a truncated Wiener--Hopf operator as above and such
that
\begin{itemize}
\item  $F$ is analytic in an open neighbourhood U  of $\R$ ;
\item $F$ goes sufficiently fast to 0 at $\pm \infty$ ;
\item $1+F\pa{\xi}$ does not vanish on $U$.
\end{itemize}
Then
\begin{equation}
\ln \ddet{}{I+K}= x\, \tau(0) + E[F] + \e{o}(1), \quad
\text{with}\quad
E[F]=\Int{0}{\infty} t \,\tau(t)\, \tau(-t)\, \dd t,
\end{equation}
in which
\begin{align}
 &\tau(t)=\f{1}{2\pi}\Int{\R}{} \ln(F(\xi)+1) \, \ex{-i t \xi}  \; \dd \xi \; .
\end{align}

\end{theorem}

\Proof
The large $x$ asymptotics of $\ddet{}{I+K}$ follows from \eqref{noyau
sinus et Wiener--Hopf} after  taking the  $q \tend +\infty$ limit
in the leading asymptotics for the corresponding generalised sine
kernel \eqref{asymptotique log det ordre zero}. This limit may seem
a little heuristic as we did not specify any estimates in $q$ for
the small o terms with respect to the leading asymptotics. However,
the validity of such a limit may either be seen by refining all the
estimates obtained in the previous section or by considering the RHP
for $\chi$ \eqref{definition RHP chi} on the whole real line from
the very beginning. We shall make the second approach more explicit
in the forthcoming subsection \ref{soussection Resolvent
Wiener--Hopf}. Here we formally take the $q\tend +\infty$ limit
in the leading asymptotics of Theorem~\ref{theorem asymptotiques
order zero log det}.

One should notice that, in the asymptotic formula $\eqref{asymptotique log det ordre zero}$, all the terms evaluated at the endpoints vanish due to the fact that $\nu\pa{\pm q} \ln q\tend 0$ when $q\tend +\infty$, which is
a consequence of the sufficiently fast decrease of $F$ at infinity. Hence, the only constant
contribution $E[F]$ to the asymptotics of $\ln \ddet{}{I+K}$ is given by the integral
\begin{equation}
E\pac{F}=\underset{q \tend +\infty}{\lim} \;\;\frac{1}{2}
\Int{-q}{q} \f{\nu'\pa{\la} \nu\pa{\mu}-\nu\pa{\la}\nu'\pa{\mu}}
{\la-\mu}\,\dd \la \,\dd \mu\; .
\end{equation}
Let us recast the constant $E\pac{F}$ in a more standard form. We
have
\begin{align*}
%
E\pac{F} &= -\f{1}{8 \pi^2} \Int{\R}{} \dd \xi\, \dd
\eta \,
   \f{ \ln'\pa{F(\xi)+1} \ln\pa{F(\eta)+1}
- \ln'\pa{ F(\eta)+1} \ln \pa{F(\xi)+1}}{\xi-\eta} \nonumber \\
&= \f{i}{16 \pi^2} \Int{\R}{} \dd \xi\, \dd \eta\, \dd x \, \dd y \, \paa{\f{1}{\xi-\eta+i0}+\f{1}{\xi-\eta-i0}}
\tau(x) \tau(y) \pa{x-y} \ex{ix \eta + i y \xi}\;. \nonumber \\
\end{align*}
Let $H$ be the Heaviside function, then
\begin{align*}
 E\pac{F} &= -\f{1}{8\pi}
\Int{\R}{}\dd \eta \,\dd x \,\dd y \; \tau\pa{x} \tau\pa{y} \pa{x-y}
\pa{\ex{i\eta\pa{x+y}} H\pa{y} -\ex{i\eta\pa{x+y}} H\pa{-y} } \nonumber \\
&= -\f{1}{4} \Int{0}{+\infty} \dd y \, \tau(y)\, \tau(-y)\, \pa{-2y}
   + \f{1}{4} \Int{-\infty}{0} \dd y \,\tau(y)\,\tau(-y)\, \pa{-2y} \nonumber \\
&= \Int{0}{+\infty} \dd y\, \tau(y)\,\tau(-y)\, y  \; ,
\end{align*}
which ends the proof of Theorem~\ref{theor-Akhiezer-Kac}.
\qed

It happens that this correspondence between truncated Wiener--Hopf
operators and generalised sine kernels can be pushed further so as
to obtain the asymptotic behaviour of Fredholm determinants of
truncated Wiener--Hopf operators with symbols having Fischer--Hartwig
type discontinuities. Considering the GSK for finite $q$ 
corresponds to the asymptotic behaviour of a determinant whose symbol
has two jumps. The case of symbols having general Fischer--Hartwig
type singularities is studied in
\cite{KozWienerHopfWithFischerHartwig,DeiftIK2008,DeiftIK2009}. The results for the case of Toeplitz, Hankel and Toeplitz + Hankel determinants with Fisher--Hartwig singularities appeared recently in \cite{DeiftIK2008,DeiftIK2009}. 
%

\subsection{The resolvent of truncated Wiener--Hopf operators}
\label{soussection Resolvent Wiener--Hopf}

\begin{prop}\label{prop-resol-Wiener--Hopf}
Let $I+K$ be a truncated Wiener--Hopf operator on $\intoo{-x}{x}\,$,
\begin{equation}
\pac{(I+K).g} (t)=g(t) + \Int{-x}{x} K(t-t')\, g(t')\, \dd t', \quad
\text{with}\quad K(t)=\mc{F}^{-1}[F](t).
\label{operateur de Wiener--Hopf a interval symetrique2}
\end{equation}
Suppose that there exists $\delta>0$ such that
\begin{itemize}
\item $F$ admits an analytic continuation to $\paa{z: \abs{\Im (z)} \leq \de}$;
\item  $\xi \mapsto F\pa{\xi \pm i\de}
\in L^{1}\pa{\R}$;
\item the analytic continuation of $1+F$ does not vanish on $U$.
\end{itemize}

\noindent  Then the resolvent $I-R$ of $I+K$ fulfills
%
%
%
\begin{equation}
R(\la,\mu)= \Int{\R}{} \f{\dd \xi \dd \eta}{4 i {\pi}^2}\,  F(\xi)
\paa{\f{\a_+(\eta)}{\a_-(\xi)} \ex{ix \pa{\xi-\eta} }
-\f{\a_+(\xi)}{\a_-(\eta)} \ex{-ix \pa{\xi-\eta}} } \f{\ex{i\pa{\mu\eta-\la\xi}}}{\xi-\eta} 
 + \e{O}\big(\ex{-2 \de x}\big),
\label{Resolvent Wiener--Hopf Asymptotique1}
\end{equation}

where
\begin{equation}\label{alpha-def}
\a(\la)=\exp\paa{\Int{\R}{} \f{\nu\pa{\mu}}{\mu-\la} \dd \mu },
\qquad \text{and}\quad \nu(\la)=\f{i}{2\pi} \ln \pa{1+F(\la)}.
\end{equation}
\end{prop}

\Proof
The GSK associated to $I+K$ through
the transformation $\mc{F}^{-1} \circ \pac{I+V} \circ \mc{F}=I+K$
acts on the whole real axis with the kernel
\begin{equation}
V\pa{\xi,\eta}= F(\xi)\,\f{\ex{i\pa{\xi-\eta} x}-\ex{i\pa{\eta-\xi} x}}{2i\pi \pa{\xi-\eta}} \; .
\end{equation}

Just as for the leading asymptotics of $\ln\ddet{}{I+K}$ (see
Section~\ref{soussection Resolvent Wiener--Hopf}), one can obtain
the leading asymptotic of the resolvent of $V$ just by taking
formally the limit $q\tend +\infty$ in all the expressions derived
in the first part of the article. Note that in this process all
power law corrections vanish: they are computed as contour integrals
around $\pm q$ and, since $F$ approaches $0$ sufficiently fast at
infinity, the residues at $\pm q$ vanish in the $q \tend +
\infty$  limit. However, in order to justify this limit, one should also
check that all the uniform estimates still hold for $q \tend
+\infty$.

An alternative way is to consider from the very beginning a RHP for
$\chi$ on the whole real axis $\R$. This is actually much
simpler, than the RHP on a finite interval. Then it is enough to
perform the first two transformations described in
Section~\ref{Section solution RHP} so as to obtain jump matrices
that are already uniformly  close to $I_2$ up to
exponentially small corrections in $x$. Moreover, the jump matrices for
this RHP are given by $M_+$ and $M_-^{-1}$ \eqref{matrixM_+M-}, so
that they approach  the identity matrix at $\lambda\tend\infty$ just
as fast as $F$ goes to zero at infinity. As expected, there is no
need for parametrices, and the corrections are immediately
exponentially decreasing with $x$. It means that, up to uniformly
exponentially small corrections, the resolvent $R_V$ of $V$ is given
by
\begin{equation}
R^{\pa{0}}_V\pa{\xi,\eta}= \f{F\pa{\xi}}{2i\pi \pa{\xi-\eta}}
\paa{\f{\a_+\pa{\eta}}{\a_-\pa{\xi}} \ex{ix\pa{\xi-\eta}}-\f{\a_+\pa{\xi}}{\a_-\pa{\eta}} \ex{-ix\pa{\xi-\eta}}} ,
\label{resolvent de V}
\end{equation}
where as usual $\a\pa{\la}$ is given  by \eqref{alpha-def}. Note
that the integral in \eqref{alpha-def} is well defined in virtue of
the assumptions made on $F$.

We should now take the Fourier/inverse Fourier of $R_V$ in order to get $R$. To this end, we must justify that
the sub-leading corrections do admit a Fourier transform in two variables. Recall the exact expression for the resolvent:
\begin{equation}
R_V\pa{\la,\mu}= R_{V}^{\pa{0}}\pa{\la,\mu} +
\bra{F^{L;\pa{0}}\pa{\la}}\f{\Pi^{-1}\pa{\la}\Pi\pa{\mu}-I_2}{\la-\mu}\ket{F^{R;\pa{0}}\pa{\mu}} \; \; .
\end{equation}
Here, the matrix $\Pi$ is defined in terms of $\Pi_+\pa{\la}$,
the limiting value of $\Pi$ on $\Sg_\Pi$ when $\la$ approaches a
point of $\Sg_\Pi$ from the $``+$'' side of the contour:
\begin{multline}
\Pi\pa{\la}
=I_2+ \Int{\Ga_+}{} \f{\dd z}{\la-z} \Pi_+\pa{z}  \pa{\ba{cc} 0 &1\\0&0 \ea }
\f{ F\pa{z} \a_{+}^{-2}\pa{z} }{1+F\pa{z}} \ex{2 ix z}  \\
+ \Int{\Ga_-}{} \f{\dd z}{\la-z} \Pi_+\pa{z}  \pa{\ba{cc} 0 &0\\1&0 \ea }
\f{ F\pa{z} \a_{-}^{2}\pa{z} }{1+F\pa{z}} \ex{2 ix z} \;\; .
\label{N integral equation at q infinite}
\end{multline}
The $L^1$ integrability of $F$ as well as the asymptotic condition $\Pi\pa{\la}\limit{\la}{+\infty} I_2$ guarantee
that the integrals are well defined. Thus one readily infers from \eqref{N integral equation at q infinite}
the asymptotics of $\Pi$ on the real axis:
\begin{equation}
\Pi\pa{\la}=I_2+\f{\ex{-2\de x}C}{\la} +\e{O}\paf{\ex{-2\de x}}{\la^2} \; ,
\end{equation}
where $C$ is some constant matrix and where we have explicitly extracted the exponential decay in $x$ of the matrix $C$.
Hence using the boundedness of $f_{\pm}^{\pa{0}}$ on the real axis we obtain that
\begin{equation}
R_V\pa{\la,\mu}=R_V^{\pa{0}}\pa{\la,\mu}+\ex{-2\de x} C \f{F\pa{\la}}{\la\mu} +\e{o}\paf{\ex{-2\de x}F\pa{\la} }{\la\mu}.
\end{equation}
Hence the corrections admit a Fourier transform in $\la$ and an inverse Fourier transform in $\mu$ as oscillatory integrals. Therefore,
taking the Fourier transform does not change the nature of the corrections.
\qed

It is clear that, up to a similarity transformation, a Wiener--Hopf
operator on $\intff{-a}{b}$ has the same generalised sine kernel
as the same operator acting on $\intff{0}{a+b}$. Therefore
our method works for any interval, of course up to a similarity
transformation on the resolvent \eqref{resolvent de V} of $V$. We
chose here to present this less standard form of Wiener--Hopf
operators as it fits better the forthcoming application.

We apply our asymptotic inversion formula for truncated Wiener--Hopf
operators acting on a symmetric interval $\intff{-x}{x}$ to
re-derive some formulas concerning the low magnetic field behaviour
of the so-called dressed charge \cite{BogoliubiovIzerginKorepinBookCorrFctAndABA}. This function, traditionally
denoted $Z(\la)$, describes the intrinsic magnetic moment of an
elementary excitation above the ground state in the XXZ
spin-$\tf{1}{2}$ model. It satisfies the following integral equation:
\begin{equation}\label{def-Z}
Z(\la)+ \Int{-x}{x}\!\! \dd \mu  \, K(\la-\mu) \, Z(\mu) = 1,
\quad \text{with}\quad
K(\la)=\f{\sin 2 \zeta}{2\pi
\sinh(\la+i\zeta)\,\sinh(\la-i\zeta)}\;.
\end{equation}
\noindent $K$ is often called the Lieb kernel and $\zeta \in \intoo{0}{\pi}$ is some real parameter describing
the coupling of the model. The large parameter $x$ is a function of the external longitudinal magnetic field; it
goes to infinity when the magnetic field vanishes.

For the study of $Z$, one should distinguish two domains in the
interval $\intff{-x}{x}$: the bulk, i.e. the region $\abs{\la} \ll
x$, and the boundaries $\la\sim\pm x$. While the asymptotic
value of $Z(\la)$ in the bulk ($\abs{\la} \ll x$) is enough to
describe the intrinsic magnetic moment of elementary excitations,
the value of $Z$ at the boundaries ($\pm x$) determines the critical
exponents of the two-point functions of the model \cite{Haldane1980,Haldane1981a,Haldane1981b,BogoluibovIzerginKorepin1986}. As we will see,
the bulk and the boundary behaviour of the dressed charge differ
fundamentally.

First, let us note that, setting directly $x = +\infty$ in \eqref{def-Z}, one can solve explicitly the integral equation for $Z$
by taking the Fourier transform: one obtains in this case that $Z(\la)$ is equal to a constant value $Z(\la)=\tf{\pi}{\pac{2\pa{\pi-\zeta}}}$ on the whole real axis.

Let us now consider the limit $x\tend\infty$ in  \eqref{def-Z} in a
more accurate way, namely, taking $x$ large but finite, and use the
method described above. Let $\wh{K}$ be the Fourier transform of
$K$,
\begin{equation}
\wh{K}\pa{\xi}\equiv\mc{F}\pac{K}\pa{\xi}=\f{\sinh\pac{\xi\pa{\zeta-\tf{\pi}{2}}}}{\sinh
\pa{\tf{\xi \zeta}{2}}} \; .
\end{equation}
Then in virtue  of Proposition~\ref{prop-resol-Wiener--Hopf},
\begin{align}
Z\pa{\la}&=1-\Int{-x}{x} R\pa{\la,\mu} \dd \mu \\
&= 1- \Int{\R}{}  \f{\dd \xi }{2i\pi} \f{\wh{K}\pa{\xi}}{\xi}
\paa{ \f{\a_+\pa{0}}{\a_{-}\pa{\xi}} \ex{i\pa{x-\la}\xi} -  \f{\a_+\pa{\xi}}{\a_{-}\pa{0}} \ex{-i\pa{x+\la}\xi}}.
\end{align}

First let us  study the bulk limit i.e. $\abs{\la}\ll x$. Using the
jump equation satisfied by $\a_{\pm}$: $\big[1+\wh{K}(\la)\big]
\a_+(\la)=\a_-(\la)$, we recast the integrand as
\begin{align}
Z\pa{\la} &= 1- \Int{\R}{}  \f{\dd \xi }{2i\pi} \f{\wh{K}\pa{\xi}}{\big[1+\wh{K}\pa{\xi}\big]\pa{\xi-i0^+}}
\paa{ \f{\a_+\pa{0}}{\a_{+}\pa{\xi}} \ex{i\pa{x-\la}\xi} -  \f{\a_-\pa{\xi}}{\a_{-}\pa{0}} \ex{-i\pa{x+\la}\xi}} \nonumber \\
&= \f{1}{1+\wh{K}\pa{0}}-
\Int{\R}{} \f{\dd \xi}{2i\pi}
\left\{
\f{\wh{K}(\xi+i\tf{\zeta}{2})}{1+\wh{K}(\xi+i\tf{\zeta}{2})  }\,
\f{\a_+(0)}{\a_+(\xi+i\tf{\zeta}{2})}\,
\f{ \ex{\pa{x-\la}(i\xi-\zeta/2)} }{\xi+\tf{\zeta}{2}} \right. \nonumber \\
& \hspace{2.8cm} -\left.
\f{\wh{K}(\xi-i\tf{\zeta}{2})}{1+\wh{K}(\xi-i\tf{\zeta}{2}) }\,
\f{\a_-(0) }{\a_-(\xi-i\tf{\zeta}{2})}\,
\f{\ex{-\pa{x+\la}(i\xi+\zeta/2)} }{ \xi-\tf{\zeta}{2} }
 \right\}  . \label{Z dans le bulk}
\end{align}
Here we have separated the integral into two parts and then moved the contour to the upper/lower half-plane. This
gives a pole contribution from $\xi=i0$. The integral appearing in \eqref{Z dans le bulk} is clearly a
$\e{O}\pa{\ex{-\pa{x-\abs{\la}}\tf{\zeta}{2} }}$.
So that, in the bulk,
\begin{equation}\label{Z-bulk}
 Z(\la)\sim\f{1}{1+\wh{K}\pa{0}}=\f{\pi}{2\pa{\pi-\zeta}}\;,
\end{equation}
up to exponentially
small corrections. As expected, we recover the value of $Z$ obtained in the case of an infinite interval. Note that
the corrections become larger and larger as we approach any of the endpoints $\pm x$.

Let us now study the behaviour of the dressed charge at the boundaries. Since the kernel $K$ is even, so is $Z$.
We can thus focus on a single boundary, say $\la=x$. We have,
\begin{align}
Z(x) &= 1- \Int{\R}{}  \f{\dd \xi }{2i\pi\pa{\xi-i0^+}}
\paa{
\wh{K}(\xi)\, \f{\a_+(0)}{\a_{-}(\xi)}
 - \f{\wh{K}(\xi)}{1+\wh{K}(\xi)}\,
   \f{\a_-(\xi)}{\a_{-}(0)}\, \ex{-2ix\xi}} .
\end{align}
As before, the integral of the second term gives an exponentially
small contribution  $\e{O}\pa{\ex{-x\zeta}}$. The integral of the
first term is explicitly computable. Using once again the jump
equation satisfied by $\a_\pm(\xi)$,
 we have,
\begin{align}
Z(x)
 &= 1- \a_+(0)  \Int{\R}{}  \f{\dd \xi }{2i\pi}
\f{\a^{-1}_+(\xi)-1+1-\a^{-1}_-(\xi)  }{\xi-i0}
+
\e{O}\pa{\ex{-x\zeta}}\nonumber\\
 &= 1- \a_+(0)  \Int{\R}{}  \f{\dd \xi }{2i\pi}
\f{\a^{-1}_+(\xi)-1 }{\xi-i0} +
\e{O}\pa{\ex{-x\zeta}}\nonumber\\
&=\a_+(0) + \e{O}\pa{\ex{-x\zeta}}\;.
\end{align}
We have computed the remaining integral by residues, since
$\a^{-1}_{\pm}(\xi)-1=\e{O}\pa{\xi^{-1}}$ for $\xi\tend \infty$ in
the respective half plane of holomorphy.

For an even kernel like the Lieb one $\wh{K}(\xi)=\wh{K}(-\xi)$, and
then it follows from the integral representation~\eqref{alpha-def}
of $\alpha$ that $\a_{+}(\xi)=\a_-^{-1}(-\xi)$. This means that
$1+\wh{K}(0)=\a_+^{-2}(0)$. Hence, for $x$ large enough
\begin{equation}
Z(x)\sim\sqrt{\f{1}{1+\wh{K}(0)}}
=\sqrt{\f{\pi}{2\pa{\pi-\zeta}}},
\end{equation}
and the value of $Z(\la)$ at the boundary is the square root of its
value in the bulk up to exponentially small corrections. In the
limit $x\to+\infty$ this correspondence becomes exact.


\section{Asymptotics of multiple integrals}\label{sec-integrales}

We have already mentioned that the
asymptotic expansion of the Fredholm determinant of the GSK can be used
for the asymptotic analysis of correlation functions of quantum
integrable models. For a relatively wide class of integrable systems,
the correlation functions can be presented as series of multiple
integrals of a special type \cite{KitKMST009}. These series can be summed up to
Fredholm determinants for the models equivalent to free
fermions. In the general case, such a reduction to determinants is
not known. However, the asymptotic behaviour of individual terms of the
series can be derived from the asymptotics of the Fredholm
determinant of the GSK. In the present section we consider this problem.

More precisely, our purpose is to derive the large $x$ asymptotic behaviour of the following type of integrals (cycle integrals):
\begin{equation}\label{integ-F}
\mc{I}_n\pac{\mc{F}_n}
=\oint\limits_{\Gamma\pa{\intff{-q}{q}}}^{}
\hspace{-3mm}\f{\dd^n z}{\pa{2i\pi}^n}
\Int{-q}{q}\f{\dd^n \la}{\pa{2i\pi}^n} \; \,
\mc{F}_n\pa{\ba{c}\paa{\la} \\ \paa{z} \ea }\;
\prod_{j=1}^n\f{\ex{ ix\pa{p\pa{z_j}-p\pa{\la_j}}  }}
  {\pa{z_j-\la_j}\pa{z_j-\la_{j+1}}}\;.
\end{equation}
In this expression, $\mc{F}_n$ is a holomorphic function of $2n$ variables $\la_1, \dots, \la_n,z_1,\dots, z_n$ in $U^n\times W^n$, in which $U$ and $W$ are open neighbourhoods of $\intff{-q}{q}$, and $\Gamma\pa{\intff{-q}{q}}$ denotes a closed counter clock-wise contour in $W$ surrounding $\intff{-q}{q}$ with index $1$. We moreover assume that $\mc{F}_n$ is symmetric separately in the $n$ variables $\la_1,\ldots,\la_n$ and in the $n$ variables $z_1,\ldots,z_n$. Finally, we agree upon $\la_{n+1}\equiv\la_1$.

\subsection{Leading asymptotic behaviour of $\mc{I}_n\pac{\mc{F}_n}$}

Let us first suppose that the function $\mc{F}_n$ is of the special (factorized) type
\begin{equation}\label{Fn-phi-psi}
 \mc{F}_n^{(\varphi,\phi)}\pa{\ba{c}\paa{\la} \\ \paa{z} \ea }
 =\prod_{i=1}^n\big[ \varphi(\la_i)\,\phi(z_i) \big] \; ,
\end{equation}
where $\varphi$ is a one-variable holomorphic function in $U$ and $\phi$ is a one-variable holomorphic function in $W$, non-vanishing on $W$.
For two such functions $\varphi$ and $\phi$ we introduce the associated GSK
$V^{(\varphi,\phi)}$ given by \eqref{generalised Sine} provided the identification $F(\la)=\varphi(\la)\phi(\la)$ and $e^{g(z)}=\phi(z)$ is made.
Then, the integral \eqref{integ-F} can be expressed in terms of 
$\log\det\big[I+V^{\pa{\varphi,\phi}}\big]$ as
\begin{align}
\mc{I}_n\big[\mc{F}_n^{(\varphi,\phi)}\big]
&=\Int{-q}{q}\dd^n\la \, \prod_{k=1}^n V^{(\varphi,\phi)}(\la_k,\la_{k+1})
              \nonumber\\
&=\f{\pa{-1}^{n-1} }{(n-1)!}\; \partial^n_{\ga}
 \ln \det\big[I+V^{(\varphi,\phi)}\big]
\pour{\ga=0}.
\label{int-phi-psi}
\end{align}
In this specific case, it is straightforward to obtain the leading asymptotic behaviour of the multiple integral \eqref{int-phi-psi} in the large $x$ limit thanks to the results of the previous sections.

This remark leads us to the following definition:

\begin{defin}\label{def-Hn-0}
Let $U$, $W$ be two open neighbourhoods of $\intff{-q}{q}$, and let $\mc{H}(U)$ (resp. $\mc{H}(W)$) be the set of holomorphic functions on $U$ (resp. on $W$).
Let also
\begin{equation}
\widetilde{\mc{S}}_n ^{\, U,W}\!
 =\paa{
  \sul{\ell=1}{p} \mc{F}_n^{(\varphi_{\ell},\phi_{\ell})}\; ; \
  p \in \mathbb{N} ,\
  \pa{\varphi_{\ell},\phi_{\ell}}\in
  \mc{H}(U)\times\mc{H}(W) \ \text{and}\ {\phi_{\ell}}\pour{\intff{-q}{q}} \not
= 0  }\; ,
\end{equation}
in which  $\mc{F}_n^{(\varphi,\phi)}$ denotes a pure factor function of $2n$ variables defined in terms of $\varphi$ and $\phi$ as in \eqref{Fn-phi-psi}.
We define the linear functional $I_n^{(0)}$ on $\widetilde{\mc{S}}_n ^{\, U,W}$ as
\begin{equation}
 I_n^{(0)}\big[\mc{F}_n^{(\varphi,\phi)} \big]
=
 \f{\pa{-1}^{n-1}}{(n-1)!}\; \partial^n_{\ga}
 \ln \det \big[I+V^{(\varphi,\phi)}\big]^{(0)}
 {}\pour{\ga=0} \, ,
\label{asymptotique logdet ordre n}
\end{equation}
and by imposing linearity on functions $\sum_{\ell=1}^{p} \mc{F}_n^{(\varphi_\ell,\phi_\ell)}$.
Here $V^{(\varphi,\phi)}$ denotes the generalised sine kernel \eqref{generalised Sine}  with
$F(\la)=\varphi(\la)\phi(\la)$ and
$\ex{g(z)}=\phi(z)$, and $\ln \det \big[I+ V^{(\varphi,\phi)}\big]^{\pa{0}}$ denotes the
leading asymptotics of the Fredholm determinant $\ln \det\big[I+V^{(\varphi,\phi)}\big]$ as in
Theorem~\ref{theorem asymptotiques order zero log det}.
\end{defin}

It is easy, using the expression \eqref{asymptotique log det ordre zero} of $\ln\det\big[I+V^{(\varphi,\phi)}\big]^{(0)}$, to obtain an explicit expression for $I_n^{(0)}\big[\mc{F}_n^{(\varphi,\phi)} \big]$:
\begin{multline}
I_n^{(0)}\big[\mc{F}_n^{(\varphi,\phi)} \big]
=  \Int{-q}{q} \f{\dd \la}{2i\pi}\,
   \varphi^{n}(\la)\,\phi^{n-1}(\la)
  \pa{i x p'(\la)\,\phi(\la)+\phi'(\la)} \\
+ \sul{\sigma =\pm}{}
\big(b_n-c_n\ln\pa{2qp'_{\sigma}x}\big)\,
[\varphi(\sigma q)\,\phi(\sigma q) ]^{n}
         \\
+
\f{n}{4\pi^2}\sul{\sigma=\pm}{} \sul{p=1}{n-1}
 \Int{-q}{q} \dd \la \, \f{[\varphi(\sigma q)\,\phi(\sigma q)
]^{n}
   - [\varphi(\sigma q)\,\phi(\sigma q) ]^{p}\,
     [\varphi(\la)\,\phi(\la) ]^{n-p} }
{p\pa{n-p}\pa{q-\sigma \la}}
  \\
+
\f{n}{8\pi^2}\sul{p=1}{n-1} \Int{-q}{q}
 \f{\dd \la \, \dd \mu }{\pa{n-p}\pa{\la- \mu}}
\Bigl\{\Dp{\la}[\varphi(\la)\, \phi(\la) ]\;
[\varphi(\la)\, \phi(\la) ]^{p-1}\,
[\varphi(\mu)\, \phi(\mu) ]^{n-p}   \\
- \Dp{\mu}[\varphi(\mu)\, \phi(\mu) ]\;
[\varphi(\mu)\, \phi(\mu) ]^{p-1}\,
[\varphi(\la)\, \phi(\la) ]^{n-p}  \Bigr\}
\; ,
\label{Demon de knorr ou H_n explicit}
 \end{multline}
where $b_n$ and $c_n$ are given by \eqref{5-bncn}.
The $x \tend +\infty$ asymptotics of the Fredholm determinant are
uniform in $\ga$ to any fixed order $n$ in
$\Dp{\ga}^{n}$. This means that
\begin{equation}
\mc{I}_n\big[\mc{F}_n^{(\varphi,\phi)}\big]
=I_n^{(0)}\big[\mc{F}_n^{(\varphi,\phi)}\big] +
\e{o}\pa{1}.
\end{equation}
In the next proposition we show that $I_n^{(0)}$ can be extended into a linear functional on the space
of holomorphic functions $\mc{F}_n$ (not necessarily of the form $\mc{F}_n^{(\varphi,\phi)}$) that are symmetric in  $n$ variables $\la_1,\ldots,\la_n$
and  $n$ variables $z_1,\ldots,z_n$ separately. This extension of $I_n^{(0)}$, as we prove below, is the good  way to evaluate cycle integrals \eqref{integ-F} with such arbitrary symmetric functions $ \mc{F}_n$.

\begin{prop}\label{Prop-dens}
Let $U$ and $W$  be open neighbourhoods of $\intff{-q}{q}$, and let
$\mc{S}\e{ym}_n(U,W)$ be the set of holomorphic functions $\mc{F}_n$
on $U^n\times W^n$ of $2n$ variables $\la_1,\ldots,\la_n,
z_1,\ldots,z_n$, symmetric in the $n$ variables $\la_1,\ldots,\la_n$
and in the $n$ variables $z_1,\ldots, z_n$ separately. Then
$I_n^{(0)}$ extends to a continuous linear functional on
$\mc{S}\e{ym}_n(U,W)$ endowed with the topology of the sup norm
convergence on compact sets.
\end{prop}

\Proof
$I_n^{(0)}\big[\mc{F}_n^{(\varphi,\phi)} \big]$ contains at most first order
derivatives of the functions $\varphi$ and $\phi$. Now recall that, for any compacts $K$, $P$ such that\footnote{Here $\overset{\circ}{P}$ is the interior of P} $K \subset \overset{\circ}{P}$ and
$P\subset U$
\begin{equation}
\forall  \, k  \in \mathbb{N},\   \exists\, c_k \ \text{such that}\
\forall\, \phi \in \mc{H}\pa{U},\
\norm{\phi^{\pa{k}}}_{0;K} \leq c_k \norm{\phi}_{0;P} \; ,
\label{continuite norme 1 par rapport norme 0}
\end{equation}
where $\norm{.}_{0;K}={\sup}_{z \in K}\abs{.}$ is the sup norm with
support on the compact $K$. In consequence,  $I_n^{(0)}$ is
continuous on $\widetilde{\mc{S}}_n^{\, U,W}$. The latter is dense
in $\mc{S}_n^{\, U,W}$, with
\begin{equation}
\mc{S}_n^{\, U,W}
 =\paa{
  \sul{\ell=1}{p} \mc{F}_n^{(\varphi_{\ell},\phi_{\ell})}\; ; \
  p \in \mathbb{N} ,\
  \pa{\varphi_{\ell},\phi_{\ell}}\in
  \mc{H}(U)\times\mc{H}(W)} .
\end{equation}
Hence $I_n^{(0)}$ extends by density to a continuous linear functional on
$\mc{S}_n^{\, U,W}$.
Due to the density Theorem~\ref{theoreme de densite} (See Appendix C), we have that $\mc{S}_n^{\, U,W}$ is dense in $\mc{S}\e{ym}_n(U, W)$.
Therefore $I_n^{(0)}$ extends to
 a linear functional on $\mc{S}\e{ym}_n(U , W)$.
\qed

\begin{cor}\label{cor-Hn de Fn}
Let $U$ and $W$  be open neighbourhoods of $\intff{-q}{q}$, and let $\mc{F}_n\!\in\mc{S}\e{ym}_n(U , W)$. Then,
\begin{multline}
I_{n}^{(0)}\big[\mc{F}_n\big]
=\f{1}{2i\pi} \Int{-q}{q} \dd\la \,
\paa{ixp'(\la) + \partial_{\eps}}
\mc{F}_n\pa{\ba{c} \paa{\la}^{n} \\ \{\la+\eps\}, \paa{\la}^{n-1} \ea}
\pour{\eps=0}
\\
+ \sul{\sg=\pm}{} \pa{b_n-c_n\ln\pa{2qp'_{\sg}x}}
\mc{F}_n\pa{\ba{c} \paa{\sg q}^{n} \\ \paa{\sg q}^{n} \ea}
 \\
 +\f{n}{\pa{2\pi}^2} \sul{\sg=\pm}{} \sul{p=1}{n-1}\Int{-q}{q} \dd \la
\f{\mc{F}_n\pa{\ba{c}\paa{\sg q}^{n} \\ \paa{\sg q}^{n} \ea }- \mc{F}_n\pa{\ba{c}\paa{\sg q}^{p} , \paa{\la}^{n-p}\\
\paa{\sg q}^{p} , \paa{\la}^{n-p}\ea } }{p\pa{n-p}\pa{q-\sg \la}} \\
+ \f{n}{2\pa{2\pi}^2}\sul{p=1}{n-1}\Int{-q}{q} \f{\dd \la \dd \mu }{\pa{n-p}\pa{\la- \mu}}
\left\{
\partial_{\eps}\mc{F}_n\pa{ \ba{c}\paa{\la+\eps},\paa{\la}^{p-1},\paa{\mu}^{n-p}\\ \paa{\la+\eps} ,\paa{\la}^{p-1},\paa{\mu}^{n-p}\ea}\right.
                 \\
\left.
-\partial_{\eps}\mc{F}_n\pa{\ba{c} \paa{\mu+\eps} ,\paa{\mu}^{p-1},\paa{\la}^{n-p} \\ \paa{\mu+\eps},\paa{\mu}^{p-1},\paa{\la}^{n-p} \ea}
\right\}\pour{\eps=0}\; .
\label{forme fonctionnelle de H1}
\end{multline}
There $\{\la\}^n$ denotes the set formed by $n$ copies of the same parameter $\la$.
\end{cor}

\Proof Apply Theorem~\ref{theoreme de densite} to \eqref{Demon de
knorr ou H_n explicit}. \qed

Finally, we have the following large $x$ asymptotic behaviour for integrals of the form \eqref{integ-F}
(which seems hardly attainable through a direct analysis of the multiple integrals):

\begin{theorem}\label{th_asympt_In}
Let $U$ and $W$  be open neighbourhoods of $\intff{-q}{q}$, and let $\mc{F}_n\in\mc{S}\e{ym}_n(U , W)$. Then, when
$x\tend+\infty$, the integral $\mc{I}_n\big[\mc{F}_n\big]$
\eqref{integ-F} behaves as
\begin{equation}\label{In-ordre-0}
\mc{I}_n\big[\mc{F}_n\big]
=I_n^{(0)}\big[\mc{F}_n\big] +
\e{O}\paf{\ln^n x }{x} ,
\end{equation}
the explicit expression of $I_n^{(0)}\big[\mc{F}_n\big]$ being given in Corollary~\ref{cor-Hn de Fn}.
\end{theorem}
The whole difficulty of the proof
is to show that the small $\e{o}\pa{1}$ in \eqref{asympt-V-ordre0} is preserved by the density procedure formulated in
Theorem~\ref{theoreme de densite}. This is nontrivial
since the series converging to $\mc{F}_n$ may not converge
absolutely.
We need therefore, so as to prove this theorem, to study more precisely the sub-leading corrections and to see how they pass through all the steps described above. This will be done in the next subsection.

\subsection{Study of sub-leading corrections}

In this subsection, we study the behaviour of the sub-leading corrections to $\ln\det[I+V]^{(0)}$ when the above procedure is applied.
In particular, we will show that they indeed remain subleading, which will prove Theorem~\ref{th_asympt_In}.
In fact we will prove an even  more general result:

\begin{theorem}\label{th_asympt_In-sub}
Let $U$ and $W$  be open neighbourhoods of $\intff{-q}{q}$, and let $\mc{F}_n\in\mc{S}\e{ym}_n(U , W)$. For any positive integer $M$, there exists a continuous linear functional $I^{\pa{M}}_n$
such that
\begin{equation}\label{In-ordre-m}
\mc{I}_n\big[\mc{F}_n\big]
=I_n^{(M)}\big[\mc{F}_n\big] +
\e{O}\paf{\ln^n x }{x^{M+1}}  \quad \e{when} \; x\tend+\infty.
\end{equation}
\end{theorem}

The explicit expression for $I_n^{\pa{M}}$ can be obtained by some perturbative computations that become more and more involved with the growth of $M$.
We will nevertheless obtain the general structure for $I_n^{(M)}$, showing that it can be decomposed in terms of non-oscillating and
oscillating contributions, with oscillating factors of the form $e^{imx(p_+-p_-)}$, $m\in\mathbb{Z}^*$:
\begin{align}
 I_n^{(M)}\big[\mc{F}_n\big]
&=I_n^{(0)}\big[\mc{F}_n\big]
+\sul{N=1}{M}\f{1}{x^N}\;I_{n}^{(N;\,\e{nosc})}\pac{\mc{F}_n}
+\sul{N=2}{M}\f{1}{x^N}\;I_{n}^{(N;\,\e{osc})}\pac{\mc{F}_n},\\
&=I_n^{(0)}\big[\mc{F}_n\big]
+\sul{N=1}{M}\f{1}{x^N}\;I_{n}^{(N;\,\e{nosc})}\pac{\mc{F}_n}
+\hspace{-2mm}
 \sum_{\substack{m\in\mathbb{Z}^*\\ |m|\le M/2}}
 \hspace{-2mm} e^{ixm(p_+-p_-)}\hspace{-2mm}
\sul{N=2|m|}{M}\f{1}{x^N}\;I_{n}^{(N;\,m)}\pac{\mc{F}_n},
\end{align}
$I_{n}^{(N;\,\e{nosc})}\pac{\mc{F}_n}$ (resp. $I_{n}^{(N;\,\e{osc})}\pac{\mc{F}_n}$) being given in terms of the function $\mc{F}_n$ and of its derivatives up to order $N$ (resp. up to order $N-2$) evaluated at $\pm q$ or integrated from $-q$ to $q$.
%

\subsubsection{General strategy}

In the previous subsection, we have defined the functional $I_n^{(0)}$ from the leading asymptotic part $\ln\det[I+V]^{(0)}$ \eqref{asymptotique log det ordre zero} of the GSK.
More precisely, we have seen in Corollary~\ref{cor-Hn de Fn} that $\Dp{\ga}^n\ln\ddet{}{I+V}^{\pa{0}}\mid_{\ga=0}$ yields the functional
$\pa{-1}^{n-1}\pa{n-1}!\, I_n^{(0)}\pac{\mc{F}_n}$ after the density procedure, as explained in Proposition~\ref{Prop-dens} and Theorem~\ref{theoreme de densite}, is applied.
In order to estimate the corrections to $I_n^{(0)}\big[\mc{F}_n\big]$ for the large $x$ behaviour of cycle integrals $\mc{I}_n\big[\mc{F}_n\big]$ of
length $n$ \eqref{integ-F}, we have to take into account the corrections $\ln\ddet{}{I+V}^{\e{sub}}$
to  $\ln\det[I+V]^{(0)}$,
\begin{equation}
\ln\ddet{}{I+V}=\ln\ddet{}{I+V}^{\pa{0}} + \ln\ddet{}{I+V}^{\e{sub}},
\end{equation}
and to analyze the effect of the density procedure on the $n$-th $\gamma$-derivative of the subleading part $\Dp{\ga}^n\ln\ddet{}{I+V}^{\e{sub}}\mid_{\ga=0}$. We will show in particular that it preserves the small $\e{o}(1)$ with respect to the $x\tend +\infty$ limit, i.e. that $\Dp{\ga}^n\ln\ddet{}{I+V}^{\e{sub}}\mid_{\ga=0}$ can only generate $\e{o}(1)$ corrections.

In the spirit of Definition~\ref{def-Hn-0}, we therefore introduce the:

\begin{defin}\label{def-Hn-sub}
Let $U$, $W$ be two open neighbourhoods of $\intff{-q}{q}$.
We define the linear functional $I_n^{\e{sub} }$ on $\widetilde{\mc{S}}_n ^{\, U,W}$ as
\begin{equation}
 I_n^{\e{sub} }\big[\mc{F}_n^{(\varphi,\phi)} \big]
=
 \f{\pa{-1}^{n-1}}{(n-1)!}\; \partial^n_{\ga}
 \ln \det \big[I+V^{(\varphi,\phi)}\big]^{\e{sub} }
 {}\pour{\ga=0} \, ,
\label{eq-def-Hn-sub}
\end{equation}
and by imposing linearity on functions $\sum_{\ell=1}^{p} \mc{F}_n^{(\varphi_\ell,\phi_\ell)}$.
Here, as in  Definition~\ref{def-Hn-0}, $\mc{F}_n^{(\varphi,\phi)}$ denotes a  factorized function of $2n$ variables defined in terms of $\varphi$ and $\phi$ as in \eqref{Fn-phi-psi}, and  $V^{(\varphi,\phi)}$ denotes the generalised sine kernel \eqref{generalised Sine}  with
$F(\la)=\varphi(\la)\phi(\la)$ and
$\ex{g(z)}=\phi(z)$.
\end{defin}

According to the scheme presented in the previous subsection, the next steps will be:
\begin{itemize}
 \item to obtain a convenient representation for $I_n^{\e{sub} }\big[\mc{F}_n^{(\varphi,\phi)} \big]$: this means in particular to obtain the form of $n$-th $\gamma$-derivatives of $\log\det [I+V]^\e{sub}$ in terms of the functions $F$ and $g$, to set $g(z)=\log \phi(z)$ and $F(z)=\varphi(z)\,\phi(z)$, and to estimate this result in the large $x$ limit;
 \item  to apply the density procedure:
one should first extend by density and continuity the functional $I_n^{\e{sub}}$ to the space $\mc{S}_n^{U,W}$;
then, for any holomorphic function $\mc{F}_n$ in $2n$ variables $\la_1,\ldots,\la_n,z_1,\ldots,z_n$, symmetric separately in the variables $\la$ and in the variables $z$, one has to consider a sequence
$\pa{\varphi_k,\phi_k}$ such that $\sum_{k=1}^{N} \mc{F}_n^{\pa{\varphi_k,\phi_k}} \tend \mc{F}_n$ so as to be able to define and characterize $I_n^{\e{sub}}\big[\mc{F}_n\big]$ and to see how it behaves in the large $x$ limit;
 \item to refine the procedure in order to get an asymptotic expansion of $I_n^{\e{sub} }\big[\mc{F}_n \big]$.
\end{itemize}
%

\subsubsection{$\gamma$-derivatives of $\ln\det[I+V]^{\e{sub}}$}

As in  Section~\ref{sec-sub-asympt}, we will obtain the corrections to $\log\det[I+V]^{(0)}$ through the $x$-derivative path, starting from formula \eqref{the x derivative path} that we recast as
\begin{equation}\label{x-path_recasted}
\Dp{x}\ln \ddet{}{I+V}
=-i\Int{-q}{q} \dd \la\, p'(\la)\, \nu\pa{\la}
+\hspace{-2mm} \Oint{\Ga\pa{ \intff{-q}{q} }}{}\hspace{-3mm}
\f{\dd \la}{4\pi}\,p(\la)\, \e{tr} \paa{\pac{\Dp{\la}\Pi(\la)}\,\sg_3\,\Pi^{-1}(\la)  }  .
\end{equation}
Here, as in Section~\ref{sec-sub-asympt}, we have chosen the counter clock-wise contour $\Ga\pa{\intff{-q}{q}}$ to lie in $U$ and to encircle $\Sg_{\Pi}$, which means that $\chi(\la)=\Pi(\la)\,\a^{-\sg_3}(\la)$ on $\Ga\pa{\intff{-q}{q}}$.
Integrating this equation with respect to $x$, we obtain
\begin{multline}\label{log-det-sub}
\ln\ddet{}{I+V}^{\e{sub}}
= \Int{+\infty}{x} \dd x'  \hspace{-3mm}\Oint{\Ga\pa{\intff{-q}{q}}}{}\hspace{-4mm}
\f{\dd \la}{4\pi}\; p(\la)\;
\Bigg\{ \e{tr}\,\Big\{\pac{\Dp{\la}\Pi\pa{\la} }\,\sg_3 \, \Pi^{-1}\pa{\la} \Big\}
 \\
+ \f{1}{x'}\Int{\Sg_{\Pi}}{}
\f{\e{tr}\,\big\{ \De^{\pa{1}}\pa{z;x'}\, \sg_3 \big\} }
  {2i\pi\pa{\la-z}^2}\,  \dd z \Bigg\} \;.
\end{multline}
The convergence of this integral will be proved later on.
We recall that the second term in \eqref{x-path_recasted} produces also, when integrated over $x$, the $\ln x$ term appearing in the definition \eqref{asymptotique log det ordre zero} of $\ln\ddet{}{I+V}^{\pa{0}}$ (see \eqref{terme-log-x}).
We have therefore substracted the corresponding contribution (second term of \eqref{log-det-sub}) in the definition of $\ln\ddet{}{I+V}^{\e{sub}}$.

In order to obtain the $n$-th $\gamma$ derivatives of this expression, we have  to compute the $\ga$-derivatives of $\partial_\la \Pi(\la)$ and of $\Pi^{-1}(\la)$, which in their turn follow from those of the jump matrix $\De(\la)$.

\paragraph{\indent $\bullet$ $\gamma$-derivatives of $\De$}

\par\indent
\smallskip

In order to determine the $n$-th $\gamma$-derivative at $\gamma=0$ of the jump matrix $\De\pa{z}$, it is convenient to express it in the following form:
\begin{equation}
\De(z)
=\kappa^{-\sg_3}(z)\; \wt{\De}(z)\; \kappa^{\sg_3}(z) \, .
\end{equation}
Here, the matrix  $\wt{\De}(z)$ depends on $\ga$ only through the combination $\ga F(z)$, whereas $\kappa^{\pm \sg_3}(z)$ depends on $\gamma$ through the combination  $\int_{-q}^{q} \dd \mu\, \pac{\nu\pa{z}-\nu\pa{\mu}}/({z-\mu})$.

It is easy to compute the multiple $\ga$-derivative of $\wt{\De}(z)$ at $\ga=0$. It is given as
\begin{equation}
\Dp{\ga}^n \,\wt{\De}(z) \pour{\ga=0}
= \Ad_{\ex{\sg_3\, g(z)/2}}\big[\partial_\ga^n \Delta_0(z)\big]\pour{\ga=0}\cdot
  F^n (z).
\end{equation}
In this expression, $\Ad_X[Y]$ stands for the usual adjoint action of the matrix $X$ on the matrix $Y$, and $\Delta_0$ denotes the jump matrix $\Delta$ at $F\equiv 1$ and $g\equiv 0$.

It remains to compute the $\ga$-derivatives of $\kappa^{\pm \sg_3}\pa{z}$.
They follow from the Faa-di-Bruno formula:
\begin{equation}
\Dp{\ga}^n\, \kappa^{\pm \sg_3}(z) \pour{\ga=0}
= \sul{\substack{p_1,\ldots,p_n=0 \\ \Sg_{s=1}^n s p_s =n } }{n}
  \f{n! \pa{\pm \sg_3}^{\Sg_{s=1}^n{p_s}}}
    {\pl{s=1}{n}p_s! }
\pl{s=1}{n} \pac{ \f{\pa{-1}^s}{2i\pi s}\Int{-q}{q} \f{F^{s}\pa{z}-F^{s}\pa{\mu}}{z-\mu} \dd \mu}^{p_{s}} \, .
\end{equation}
Therefore, gathering these informations and applying Leibnitz's rule, we obtain that
\begin{align}
\Dp{\ga}^n\, \De\pa{z} \pour{\ga=0}
 &= \sul{\substack{p+p_0+q=n \\ p_0 \geq 1} }{} \,
    \sul{\substack{ p_1,\ldots,p_n=0 \\ \Sg_{s=1}^n s p_{s}=p } }{n} \,
    \sul{\substack{ q_1,\ldots,q_n=0 \\ \Sg_{s=1}^n s q_{s}=q } }{n}
\f{C^p_n\, C^q_{n-p}\, p!\, q!}{\pl{s=1}{n}\pa{p_s}!\pa{q_s}!} \,
  \nonumber \\
& \hspace{3cm} \times
 \Ad_{\ex{\sg_3\, g(z)/2}}\bigg[
 \pa{-\sg_3}^{\Sg_{s=1}^n p_s }\cdot
 \partial_\ga^{p_0} \Delta_0(z)\pour{\ga=0} \cdot
\pa{ \sg_3}^{\Sg_{s=1}^n q_s }  \bigg] \nonumber \\
& \hspace{3cm} \times F^{p_0}(z)\;
\pl{s=1}{n} \pac{
 \f{\pa{-1}^s}{2i\pi s} \Int{-q}{q} \f{F^s(z)-F^s(\mu) }{z-\mu } \dd \mu
 }^{p_s+q_s}
  \nonumber\\
&= \sul{ \substack{p_0 + \Sg s p_s =n \\  p_0 \geq 1 } }{}
  \dep^{ \pa{\paa{p_i}} }(z;x)\
  F^{p_0}(z) \;
  \pl{s=1}{n}
  \pac{ \Int{-q}{q} \f{F^{s}\pa{z}-F^{s}\pa{\mu}}{z-\mu} \dd \mu}^{p_{s}}
   .
\label{deriveeDelta}
\end{align}
Note that, in the first line, we have extended for convenience the sum over parameters $p_s$ and $q_s$ up to $n$, since anyway, due to the constraint, $p_s\le p$ and $p_s=0$ if $s>p$ (resp. $q_s\le q$ and $q_s=0$ if $s>q$).

In the last line, we have changed the order of summations and incorporated all the $F$ independent prefactors  into
the definition of the matrix $\dep^{\pa{\paa{p_i}}}\pa{z;x}$. More precisely,
\begin{multline}
\dep^{\pa{\paa{p_i}}}\pa{z;x}
= \pl{s=1}{n}\paf{\pa{-1}^s}{2i\pi s}^{p_s}
  \sul{p+q=n-p_0}{} \, \sul{\Sg_{s=1}^n s q_s=q}{}
  \f{C^p_n\, C^q_{n-p}\, p!\, q!}{\pl{s=1}{n}\pa{q_s}!\pa{p_s-q_s}!}
          \\
\times
 \Ad_{\ex{\sg_3\, g(z)/2}}\bigg[
 \pa{-\sg_3}^{\Sg_{s=1}^n (p_s-q_s) }\cdot
 \partial_\ga^{p_0} \Delta_0(z)\pour{\ga=0} \cdot
\pa{ \sg_3}^{\Sg_{s=1}^n q_s }  \bigg]\, .
\label{definition des matrices delta avce plusieurs indices}
\end{multline}
From the properties of the jump matrix $\De_0(z)$, it is easy to see that the diagonal entries of the matrices $\dep^{(\paa{p_i})}\pa{z;x}$ are a $\e{O}\big(x^{-1}\big)$, whereas their off-diagonal ones are
a $\e{O}\pa{\tf{\ln^{p_0} x}{x}}$ uniformly on the
contours $\Sg_\Pi$.

\paragraph{\indent $\bullet$ $\ga$-derivatives of $\partial_\la \Pi$}

\par\indent
\smallskip

Let us recall the integral representation for $\partial_\la \Pi\pa{\la}$, which is a direct consequence of \eqref{int-Pi},
\begin{equation}\label{int-deriv-Pi}
\partial_\la \Pi\pa{\la}= -\frac 1 {2i\pi} \Int{\Sg_\Pi}{} \f{\dd z}{(\la-z)^2}  \Pi_+\pa{z} \De\pa{z}.
\end{equation}
Therefore, the $\ga$-derivatives of $\partial_\la \Pi(\la)$
can be directly obtained from the ones of $\De(\la)$ and of $\Pi_+(\la)$.

Recall that $\Pi_+(\la)$
satisfies the following integral equation on $\mc{M}_2\pa{L^2\pa{\Sg_\Pi}}$:
\begin{equation}\label{int-eq-Pi+}
\pa{I-C^{\De}_{\Sg_\Pi}}\pac{\Pi_+}=I_2 \, ,
\end{equation}
where the operator $C_{\Sg_\Pi}^{\De}$ is defined by
\begin{equation}
C_{\Sg_\Pi}^{\De}\pac{M}\equiv C_{\Sg_\Pi}^{+}\pac{M \De}, \quad
\forall M\in \mc{M}_2\pa{L^2\pa{\Sg_\Pi}} .
\end{equation}
This matrix Cauchy operator is invertible, at least for $x$ large enough. Indeed, using the continuity
of the scalar Cauchy operator:
\begin{equation}
 \exists\, c_2>0\ \text{such that}, \ \forall g \in L^{2}\pa{\Sg_{\Pi}},\quad
 \norm{C^{+}_{\Sg_\Pi}\pac{g}}_{L^2\pa{\Sg_\Pi}} \leq c_2 \norm{g}_{L^2\pa{\Sg_\Pi}},
\end{equation}
one gets that the operator norm $\abs{\norm{C^{\De}_{\Sg_\Pi}}}$ fulfills:
\begin{equation}
 \abs{\norm{C^{\De}_{\Sg_\Pi}}}\leq c_2 \norm{\De}_{\mc{M}_2\pa{L^2\pa{\Sg_\Pi}}} \limit{x}{+\infty} 0 .
\end{equation}
Moreover, $C^{\De}_{\Sg_\Pi}$ being a holomorphic function of  $\ga$ we have that, for $x$ large enough, $(I-C^{\De}_{\Sg_\Pi})$ is invertible and that its inverse is also a holomorphic function of $\ga$. In particular, \eqref{int-eq-Pi+} implies
\begin{equation}
\Dp{\ga}\Pi_+
= \pa{I-C^{\De}_{\Sg_\Pi}}^{-1} \circ \pa{\Dp{\ga} C^{\De}_{\Sg_\Pi}}\! \pac{\Pi_+ }\; .
\end{equation}
A straightforward induction shows that there exist some coefficients $\mf{c}_r^{(\{p_i\})} \in \mathbb{Z}$ such that
\begin{multline}
\Dp{\ga}^n \Pi_+
=\sul{r=1}{n}\sul{\Sg_{i=1}^r p_i=n}{} \mf{c}_r^{(\{p_i\})}\;
\pa{I-C^{\De}_{\Sg_\Pi}}^{-1} \circ \pa{\Dp{\ga}^{p_1} C^{\De}_{\Sg_\Pi}} \\
\circ
\pa{I-C^{\De}_{\Sg_\Pi}}^{-1} \circ \pa{\Dp{\ga}^{p_2} C^{\De}_{\Sg_\Pi}}
\circ \cdots
\circ
\pa{I-C^{\De}_{\Sg_\Pi}}^{-1} \circ \pa{\Dp{\ga}^{p_r} C^{\De}_{\Sg_\Pi}} \pac{\Pi_+}.
\end{multline}
This expression simplifies at $\ga=0$ as $\De_{\mid_{\ga=0}}=0$ and $\Pi_{+\, {\mid_{\ga=0}}}=I_2$. Hence,
\begin{equation}
{\Dp{\ga}^n \Pi_+}\pour{\ga=0}=\sul{r=1}{n}\sul{\Sg_{i=1}^r p_i=n}{}  \mf{c}_r^{(\{p_i\})}\,
\pa{\Dp{\ga}^{p_1} C^{\De}_{\Sg_\Pi}} \circ \pa{\Dp{\ga}^{p_2} C^{\De}_{\Sg_\Pi}} \circ \cdots
 \circ \pa{\Dp{\ga}^{p_r} C^{\De}_{\Sg_\Pi}} \pac{I_2}.
\end{equation}
We can slightly deform the different contours $\Sg_{\Pi}$, so as to regularize the explicit integral representation for the above chain of operators. Namely, recalling the construction of the jump contours for the matrices occurring in the different transformations applied to  the RHP for $\chi$, we are able to write
\begin{equation}
{\Dp{\ga}^n \Pi_+}\pour{\ga=0}=\sul{r=1}{n}\sul{\Sg_{i=1}^r p_i=n}{}  \mf{c}_r^{(\{p_i\})}\,
\pa{\Dp{\ga}^{p_1} C^{\De}_{\Sg^{\pa{1}}_\Pi}} \circ
\pa{\Dp{\ga}^{p_2} C^{\De}_{\Sg^{\pa{2}}_\Pi}}
\circ \cdots \circ
\pa{ \Dp{\ga}^{p_r} C^{\De}_{\Sg^{\pa{r}}_\Pi}} \pac{I_2}.
\label{chaine d'operateur de Cauchy}
\end{equation}
There, the contours $\Sg_{\Pi}^{\pa{i}}$ are such that the $-$ side of $\Sg_{\Pi}^{\pa{i-1}}$ is at small but non vanishing distance from the $+$ side of $\Sg_{\Pi}^{\pa{i}}$, with the exception of a finite number of points of intersection (cf. Fig.~\ref{Contour encased for proof of asymp DA}). The matrix $\De$ corresponding to the contour $\Sg_{\Pi}^{\pa{i}}$ is equal to
$M_{\pm}^{\pm 1}-I_2$ on $\Ga^{' \pa{i}}_{\pm}$, to $\mc{P}-I_2$ on $\Dp{}D_{-q,\de_i}$, and to $\wt{\mc{P}}-I_2$ on $\Dp{}D_{q,\de_i}$. We emphasize that, already in \eqref{chaine d'operateur de Cauchy}, one can use the integral representation for the Cauchy operators without  turning to boundary values. Indeed, the integrand appearing
in \eqref{chaine d'operateur de Cauchy} is already integrable on
$\Sg_{\Pi}^{\times \pa{r}} \equiv \Sg_{\Pi}^{\pa{1}}\times\dots \times\Sg_{\Pi}^{\pa{r}} $.
\begin{figure}[h]
\begin{center}

\begin{pspicture}(10,5.5)


\pscircle(2,3){1}
\psdots(2,3)
\rput(1.6,3){$-q$}
\psline[linewidth=2pt]{->}(3,3)(3,3.1)

\pscircle(8,3){1}
\psdots(8,3)
\rput(8.2,3){$q$}
\psline[linewidth=2pt]{->}(9,3)(9,3.1)

\psline[linearc=.25]{-}(2,4)(2,5)(8,5)(8,4)
\psline[linewidth=2pt]{->}(5,5)(5.1,5)
\rput(5.3,5.25){$\Gamma_+^{'\pa{i-1}}$}

\psline[linearc=.25]{-}(2,2)(2,1)(8,1)(8,2)
\psline[linewidth=2pt]{<-}(4,1)(4.1,1)
\rput(4.5,0.5){$\Gamma_-^{'\pa{i-1}}$}


\pscircle[linestyle=dashed](2,3){1.3}
\psline[linewidth=2pt]{->}(3.3,3)(3.3,3.1)
\psbezier[linestyle=dashed](2,4.3)(4,4.7)(6,4.7)(8,4.3)
\pscircle[linestyle=dashed](8,3){1.3}
\psline[linewidth=2pt]{->}(9.3,3)(9.3,3.1)
\psbezier[linestyle=dashed](8,1.7)(6,1.2)(4,1.2)(2,1.7)

\psline[linewidth=2pt]{->}(4,4.55)(4.2,4.55)
\rput(4,4.2){$\Gamma_+^{'\pa{i}}$}

\psline[linewidth=2pt]{<-}(5.9,1.4)(6,1.4)

\rput(6.4,1.7){$\Gamma_-^{'\pa{i}}$}


\psline(2,3)(2.7,3.7)
\rput(2.2,3.65){$\de_{i-1}$}

\psline[linestyle=dashed](2,3)(2.9,2.15)
\rput(2.3,2.5){$\de_{i}$}
\end{pspicture}

\caption{Encased Contours $\Sg^{\pa{i}}_\Pi$ (in the case $p=\id$).\label{Contour encased for proof of asymp DA}}
\end{center}
\end{figure}
%
%
%
\medskip

Finally, one infers from \eqref{chaine d'operateur de Cauchy}, from \eqref{deriveeDelta} and from the integral representation for $\partial_\la\Pi$ \eqref{int-deriv-Pi}
that, when  $\la \in \Ga\pa{\intff{-q}{q}}$, there exist some recursively computable coefficients $\tilde{\mf{c}}_r^{(\{ p_{\ell i} \})}$ such that
\begin{multline}\label{Derivee-Pi}
\Dp{\ga}^n \Dp{\la}\Pi\pa{\la} \pour{\ga=0}
=-\sul{r=1}{n}
  \sul{ \substack{p_{\ell 0}\geq 1,\, p_{\ell 1},\ldots,p_{\ell n}\geq 0\\
                 \Sg_{\ell=1}^r \bar{p}_\ell =n
                   } }{}
\hspace{-2mm}
  \tilde{\mf{c}}_r^{(\{ p_{\ell i} \})}
\hspace{-1mm}
  \Int{\Sg^{\times \pa{r}}_\Pi }{}
\hspace{-1mm}
  \f{\dd^r z }{\pa{2i\pi}^{r}}\;
  \f{\dep^{ \pa{\paa{p_{r i}}} }\pa{z_r;x} \dots
                        \dep^{ \pa{\paa{p_{1 i}}} }\pa{z_1;x} }
    {\pa{\la-z_1}^2\pl{\ell=2}{r}{\pa{z_{\ell-1}-z_{\ell}}}  }
          \\
\times
  \pl{\ell=1}{r}\paa{ F^{p_{\ell 0}}\pa{z_{\ell}}
  \pl{m=1}{ n }
  \pac{\Int{-q}{q}
     \f{F^{m}\pa{z_{\ell}} - F^{m}\pa{\mu}}{z_{\ell}-\mu} }^{p_{\ell m}}
       }.
\end{multline}
In this expression, the integration is performed over the skeleton $\Sg_{\Pi}^{\times\pa{r}}=\Sg_{\Pi}^{\pa{1}}\times \dots \times \Sg_{\Pi}^{\pa{r}}$, and the second summation is performed over integers $p_{\ell j}$, $1\le \ell \le r$, $1 \le j \le n$, with $1\le p_{\ell 0}\le n$ and $0\le p_{\ell j}\le n$ for $j\ge 1$, and such that
$\Sg_{\ell=1}^r \bar{p}_\ell=n$, in which we have introduced the notation
$\bar{p}_\ell=p_{\ell 0}+ \Sg_{j=1}^n j p_{\ell j}$.

\paragraph{\indent $\bullet$ $\ga$-derivatives of $\Pi^{-1}$}

\par\indent
\medskip

All the above observations also hold for the inverse matrix $\Pi^{-1}\pa{\la}$. Indeed
$\Pi_+^{-1}$ satisfies the integral equation
\begin{equation}
\Pi_+^{-1}\pa{\la}=I_2+ \, ^t C^{\nabla}_{\Sg_\Pi}\pac{\Pi_+^{-1}} ,
\end{equation}
in which
\begin{equation}
^t C^{\nabla}_{\Sg_\Pi}\pac{M}\equiv
C^+_{\Sg_\Pi}\pac{\nabla M},\qquad
\forall M \in \mc{M}_2(L^2(\Sigma_\Pi))\, ,
\end{equation}
and the matrix $\nabla$ is defined\footnote{We stress that this matrix $\nabla$ has nothing to do with the differential operator usually denoted by $\nabla$.} by the equation $I_2+\nabla=\pa{I_2 +\De}^{-1}$. In other words, $\nabla$ is the adjugate of $\Delta$ (we remind that we consider $2\times2$ matrices and that $\ddet{}{I+\De}=1$). Hence, one easily sees that, for $n\ge 1$,
\begin{multline}\label{Derivee-invPi}
\Dp{\ga}^n\pac{\Pi^{-1}}\pa{\la} \pour{\ga=0}
=\sul{t=1}{n}
\sul{ \substack{\Sg_{\ell=1}^t \bar{p}_\ell=n \\ p_{\ell 0} \geq 1 }} {}
\hspace{-2mm}\tilde{\mf{c}}_t^{(\{ p_{\ell i} \})}
\hspace{-1mm}
\Int{\Sg^{\times \pa{t}}_\Pi }{}
\hspace{-1mm}
 \f{\dd^t z }{\pa{2i\pi}^{t}}\;
\f{\varrho^{ \pa{\paa{p_{1 i}}} }\pa{z_1;x} \dots \varrho^{ \pa{\paa{p_{t i}}} }\pa{z_t;x} } {\pa{\la-z_1}\pl{\ell=2}{t}{\pa{z_{\ell-1}-z_{\ell}}}  }\\
\times
\pl{\ell=1}{t} \paa{ F^{p_{\ell 0} }\pa{z_{\ell}} \pl{m=1}{n}
\pac{\Int{-q}{q} \f{F^{m}\pa{z_{\ell}} - F^{m}\pa{\mu}}{z_{\ell}-\mu} }^{p_{\ell m}} },
\end{multline}
where $\varrho^{ \pa{\paa{p_{\ell i}}} }\pa{z;x}$ is the adjugate matrix of $\dep^{\pa{\paa{p_{\ell i}}}}\pa{z;x}$.
%
%
\paragraph{\indent $\bullet$ $\ga$-derivatives of $\ln \ddet{}{I+V}^{\e{sub}}$}

\par\indent
\medskip

From the expressions \eqref{log-det-sub}, \eqref{Derivee-Pi} and \eqref{Derivee-invPi}, it is easy to see that there exist some combinatorial coefficients $\mf{C}_{r,t}^{(\{p_{\ell i}\})}\in\mathbb{Z}$ (with $\mf{C}_{1,0}^{(\{p_{1 i}\})}=-\delta_{p_{10},n}$) such that
\begin{multline}
  \Dp{\ga}^n\ln \ddet{}{I+V}^{\e{sub}} {}\pour{\ga=0}
 = \sul{  \substack{1 \leq r+t \leq n \\ r\geq1 ,\; t\geq 0 }  }{}\;
\sul{ \substack{p_{\ell 0}\geq 1,\, p_{\ell 1},\ldots,p_{\ell r+t}\geq 0 \\ \Sg_{\ell=1}^{r+t} \bar{p}_\ell=n  } }{}
\hspace{-2mm}
 \mf{C}_{r,t}^{(\{p_{\ell i}\})}
\hspace{-2mm}
\Oint{\Ga\pa{\intff{-q}{q}}}{}
\hspace{-3mm}
\f{\dd \la }{4\pi}\, p(\la)
\hspace{-1mm}
\Int{\Sg^{\times \pa{r+t}}_\Pi }{}
\hspace{-2mm}
 \f{\dd^{r+t} z}{\pa{2i\pi}^{r+t}}\; \\
\times
\f{\Int{+\infty}{x}\dd x'\;
\mf{tr}_{r,t}^{(\{p_{\ell j}\})}(\{z_j\};x') [g] }
{\pa{\la-z_1}^2 \pa{\la-z_{r+1}}\pl{\ell=2}{r}{\pa{z_{\ell-1}-z_{\ell}}}\pl{\ell=r+2}{r+t}{\pa{z_{\ell-1}-z_{\ell}}} } \\
\times \pl{\ell=1}{r+t}\paa{ F^{p_{\ell 0}}\pa{z_{\ell}} \pl{m=1}{ n}
\pac{\Int{-q}{q} \f{F^{m}\pa{z_{\ell}} - F^{m}\pa{\mu}}{z_{\ell}-\mu} \,\dd\mu}^{p_{\ell m}} },
\label{der-log-det-sub}
\end{multline}
in which, in the terms $t=0$, the empty product $\pa{\la-z_{r+1}}\prod_{\ell=r+2}^{r+t}{\pa{z_{\ell-1}-z_{\ell}}}$ should be understood as $1$.
In this expression, $\mf{tr}_{r,t}^{(\{p_{\ell j}\})}(\{z_j\};x') [g]$ corresponds to the following trace:
\begin{align}
 &\mf{tr}_{1,0}^{(\{p_{1 j}\})}(z;x') [g]
     =\e{tr}\paa{ \sg_3\,
\partial_\ga^n\Big[  \De_0(z;x')-\frac{1}{x'}\De_0^{(1)}(z;x')\Big]\pour{\ga=0} },
 \label{def-trace-10}\\
 &\mf{tr}_{r,t}^{(\{p_{\ell j}\})}(\{z_j\};x') [g]
    =\e{tr}\,\bigg\{
   \dep^{ \pa{\paa{p_{r i}}} }(z_r;x') \dots
   \dep^{ \pa{\paa{p_{1 i}}} }(z_1;x')\;
   \sg_3\; \nonumber\\
 &\hspace{4cm}\times
   \varrho^{ \pa{\paa{p_{r+1 i}}} }(z_{r+1};x') \dots
   \varrho^{ \pa{\paa{p_{r+t i}}} }(z_{r+t};x')  \bigg\}
  \quad\text{if $r+t>1$.} \label{def-trace-rt}
\end{align}
In \eqref{def-trace-10}, $\De_0^{(1)}(z;x')$ corresponds to the first term in the asymptotic expansion \eqref{expansion-Delta} of $\Delta_0(z)$.

\smallskip

\begin{rem}
We have gathered in the term $r=1,t=0$ the contribution of the second term in \eqref{log-det-sub}, as well as the term that would correspond to the contribution of only one jump matrix $\De$ in the first term of \eqref{log-det-sub}. Note that, in this term $r=1,t=0$, the only non-zero contribution comes from the diagonal elements of $\De$, hence from the sequence $p_{10}=n$, $p_{1j}=0$ for $j\geq 1$ (indeed we have $\mf{C}_{1,0}^{(\{p_{1 i}\})}=-\delta_{p_{10},n}$).
\end{rem}


\begin{rem}
It is easy to see from these expressions that the integrals over $x'$ are convergent. Indeed, it follows from the asymptotic expansion of the matrices $\De_0$ that
\begin{equation}
\partial_\ga^n\Big[  \De_0(z;x)-\frac{1}{x}\De_0^{(1)}(z;x)\Big]\pour{\ga=0}
=\e{O}\paf{\ln^n x}{x^2}
\end{equation}
uniformly on the integration contour, so that an integration of the trace~\eqref{def-trace-10} is convergent. We emphasize that  the trace \eqref{def-trace-rt} is at least
$\e{O}\,\big(\tf{ (\ln x)^n}{x^2 }\big)$ uniformly on the integration contour: indeed, each of the matrices $\dep^{ ( \{p_{\ell j}\} ) }$ or
$\varrho^{ (  \{p_{\ell j}\}  ) }$ is uniformly a $\e{O}\pa{ \pa{\ln x}^{p_{\ell 0}} / x  }$; in addition,  the trace \eqref{def-trace-rt} involves a product of at least two such matrices since $r+t\geq 2$. These estimates guarantee that the integrals over $x'$ in \eqref{der-log-det-sub} are well defined.
\end{rem}

\subsubsection{Application of the density procedure and proof of Theorem~\ref{th_asympt_In} }

In order to be able to apply the density procedure, we should express more explicitly the functional dependence of $\partial_\ga^n \log\det\big[I+V^{(\varphi,\phi)}\big]^{\e{sub}}\mid_{\ga=0}$ on $\mc{F}_n^{(\varphi,\phi)}$.

The $F$-dependence of $\partial_\ga^n \log\det\big[I+V\big]^{\e{sub}}\mid_{\ga=0}$ has already been explicitly extracted in \eqref{der-log-det-sub}, and all the $g$-dependence is contained in the traces $\mf{tr}_{r,t}^{(\{p_{\ell j}\})}(\{z_j\};x') [g]$. Using the structure of the matrices $\dep^{ \pa{  \paa{p_{\ell i}}  } } \pa{z;x}$ and $\rho^{ \pa{  \paa{p_{\ell i}} } } \pa{z;x}$, one can be more precise concerning this $g$-dependence. Indeed, it follows from~\eqref{definition des matrices delta avce plusieurs indices} that there exist some coefficients
$D_{r,\, t}^{[\{p_{\ell j}\},\{\eps_j\}]}\big(\big\{z_j\big\}; x \big)$ which are piecewise smooth on the integration contour such that
\begin{equation}
\mf{tr}_{r,t}^{(\{p_{\ell j}\})}(\{z_j\};x) [g]
%
%
=\sul{\substack{\eps_1,\ldots,\eps_{r+t}  \in \paa{\pm 1, 0} \\
                \Sg \eps_i =0 }}{}
D_{r,\, t}^{[\{p_{\ell j}\},\{\eps_j\}]}\big(\big\{z_j\big\}; x \big)
\;
\exp\paa{\sul{\ell=1}{r+t}\eps_{\ell} \, g\pa{z_{\ell}}  }.
\label{trace comme somme}
\end{equation}
Note that these coefficients $D_{r,\, t}^{[\{p_{\ell j}\},\{\eps_j\}]}\big(\big\{z_j\big\}; x \big)$ are at least $\e{O}\pa{\tf{ \pa{\ln x}^n}{x^2 }}$ uniformly on the integration contour.
Integrating these coefficients with respect to $x$, and defining
\begin{equation}
\wt{D}_{r,\, t}^{[\{p_{\ell j}\},\{\eps_j\}]}\big(\big\{z_j\big\}; x \big)
  = \mf{C}_{r,t}^{(\{p_{\ell j}\})} \Int{+\infty}{x} \dd x'
 D_{r,\, t}^{[\{p_{\ell j}\},\{\eps_j\}]}\big(\big\{z_j\big\}; x' \big) ,
\label{primitive des coeffs du DA de integral cyclique}
\end{equation}
which are at least
$\e{O}\pa{ \tf{\pa{\ln x}^n}{x} }$ uniformly on the integration contours $\Sg^{\pa{i}}_{\Pi}$, one gets
\begin{multline}
\Dp{\ga}^n\ln \ddet{}{I+V}^{\e{sub}} {}\pour{\ga=0}
=
\sul{  \substack{1 \leq r+t \leq n \\ r\geq1 ,\; t\geq 0 }  }{}\;
\sul{ \substack{\Sg_{\ell=1}^{r+t} \bar{p}_\ell=n \\
                p_{\ell 0}\geq 1 } }{}\;
\Oint{\Ga\pa{\intff{-q}{q}}}{}
 \hspace{-2mm}
\f{\dd \la }{4\pi}\, p(\la)
\hspace{-1mm}
\Int{\Sg^{\times \pa{r+t}}_\Pi }{}
\hspace{-2mm}
 \f{\dd^{r+t} z}{\pa{2i\pi}^{r+t}} \\
\times
\sul{ \substack{\eps_1,\ldots,\eps_{r+t}\in\{\pm 1,0\}\\
     \sum\eps_i=0}}{}
\f{\wt{D}_{r,\, t}^{[\{p_{\ell j}\},\{\eps_j\}]}\big(\big\{z_j\big\}; x \big) }{\pa{\la-z_1}^2 \pa{\la-z_{r+1}}\pl{\ell=2}{r}{\pa{z_{\ell-1}-z_{\ell}}}\pl{\ell=r+2}{r+t}{\pa{z_{\ell-1}-z_{\ell}}} } \\
\times \pl{\ell=1}{r+t}\paa{ F^{p_{\ell 0}}\pa{z_{\ell}} \ex{\eps_{\ell} g\pa{z_{\ell}}} \pl{m=1}{ n}
\pac{\Int{-q}{q} \f{F^{m}\pa{z_{\ell}} - F^{m}\pa{\mu}}{z_{\ell}-\mu}\; \dd\mu }^{p_{\ell m}} }.
\label{derivee gamma log det sub}
\end{multline}
We stress that each $\ex{\pm g\pa{z_{\ell}}}$ may only appear in combination with
at least one $F\pa{z_{\ell}}$ (as $p_{\ell 0}\geq 1$): $F\pa{z_{\ell}}\ex{\pm g\pa{z_{\ell}}}$. This guarantees that the functional above is continuous with respect to the sup norm on the
space of symmetric functions in $n$ variables $z$ and $n$ variables
$\la$.

Before applying the density procedure, let us introduce one more useful notation.
Define the finite difference operator $\eth^{\pa{m}}_{z}(\mu)$ by its action on pure product functions
\begin{equation}
\eth^{\pa{m}}_{z}(\mu)\cdot F^k(z)=F^{k}(z)-F^{k-m}(z)\,F^m(\mu) \, .
\end{equation}
This action naturally extends to symmetric functions of $n$ variables
\begin{multline}
\eth^{\pa{m}}_{z_\ell}(\mu) \cdot
\mc{F}_n\pac{ \paa{ \paa{z_{i}}^{\bar{p}_i } }_{i=1,\ldots,r} }
=
\mc{F}_n\pac{ \paa{ \paa{z_{i}}^{\bar{p}_i } }_{i=1,\ldots,r} }
-
\mc{F}_n\pac{ \paa{ \paa{z_{k}}^{\bar{p}_k } }_{k\not= \ell} , \paa{\mu}^{m}, \paa{z_{\ell}}^{\bar{p}_\ell-m}} \, .
\end{multline}
We remind here that $\bar{p}_\ell=p_{\ell 0 }
+ \Sg_{s=1}^n s p_{\ell s}$, with $\sum_{\ell=1}^{r+t} \bar{p}_\ell=n$. We have moreover used the notation
$\paa{z_{\ell}}^{\bar{p}_\ell }$, which means that the variable $z_{\ell}$ is repeated $\bar{p}_\ell$
times, and $\paa{ \paa{z_{i}}^{\bar{p}_i } }_{i=1,\ldots,r}$, which means that the variable $z_1$ is repeated $\bar{p}_1$ times, $z_2$ is repeated $\bar{p}_2$ times, \ldots, $z_r$ is repeated $\bar{p}_r$ times. The purpose of introducing such finite difference operator is to recast  products of functions
$F^{m}\pa{z_\ell}-F^{m}\pa{\mu}$ appearing in \eqref{derivee gamma log det sub} into a more compact form. Namely,
\begin{multline}
\pl{m=1}{ n }
\pac{\Int{-q}{q} \f{F^{m}\pa{z_{\ell}} - F^{m}\pa{\mu}}{z_{\ell}-\mu}\,
\dd\mu }^{ p_{\ell m} } =
\pl{m=1}{ n}
\pac{\Int{-q}{q}
\f{ \eth^{\pa{m}}_{z_\ell}(\mu)\cdot  F^{m}\pa{z_{\ell}} }
  {z_{\ell}-\mu}
 \dd \mu }^{p_{\ell m}} \nonumber \\
= \pl{m=1}{n } \pl{j=1}{p_{\ell m}} \Int{-q}{q} \f{\dd \mu_{\ell, m, j}}{z_{\ell}-\mu_{\ell,m,j}}
\paa{\pl{m=1}{ n } \pl{j=1}{p_{\ell m}} \eth^{\pa{m}}_{z_\ell}(\mu_{\ell,m,j})  }\cdot
F^{\bar{p}_\ell-p_{\ell 0} }\pa{z_{\ell}} \, .
\end{multline}

Therefore, setting $e^{g(z)}=\phi(z)$ and $F(z)=\varphi(z)\,\phi(z)$, we get
\begin{multline}
\Dp{\ga}^n\ln \det\big[I+V^{(\varphi,\phi)}\big]^{\e{sub}}
{}\pour{\ga=0}
= 
\sul{  \substack{1 \leq r+t \leq n \\ r\geq1 ,\; t\geq 0 }  }{}\;
\sul{ \substack{\Sg_{\ell=1}^{r+t} \bar{p}_\ell=n \\
                p_{\ell 0}\geq 1 } }{}\;
\Oint{\Ga\pa{\intff{-q}{q}}}{}
 \hspace{-2mm}
\f{\dd \la }{4\pi}\, p(\la)
\hspace{-1mm}
\Int{\Sg^{\times \pa{r+t}}_\Pi }{}
\hspace{-2mm}
 \f{\dd^{r+t} z}{\pa{2i\pi}^{r+t}} \\
\times
\sul{ \substack{\eps_1,\ldots,\eps_{r+t}\in\{\pm 1,0\}\\
     \sum\eps_i=0}}{}
\f{\wt{D}_{r,\, t}^{[\{p_{\ell j}\},\{\eps_j\}]}\big(\big\{z_j\big\}; x \big) }
{\pa{\la-z_1}^2 \pa{\la-z_{r+1}}\pl{\ell=2}{r}{\pa{z_{\ell-1}-z_{\ell}}}\pl{\ell=r+2}{r+t}{\pa{z_{\ell-1}-z_{\ell}}} } \\
\times \pl{\ell=1}{r+t}  \pl{m=1}{ n } \pl{j=1}{p_{\ell m}}
\paa{
\Int{-q}{q} \f{\dd \mu_{\ell, m, j}  }{z_{\ell}-\mu_{\ell,m,j}}\; \eth^{\pa{m}}_{z_\ell}(\mu_{\ell,m,j})
}
\cdot
\prod_{\ell=1}^{r+t}
 \paa{ \varphi^{\bar{p}_\ell }(z_{\ell})\;
       \phi^{\bar{p}_\ell +\eps_{\ell}}(z_{\ell})} .
\label{formule-avant-densite}
\end{multline}

It follows immediately from the density procedure formulated in
Theorem \ref{theoreme de densite} that $I_{n}^{\e{sub}}$ can be extended into a linear functional on $\mc{S}{\rm ym}_n(U,W)$.
Its action on a holomorphic function $\mc{F}_n\in\mc{S}{\rm ym}_n(U,W)$ is given as
\begin{multline}
 I_{n}^{\e{sub}}\pac{\mc{F}_n}
=\frac{\pa{-1}^{n-1}}{ \pa{n-1}!}\;
\sul{  \substack{1 \leq r+t \leq n \\ r\geq1 ,\; t\geq 0 }  }{}\;
\sul{ \substack{\Sg_{\ell=1}^{r+t} \bar{p}_\ell=n \\
                p_{\ell 0}\geq 1 } }{}\;
\Oint{\Ga\pa{\intff{-q}{q}}}{}
 \hspace{-2mm}
\f{\dd \la }{4\pi}\, p(\la)
\hspace{-1mm}
\Int{\Sg^{\times \pa{r+t}}_\Pi }{}
\hspace{-2mm}
 \f{\dd^{r+t} z}{\pa{2i\pi}^{r+t}} \\
\times
\sul{ \substack{\eps_1,\ldots,\eps_{r+t}\in\{\pm 1,0\}\\
     \sum\eps_i=0}}{}
\f{\wt{D}_{r,\, t}^{[\{p_{\ell j}\},\{\eps_j\}]}\big(\big\{z_j\big\}; x \big) }
{\pa{\la-z_1}^2 \pa{\la-z_{r+1}}\pl{\ell=2}{r}{\pa{z_{\ell-1}-z_{\ell}}}\pl{\ell=r+2}{r+t}{\pa{z_{\ell-1}-z_{\ell}}} } \\
\times \pl{\ell=1}{r+t}  \pl{m=1}{ n } \pl{j=1}{p_{\ell m}}
\paa{
\Int{-q}{q} \f{\dd \mu_{\ell, m, j}  }{z_{\ell}-\mu_{\ell,m,j}}\; \eth^{\pa{m}}_{z_\ell}(\mu_{\ell,m,j})
}
\cdot
\mc{F}_n\pab{
\big\{ \paa{z_{\ell}}^{\bar{p}_\ell } \big\}_{1\le \ell \le r+t} }
{ \big\{ \paa{z_{\ell}}^{\bar{p}_\ell+\eps_{\ell}  }   \big\}_{1\le \ell \le r+t}  }\, .
\label{formule apres passage par densite}
\end{multline}
The sum appearing in \eqref{formule apres passage par densite} is finite,
and since each integrand is a $\e{O}\pa{\tf{\ln^n x}{x}}$, $I_n^{\e{sub}}\pac{\mc{F}_n}$ is itself  a $\e{O}\pa{\tf{\ln^n x}{x}}$.
Hence Theorem~\ref{th_asympt_In} follows directly, since
\begin{equation}
\mc{I}_n\pac{ \mc{F}_n }
=I_n^{(0)}\pac{\mc{F}_n}
+I_{n}^{\e{sub}}\pac{\mc{F}_n} \, .
\end{equation}
%

\subsubsection{Asymptotic expansion of $I_n^\e{sub}[\mc{F}_n]$ and proof of Theorem~\ref{th_asympt_In-sub} }

In order to prove the existence of an asymptotic series of for $\mc{I}_{n}\pac{\mc{F}_n}$, i.e. for $I_n^\e{sub}[\mc{F}_n]$, we should be more precise
on the structure of the coefficients $\wt{D}_{r,t}$, i.e. show that they themselves admit an asymptotic expansion. Let us recall that these coefficients are obtained from the traces \eqref{trace comme somme} involving
the matrices $\dep^{\pa{\paa{p_{\ell j}}}}\pa{z_{\ell};x}$ and $\varrho^{\pa{\paa{p_{\ell j}}}}\pa{z_{\ell};x}$. The latter being obtained from the jump matrix $\De_0$.

Clearly, all terms corresponding to an integration on the contours $\Ga^{' \pa{i}}_{\pm}$ yield exponentially small corrections. Thus in what concerns the proof of an asymptotic expansion we can only focus on integrations
along the contours $\Dp{}D_{\pm q, \de_i}$.  We decompose the relevant contour
 $\big[\Dp{}D_{q,\de}\cup\Dp{}D_{-q,\de}\big]^{\times \pa{r+t}}$ into sums of elementary skeletons
 $\Dp{}D_{\sg q,\de}^{\times \pa{r+t}}\equiv\Dp{}D_{\sg_1 q,\de_1}\times \dots \times \Dp{}D_{\sg_{r+t} q, \de_{r+t}}$, where
 each $\sg_i$ takes values in  $\paa{\pm 1}$:
\begin{equation}
\Int{\Sg_{\Pi}^{\pa{r+t}}}{} \f{\dd^{r+t} z}{ \pa{2i\pi}^{r+t} }=\sul{\sg_i=\pm}{} \Int{\Dp{}D_{\sg q, \de}^{\pa{r+t}}}{} \f{\dd^{r+t} z}{ \pa{2i\pi}^{r+t} } + \e{O}\pa{x^{-\infty}}\, .
\end{equation}

The matrices $\dep^{(\{p_{ j}\})}\pa{z;x}$ 
\eqref{definition des matrices delta avce plusieurs indices}
admit an asymptotic expansion into inverse powers of $x$ on $\Dp{}D_{q,\de}\cup\Dp{}D_{-q,\de}$. This
fact follows from the asymptotic expansion of $\Delta_0(z)$.
The latter is obtained by taking adequate $a$-derivatives at $a=0$ or $1$ of the asymptotics series \eqref{asy-Psi}
for $\Psi\pa{a,1;z}$ when  $z\tend \infty$.  This is licit as, for fixed $M$, the $\e{O}\big(z^{-M-1}\big)$ estimate in the asymptotic series \eqref{asy-Psi} is uniform with respect to $a$ and since we perform a finite number of derivative with respect to $a$. This asymptotic expansion takes the following form:
\begin{equation}
\dep^{(\{p_j\})}\pa{z;x}= \left\{ \ba{ll}
\sul{k=1}{M}
\f{ \Ad_{\ex{\sg_3\,[ixp_+ + g(z)]/2}}
               \Big[ \dep_+^{ (\{p_{j}\},k) } \pa{z;X_+} \Big]}
  {x^k \pa{p\pa{z}-p_+}^k}
              +\e{O}\paf{\ln^{p_0}x}{x^{M+1}}\, ,
                     &\quad 
                      z\in\Dp{}D_{q,\de}\, ,
         \vspace{1mm}\\
\sul{k=1}{M}  \f{ \Ad_{\ex{\sg_3\,[ixp_- + g(z)]/2}}
               \Big[ \dep_-^{ (\{p_{j}\},k) } \pa{z;X_-} \Big]}
                {x^k \pa{p\pa{z}-p_-}^k}
              +\e{O}\paf{\ln^{p_0}x}{x^{M+1}}\, ,
                      &\quad 
                        z\in\Dp{}D_{-q,\de}\, ,
                        \ea \right.
\end{equation}
the corrections being uniform on the contours.
There the diagonal entries of the matrices $\dep_{\pm}^{(\{p_{j}\},k)}\pa{z;X_\pm}$ are some constants (i.e. $x$ and $z$ independent),
whereas the off-diagonal ones are polynomials of degree $p_0$ in the variable
$X_{\pm}=\ln\pac{\pm x  \pa{p\pa{z}-p_{\mp}}}$.

An exactly similar structure holds for $\varrho^{(\{p_{ j}\})}\pa{z;x}$ as it is adjunct to  $\dep^{(\{p_{ j}\})}\pa{z;x}$. Hence,  on the skeleton
$\Dp{}D_{\sg q,\de}=\Dp{}D_{\sg_1 q, \de_1}\times \dots \times \Dp{}D_{\sg_{r+t} q, \de_{r+t}}$, the trace~\eqref{trace comme somme} can be expanded in the following form:
\begin{multline}
\mf{tr}_{r,t}^{(\{p_{\ell j}\})}(\{z_j\};x) [g]
%
%
=\sul{N=2}{M+1} \f{1}{x^{N}}
\sul{\substack{k_1,\ldots,k_{r+t}=1 \\ \Sg k_i =N}}{N} \,
\sul{\substack{\eps_1,\ldots,\eps_{r+t} \in \paa{\pm 1 ,0} \\ \Sg \eps_i=0} }{} \hspace{-3mm}
\mf{D}_{r,\,t}\begin{bmatrix}
              \{p_{\ell j}\},\{\eps_i\} \\ \{\sg_i\}, \{k_i\}
              \end{bmatrix}
\pa{\paa{z_i} ; \ln x}\;
\\
\times
\f{ \exp\paa{\sul{\ell=1}{r+t} \eps_{\ell}\, \big[ g\pa{z_{\ell}} +i x  p_{\sg_{\ell}}\big] } }
{\pl{\ell=1}{r+t} \pa{p\pa{z_{\ell}}-p_{\sg_{\ell}}  }^{k_{\ell}}    }
 + \e{O}\paf{\ln^{n}x}{x^{M+2}}.
\end{multline}
There we have explicitly factored out the dependence on the oscillating factor $\exp\paa{i x \Sg_{\ell}\eps_{\ell} p_{\sg_{\ell}} }$. In this expression, the coefficients $\mf{D}_{r,\, t}$ are piecewise smooth on the integration contour and are polynomials of degree $\Sg_{\ell} \abs{\eps_{\ell}} p_{\ell 0} $ in $\ln x$.

We set, for $N\geq 2$,
\begin{equation}
x^{1-N} \wt{\mf{D}}^{\pa{0}}_{r,\, t}\begin{bmatrix}
              \{p_{\ell j}\},\{\eps_i\} \\ \{\sg_i\}, \{k_i\}
              \end{bmatrix}
\pa{\paa{z_i} ; \ln x}
=
\mf{C}_{r,t}^{(\{p_{\ell j}\})} \Int{+\infty}{x} \f{\dd x'}{\pa{x'}^N} \mf{D}_{r,\,t}\begin{bmatrix}
              \{p_{\ell j}\},\{\eps_i\} \\ \{\sg_i\}, \{k_i\}
              \end{bmatrix}
\pa{\paa{z_i} ; \ln x'}
\end{equation}
when $\sum{}{}\eps_{\ell} p_{\sg_{\ell}}=0$, and
\begin{multline}
\ex{ix \sum{}{}p_{\sg_{\ell}}\eps_{\ell} }\sul{k=N}{M}x^{-k}\, \wt{\mf{D}}^{\pa{k}}_{r,\, t}\begin{bmatrix}
              \{p_{\ell j}\},\{\eps_i\} \\ \{\sg_i\}, \{k_i\}
              \end{bmatrix}
\pa{\paa{z_i} ; \ln x} +\e{O}\paf{\ln^{n} x}{X^{M+1}}\hspace{3cm} \\
= \mf{C}_{r,t}^{(\{p_{\ell j}\})} \Int{+\infty}{x} \f{\dd x'}{\pa{x'}^N} \mf{D}_{r,\,t}\begin{bmatrix}
              \{p_{\ell j}\},\{\eps_i\} \\ \{\sg_i\}, \{k_i\}
              \end{bmatrix}
\pa{\paa{z_i} ; \ln x}
\ex{ix' \sum{}{}p_{\sg_{\ell}}\eps_{\ell} }
\end{multline}
otherwise. Note at this stage that, due to the constraint $\sum\eps_i=0$, there exist some integer $m\not=0$ such that $\ex{ix \sum{}{}p_{\sg_{\ell}}\eps_{\ell} } =\ex{ixm(p_+-p_-)}$.

We then insert the result of integration into the expression for $I^{\e{sub}}_n\pac{\mc{F}_n}$,  rearrange the
asymptotic expansion into decreasing powers of $x$ and separate the oscillating and non-oscillating parts. We obtain
\begin{equation}\label{series-Hn-sub}
I_{n}^{\e{sub}}\pac{\mc{F}_n}
=\sul{N=1}{M}\f{1}{x^N}\;I_{n}^{(N;\,\e{nosc})}\pac{\mc{F}_n}
+\sul{N=2}{M}\f{1}{x^N}\;I_{n}^{(N;\,\e{osc})}\pac{\mc{F}_n}
+ \e{O}\paf{\ln^n x}{x^{M+1}}\, ,
\end{equation}
where
\begin{align}
&I_{n}^{(N;\,\e{nosc})}\pac{\mc{F}_n}
 =\sum_{\substack{r,t \\ \{p_{\ell j}\},\{\eps_i\} } }
  \sul{ \substack{\sg_1,\ldots,\sg_{r+t}=\pm \\
       \Sg \eps_{\ell}p_{\sg_{\ell}}=0 } }{} \;
 \sul{  \substack{ k_1,\ldots, k_{r+t}=1 \\ \Sg k_i =N+1 } }{N+1}
 \mf{I}_{r,\, t}^{\pa{0}}
              \begin{bmatrix}
              \{p_{\ell j}\},\{\eps_i\} \\ \{\sg_i\}, \{k_i\}
              \end{bmatrix}
\pa{ \ln x}  \pac{\mc{F}_n},
                \label{H non oscillant}\\
&I_{n}^{(N;\,\e{osc})}\pac{\mc{F}_n}
=\sum_{\substack{r,t \\ \{p_{\ell j}\},\{\eps_i\} } }
  \sul{ \substack{\sg_1,\ldots,\sg_{r+t}=\pm \\
       \Sg \eps_{\ell}p_{\sg_{\ell}}\not= 0 } }{} \;
\sul{s=2}{N} \;\sul{  \substack{ k_1,\ldots, k_{r+t}=1 \\ \Sg k_i =s } }{s}
\hspace{-3mm}\ex{i x \sul{\ell=1}{r+t} \eps_{\ell} p_{\sg_{\ell}} }\;
 \mf{I}_{r,\, t}^{\pa{s}}
              \begin{bmatrix}
              \{p_{\ell j}\},\{\eps_i\} \\ \{\sg_i\}, \{k_i\}
              \end{bmatrix}
\pa{ \ln x}  \pac{\mc{F}_n}.
              \label{H oscillant}
\end{align}
In these expressions,
\begin{equation}
 \sum_{\substack{r,t \\ \{p_{\ell j}\},\{\eps_i\} } }
 \equiv
 \sul{  \substack{1 \leq r+t \leq n \\ r\geq1 ,\, t\geq 0 }  }{}
 \sul{ \substack{p_{\ell 0}, \ldots, p_{\ell n}\\
                 p_{\ell 0}\ge 1,\,\Sg_{\ell=1}^{r+t} \bar{p}_\ell=n }}{}
 \sul{ \substack{ \eps_1,\ldots,\eps_{r+t} \in \paa{\pm 1, 0} \\
                  \Sg{\eps_i=0} } }{}
\end{equation}
and the functional $\mf{I}_{r,\, t}$ is given by
\begin{multline}\label{H_calligraphique}
  \mf{I}_{r,\, t}^{\pa{s}}
              \begin{bmatrix}
              \{p_{\ell j}\},\{\eps_i\} \\ \{\sg_i\}, \{k_i\}
              \end{bmatrix}
\pa{ \ln x}  \pac{\mc{F}_n}
=
 \frac{(-1)^{n-1}}{(n-1)!}
\hspace{-1mm}
   \Oint{\Ga\pa{\intff{-q}{q}}}{} \hspace{-3mm}
   \f{\dd \la}{4\pi}\, p(\la)
\hspace{-1mm}
   \Int{\Dp{}D_{\sg q,\de}^{\times(r+t)} }{}
   \hspace{-2mm}
   \f{\dd^{r+t} z}{\pa{2i\pi}^{r+t}}   \\
 \times
\f{ \wt{\mf{D}}^{\pa{s}}_{r,\, t}\begin{bmatrix}
              \{p_{\ell j}\},\{\eps_i\} \\ \{\sg_i\}, \{k_i\}
              \end{bmatrix}
\pa{\paa{z_i} ; \ln x}  }
{ \pa{\la-z_1}^2 \pa{\la-z_{r+1}}\pl{\ell=2}{r}\pa{z_{\ell-1}-z_{\ell}} \pl{\ell=r+2}{r+t}\pa{z_{\ell-1}-z_{\ell}} }
\pl{\ell=1}{r+t} \f{1}{\pa{p(z_{\ell})-p_{\sg_{\ell}} }^{k_{\ell}}}  \; \\
\times
\pl{\ell=1}{r+t}  \pl{m=1}{ n } \pl{j=1}{p_{\ell m}}
\paa{
\Int{-q}{q} \f{\dd \mu_{\ell, m, j}  }{z_{\ell}-\mu_{\ell,m,j}}\; \eth^{\pa{m}}_{z_\ell}(\mu_{\ell,m,j})
}
\cdot
\mc{F}_n\pab{
\big\{ \paa{z_{\ell}}^{\bar{p}_\ell } \big\}_{1\le \ell \le r+t} }
{ \big\{ \paa{z_{\ell}}^{\bar{p}_\ell+\eps_{\ell}  }   \big\}_{1\le \ell \le r+t}  }\, .
\end{multline}

The coefficients $\wt{\mf{D}}_{r,\, t}$ being polynomials of degree
$\sum_\ell\abs{\eps_{\ell}} p_{\ell 0}$ in $\ln x$, this ends the proof
of Theorem~\ref{th_asympt_In-sub} concerning the existence of the asymptotic expansion of cyclic integrals to any order in $\tf{1}{x}$.

As it is presented, the form of this asymptotic expansion may look quite involved. Note however that the integrals over the contours
$\Dp{}D_{\sg q,\de}^{\times(r+t)}$ in \eqref{H_calligraphique} can be computed; they are expressible in terms of partial derivatives of the function $\mc{F}_n$ at $\pm q$ (see Appendix~\ref{append-subleading}). It is proved in Appendix~\ref{append-subleading} that the non-oscillating term $I_{n}^{(N;\,\e{nosc})}\pac{\mc{F}_n}$ of order $N$ can be expressed in terms of derivatives of $\mc{F}_n$ of total order not higher than $N$, whereas the order of such derivatives does not exceed $N-2$ in the case of $I_{n}^{(N;\,\e{osc})}\pac{\mc{F}_n}$. This property is useful in \cite{KitKMST009}, when we sum up the asymptotic behaviour of a whole class of cycle integrals of the form \eqref{integ-F}  to obtain the asymptotic behaviour of correlation functions. To perform this summation we use the knowledge of the number of partial derivatives applied to $\mc{F}_n$. We finally point out that the integral over $\la$ produces derivatives of the function $p(\la)$ evaluated at $\pm q$.

\section{More general kernels}\label{sec-general-kernel}

In the applications to quantum integrable models, one sometimes needs
 to use some modified versions of the GSK.

Consider the operator $I+V_{\th}$ acting on $\intff{-q}{q}$ with kernel
\begin{equation}
V_{\th}\pa{\la,\mu}=\sqrt{F\pa{\la}F\pa{\mu}
\theta'\pa{\la}\theta'\pa{\mu}}\;
\f{e_+\pa{\la}e_{-}\pa{\mu}-e_-\pa{\la}e_{+}\pa{\mu}} {2i\pi \,
[\theta\pa{\la}-\theta\pa{\mu}] },
\end{equation}
where $e_{\pm}$ and $F$ are defined as in \eqref{epm}. We assume in
addition that $\theta$ is a biholomorphism of $U$ onto its image,
that $\theta\pa{\intff{-q}{q}} \subset \R$ and $\theta\pa{U\cap
{\mathcal H}_{\pm}}\subset {\mathcal H}_{\pm}$.

Then the asymptotic behaviour of $\ln\det[I+V_{\th}]$ when $x\tend\infty$
follows from Theorem~\ref{theorem asymptotiques order zero log det}.
More precisely, we have the following corollary:

\begin{cor}
Let $V_{\th}$ be as above. Then
\begin{multline}
\hspace{-2mm} \ln\ddet{}{I+V_{\th}} =2 \Int{-q}{q}\dd\la\, \nu(\la)
\ln'[e_-(\la)] + \sul{\sigma=\pm}{}\ln\pac{
   \f{G(1,\nu_{\sigma})\; \theta'(\sigma q)^{\nu_{\sigma}^2}\;
   \mc{K}^{\sigma \nu_{\sigma}}(\sigma q;q) }
{\pac{ \big(\theta(q)-\theta(-q)\big)\, p'_{\sigma} x}^{\nu_{\sigma}^2}  }}   \\
+  \f{1}{2} \Int{-q}{q} \dd \la\, \dd \mu\, \f{\nu'(\la)\,
\theta'(\mu)\, \nu(\mu) - \nu(\la)\, \theta'(\la)\, \nu'(\mu)
}{\theta(\la)-\theta(\mu) } +\e{o}\pa{1},
\label{asymptotique log det general ordre zero}
\end{multline}
where
\begin{equation}
\nu(\la)=\f{-1}{2i\pi} \ln \pa{1+\ga F(\la)}, \quad
\mc{K}(\la;q)=\exp\paa{ \Int{-q}{q} \f{ \nu(\la)-\nu(\mu) }{
\theta(\la)-\theta(\mu) }\, \theta'(\mu)\,\dd \mu  },
\end{equation}
and, as before, $p'_{\pm}=[\Dp{\la}p(\la)]\mid_{\la=\pm q}$,
$\nu_{\pm}=\nu\pa{\pm q}$.
\end{cor}
\Proof
The change of variables $\theta\pa{\la}=\xi$ maps the kernel $V_{\th}$ on
the one of the GSK
\begin{equation}
V(\xi,\eta) =\sqrt{F\circ\theta^{-1}(\xi)\; F\circ\theta^{-1}(\eta)
}\; \f{e_+\circ\theta^{-1}(\xi)\; e_{-}\circ\theta^{-1}(\eta)
-e_-\circ\theta^{-1}(\xi)\; e_{+}\circ\theta^{-1}(\eta)} {2i\pi
\pa{\xi-\eta} }.
\nonumber \end{equation}
This kernel acts on $\intff{\theta\pa{-q}}{\theta\pa{q}}$ which is,
a priori, a non symmetric interval. However, it is enough to apply
the transformation $\la \mapsto
\la-\tf{\pa{\theta\pa{q}+\theta\pa{-q}}}{2}$ so as to recover the
symmetry of the interval. Then, it remains to enforce the inverse transformations on
the asymptotic formula for the Fredholm determinant of $V$. \qed

Let us write explicitly the asymptotics $\eqref{asymptotique log det
general ordre zero}$ in the case of the kernel
\begin{equation}
V_{\e{sh}}(\la,\mu)=\ga \sqrt{F(\la)\, F(\mu) }\; \f{e_+(\la)\,
e_{-}(\mu) -e_-(\la)\, e_{+}(\mu)}{2i\pi \sinh(\la-\mu)}, \quad
e_{\pm}(\la)=\ex{\pm [i\tf{xp(\la)+g(\la)]}{2}},
\nonumber
\end{equation}
\noindent as it plays a crucial role in the analysis of the
asymptotic behaviour of the two-point  functions in the massless phase of
the $XXZ$ Heisenberg chain \cite{KitKMST009}. In this case,
equation \eqref{asymptotique log det general ordre zero} reads
\begin{multline}
\ln\ddet{}{I+V_{\e{sh}}}= 2 \Int{-q}{q} \dd\la\, \nu(\la)
\ln'[e_-(\la)]
+  \f{1}{2} \Int{-q}{q} \dd \la\, \dd \mu \,  \f{\nu'(\la)\, \nu(\mu)-
\nu(\la)\, \nu'(\mu) }{\tanh \pa{\la-\mu}}
 \\
+ \sul{\sigma=\pm}{} \left\{ \ln \f{ G(1,\nu_{\sigma})  }
{ \pac{ \sinh(2q)\, p'_{\sigma} x}^{\nu_{\sigma}^2}  } + \sigma
\nu_{\sigma} \Int{-q}{q} \dd \la \, \f{ \nu_\sigma-\nu(\la) }{
\tanh\pa{\sigma q-\la} } \right\} +\e{o}\pa{1}  .
\label{asymptotique sh}
\end{multline}
It is clear that the last equation can be used in order to obtain an
analog of the asymptotic expansion for multiple integrals of the
type \eqref{integ-F} where the rational functions $z-\lambda$ are
replaced by the hyperbolic $\sinh(z-\lambda)$. Namely, let
\begin{equation}\label{integ-Fsh}
\mc{I}_n^{\e{sh}}\pac{\mc{F}_n}
=\oint\limits_{\Gamma\pa{\intff{-q}{q}}}^{} \hspace{-3mm}\f{\dd^n
z}{\pa{2i\pi}^n}
\Int{-q}{q}\f{\dd^n \la}{\pa{2i\pi}^n} \; \,
\mc{F}_n\pa{\ba{c}\paa{\la} \\ \paa{z} \ea }\,
\prod_{j=1}^n
\f{\ex{ ix\pa{p\pa{z_j}-p\pa{\la_j}}  }}
  {\sinh\pa{z_j-\la_j}\sinh\pa{z_j-\la_{j+1}}}\;.
\end{equation}

Then under the conditions of Corollary~\ref{cor-Hn de Fn} one has
the following asymptotic estimate
\begin{multline}
\mc{I}_n^{\e{sh}}\pac{\mc{F}_n} =\f{1}{2i\pi} \Int{-q}{q} \dd\la \,
\paa{ixp'(\la) +
\partial_{\eps}} \mc{F}_n\pa{\ba{c} \paa{\la}^{n}
\\ \{\la+\eps\}, \paa{\la}^{n-1} \ea} \pour{\eps=0}
\\
+ \sul{\sg=\pm}{} \pa{b_n-c_n\ln\pa{\sinh(2q)p'_{\sg}x}}
\mc{F}_n\pa{\ba{c} \paa{\sg q}^{n} \\ \paa{\sg q}^{n} \ea}
 \\
 +\f{n}{\pa{2\pi}^2} \sul{\sg=\pm}{} \sul{p=1}{n-1}\Int{-q}{q} \dd \la
\f{\mc{F}_n\pa{\ba{c}\paa{\sg q}^{n} \\ \paa{\sg q}^{n} \ea }- \mc{F}_n\pa{\ba{c}\paa{\sg q}^{p} , \paa{\la}^{n-p}\\
\paa{\sg q}^{p} , \paa{\la}^{n-p}\ea } }{p\pa{n-p}\tanh\pa{q-\sg \la}} \\
+ \f{n}{2\pa{2\pi}^2}\sul{p=1}{n-1}\Int{-q}{q} \f{\dd \la \dd \mu
}{\pa{n-p}\tanh\pa{\la- \mu}}
\left\{
\partial_{\eps}\mc{F}_n\pa{ \ba{c}\paa{\la+\eps},\paa{\la}^{p-1},\paa{\mu}^{n-p}\\ \paa{\la+\eps} ,\paa{\la}^{p-1},\paa{\mu}^{n-p}\ea}\right.
                 \\
\left. -\partial_{\eps}\mc{F}_n\pa{\ba{c} \paa{\mu+\eps}
,\paa{\mu}^{p-1},\paa{\la}^{n-p} \\
\paa{\mu+\eps},\paa{\mu}^{p-1},\paa{\la}^{n-p} \ea}
\right\}\pour{\eps=0}+\e{o}(1)\; .
\label{Asy-sinh}
\end{multline}
%


\section*{Conclusion\label{C-n}}

We have obtained in this article the leading asymptotic expansion of the
Fredholm determinant of the GSK. As we have mentioned, our main
motivation  is to apply this result to the asymptotic analysis
of the correlation functions of quantum integrable models, using in particular the asymptotic study of multiple integrals performed in Section~\ref{sec-integrales}. This is done in  \cite{KitKMST009}.

Another development is to extend the above
analysis so as to handle truncated Wiener--Hopf operators with
symbols having Fischer--Hartwig type discontinuities. The
corresponding results are published in
\cite{KozWienerHopfWithFischerHartwig}. The results for the case of Toeplitz, Hankel and Toeplitz + Hankel determinants with Fisher-Hartwig singularities appeared recently in \cite{DeiftIK2008,DeiftIK2009}.

Let us also point out some unsolved problems. One of them concerns
the derivation of the asymptotics of the Fredholm determinant of the GSK
via the method based on its derivative over endpoint $q$. It would
be important to obtain a complete justification of this method,
since it is rather powerful and at the same time relatively simple.

Another problem is to prove the conjecture on the $\mathbb{Z}-\nu$ periodicity for the asymptotic expansion of the Fredholm determinant. If
this property does hold, then all oscillating corrections can be
obtained from the non-oscillating ones via a simple shift of $\nu$ by
integer numbers. This could lead to a much simpler way to compute sub-leading corrections for such determinants.

\section*{Acknowledgements}

We are very grateful to A.~R.~Its for useful and numerous
discussions. J. M. M., N. S. and V. T. are supported by CNRS.
N. K., K. K. K., J. M. M. and V. T. are supported by the ANR program GIMP
ANR-05-BLAN-0029-01.
N. K. and V. T. are supported by the ANR program  MIB-05 JC05-52749.
We acknowledge support from the French-Russian
Exchange Program GDRI-471 of CNRS and RFBR-09-01-93106-CNRSa. N. S. is also supported by the Program of RAS Mathematical Methods of the
Nonlinear Dynamics, RFBR-08-01-00501a, Scientific Schools
795.2008.1.
N. K and N. S. would like to thank the Theoretical
Physics group of the Laboratory of Physics at ENS Lyon for
hospitality, which makes this collaboration possible.


\appendix

\section{Some properties of confluent hypergeometric function}
\label{App-A}

For generic parameters $\pa{a,c}$ the Tricomi confluent hypergeometric
function $\Psi\pa{a,c;z}$ is one of the  solutions to the
differential equation
\begin{equation}
z y'' + \pa{c-z} y' -a y=0\;.
\end{equation}
It satisfies the properties:

  $\bullet$ Differentiation:
\begin{align}
 \Psi'(a,c;z)&=\frac az\Bigl[(a-c+1)\Psi(a+1,c;z)-\Psi(a,c;z)\Bigr]
             \nonumber\\
  & =\frac 1z\Bigl[(a-c+z)\Psi(a,c;z)-\Psi(a-1,c;z)\Bigr].
\label{differentiation de CHF}
\end{align}

 $\bullet$ Monodromy:
\begin{multline}
 \Psi(a,1;ze^{2mi\pi})= \Psi(a,1;z)\left(1-me^{i\pi a(\epsilon+1)}
 +me^{i\pi a(\epsilon-1)}\right)
 \\
 +\frac{2\pi ime^{i\pi a\epsilon+z}}{\Gamma^2(a)}
 \Psi(1-a,1;-z),
\label{Monodromie de Psi}
\end{multline}
where $\eps=\e{sgn}\pa{\Im (z)}$. In particular,
\begin{align}
 &\Psi(a,1;ze^{2i\pi})= \Psi(a,1;z)e^{-2i\pi a}+
 \frac{2\pi ie^{-i\pi a+z}}{\Gamma^2(a)}
 \Psi(1-a,1;-z),  &\Im (z)<0,
\label{cut-Psi-1}\\
 &\Psi(a,1;ze^{-2i\pi})= \Psi(a,1;z)e^{2i\pi a}-
 \frac{2\pi ie^{i\pi a+z}}{\Gamma^2(a)}
 \Psi(1-a,1;-z),  &\Im (z)>0.
\label{cut-Psi-2}
\end{align}

 $\bullet$ Asymptotic expansion:
\begin{equation}
 \Psi(a,c;z) \sim \sum_{n=0}^\infty(-1)^n\frac{(a)_n(a-c+1)_n}{n!}z^{-a-n},
 \quad z\to\infty,
 \quad -\frac{3\pi}2<\arg(z)<\frac{3\pi}2,
\label{asy-Psi}
\end{equation}
with $\pa{a}_n= \tf{\Ga\pa{a+n}}{\Ga\pa{a}}$.

We have the following recombination between the Tricomi CHF
$\Psi\pa{a,c;z}$ and the Humbert CHF $\Phi\pa{a,c;z}$
\begin{equation}
\Phi\pa{a,c;z}
 =\f{\Gamma\pa{c}}{\Gamma\pa{c-a}} \ex{i\eps a \pi} \Psi\pa{a,c;z}
 +\f{\Gamma\pa{c}}{\Gamma\pa{a}}\ex{i\eps \pi \pa{a-c}+z} \Psi\pa{c-a,c;-z},
\label{Phi s'ecrit comme Psi}
\end{equation}
where $\eps= \e{sgn}\pa{\Im (z)}$, and
\begin{equation}
 \Phi\pa{a,c;z}=\sum_{n=0}^\infty \frac{(a)_n}{(c)_n}\frac{z^n}{n!}.
\end{equation}
Such a recombination formula allows to obtain the asymptotic
expansion of the Humbert CHF:
\begin{multline}
\Phi\pa{a,c;z} = \f{\Ga\pa{c}}{\Ga\pa{c-a}} \paf{\ex{i\pi \eps}}{z}^{a}
\sul{n=0}{M} \f{\pa{a}_n\pa{a-c+1}_n}{n!\pa{-z}^n}  + \e{O}\pa{\abs{z}^{-a-M-1}}\\
 +\f{\Ga\pa{c}}{\Ga\pa{a}} \ex{z} z^{a-c} \sul{n=0}{N} \f{\pa{c-a}_n \pa{1-a}_n}{n! z^n}
+ \e{O}\pa{\abs{\ex{z}z^{a-1-c-N}}}.
\label{asymptotiques phi}
\end{multline}

One can estimate  integrals involving a product of two CHF as below, either by using La\-pla\-ce--type integral representations for the functions $\Phi\pa{a,c;z}$ and $\Psi\pa{a,c;z}$ or applying the
method given in \cite{KozWienerHopfWithFischerHartwig}.
The latter uses Erdelyi's
representation of Laplace transforms of products of CHF in terms of
Lauricella function adjoint to some asymptotic expansion of
Lauricella function. In any case, the result reads:
\begin{equation}
\Int{0}{+\infty} \dd t \paa{\ex{-i\pi a} \varphi\pa{a; t} - 1 }
 = -2i a,\label{integrales tau et phi0}
\end{equation}
\begin{equation}
\Int{0}{+\infty} \dd t \paa{\ex{-i\pi a} \tau\pa{a; t} + 1 +
\f{2ia}{t+1}}
 = 2i a -a \pac{\psi\pa{a}+\psi\pa{-a}} ,
\label{integrales tau et phi}
\end{equation}
and the Riemann integrability of the integrands is part of the conclusion. We remind the definition of the
functions $\tau\pa{a;t}$ and $\varphi\pa{a;t}$:
\begin{align}
&\varphi\pa{\nu;t}= \Phi\pa{-\nu,1;-i t}\Phi\pa{\nu,1;i t},  \\
&\tau\pa{\nu;t}
 =-\Phi\pa{-\nu,1;-i t}\Phi\pa{\nu,1;i t}
  +  \pa{\Dp{z}\Phi}\pa{-\nu,1;-i t}\Phi\pa{\nu,1;i t}\nonumber\\
&\hspace{6.5cm}
  +  \Phi\pa{-\nu,1;-i t}\pa{\Dp{z}\Phi}\pa{\nu,1;i t}.
\end{align}

\section{Three preparatory Lemmas}\label{Two lemme preparatopire}

Here we prove three preparatory integration lemmas used in
Section~\ref{Section asymptotique log det}.
\begin{lemme}
\label{lemme preparatoire}
Let $\mc{R}\pa{u,t}$ be a function of two variables defined on
$I\times \R^+ $, where $I$ is an open interval of $\R$ containing
$0$. Suppose that the partial applications $u \tend \mc{R}\pa{u,t}$
are $\msc{C}^{1}\pa{I}$ for  all but finitely many $t$'s and that
$t\tend \mc{R}\pa{u,t}$ is Riemann integrable uniformly in $u$,
i.e.:
\begin{align}
 &\forall \rho >0 , \  \forall M > 0  , \  \forall u_0 \in I   , \  \exists \upsilon >0  \quad\text{such that}
              \nonumber\\
& u \in \intoo{-\upsilon + u_0}{\upsilon+u_0} \cap I, k \in
\paa{0,1} \Rightarrow
\abs{\Int{M}{+\infty} \dd t \pac{
\Dp{1}^k\mc{R}\pa{u,t}-\Dp{1}^k\mc{R}\pa{u_0,t}}  } \leq \rho.
\end{align}
Then for $g \in \msc{C}^{1}\pa{I}$
\begin{equation}
\Int{0}{\delta} x  g\pa{t} \mc{R}\pa{t,x t}\,\dd t= g\pa{0}
\Int{0}{+\infty} \mc{R}\pa{0,t} \dd t + \e{o}\pa{1}
\end{equation}
where the small  $\e{o}\pa{1}$ is with respect to the successive
limit $x\delta \tend +\infty$ and $\delta \tend 0$.
\end{lemme}

\Proof
One has
\begin{equation}
\Int{0}{\delta} x  \bigl(g\pa{t}\mc{R}\pa{t,x
t}-g\pa{0}\mc{R}\pa{0,x t}\bigr)\,\dd t= \Int{0}{\delta} \dd y
\Int{x y}{x \delta}
\Dp{y}\pac{g\pa{y} \mc{R}\pa{y,t}} \dd t  \; .
\end{equation}
Consider a function
\begin{equation}
g:\pa{y,a,b} \mapsto \Int{a}{b} \dd t \Dp{y}\pac{g\pa{y}
\mc{R}\pa{y,t}}
\end{equation}
on the compact set $\intff{0}{\delta}\times \ov{\R^+}\times
\ov{\R^+}$.  $g$ is clearly continuous on the interior and the
uniform Riemann-integrability of $\mc{R}\pa{y,t}$ guarantees that it
is continuous in an neighbourhood of $\pa{*,+ \infty,*}$, $\pa{*,*,+
\infty}$  and $\pa{*,+\infty,+\infty}$. Hence, $\abs{g}$ is bounded,
say by $B$, as continuous function on a compact set. Thus,
\begin{equation}
\abs{\Int{0}{\delta} \dd y \Int{x y}{x \delta}
\Dp{y}\pac{g\pa{y} \mc{R}\pa{y,t}} \dd t  } \leq \delta B,
\end{equation}
which ends the proof of Lemma~\ref{lemme preparatoire}. \qed

\begin{lemme}
\label{lemme preparatopire 2}
Let $g\in \msc{C}^1\pa{I}$ for some open interval $I$ containing
$0$, then
\begin{equation}
\Int{0}{\de}  \f{g\pa{ t } x \dd t}{1+ x t} = g\pa{0} \ln x \de
+\e{o}\pa{1},
\end{equation}
where $\e{o}\pa{1}$ stands with respect to the successive limits
$x\de \tend +\infty$ and $\de \tend 0$.
\end{lemme}

\Proof
We have
\begin{align}
\Int{0}{\de} g\pa{t} \f{x \dd t}{1+x t} &=  g\pa{0}\ln \pa{x\de +1} +\Int{0}{\de} \dd t \Int{0}{t}  \dd y \f{g'\pa{y} x}{1+x t} \\
&= g\pa{0} \ln \de x+ \e{o}\pa{1} + \Int{0}{\de} \dd y g'\pa{y} \ln
\paf{x\de+1}{xy +1} \; .
\end{align}
But,
\begin{equation}
\abs{ \Int{0}{\de} \dd y g'\pa{y} \ln\paf{\de+\tf{1}{x} }{y +
\tf{1}{x}}  } \leq \underset{\intff{0}{\de}}{\sup}{\abs{g'}}
 \times \pa{\de -\tf{\ln\pa{\de+\tf{1}{x}}}{x} } \limit{\de}{0} 0 \; ,
\end{equation}
which ends the proof of Lemma~\ref{lemme preparatopire 2}. \qed

\begin{lemme}
\label{lemme h et derivee kappa NLS}
Let $\kappa$ be defined in terms of $\nu$ as in \eqref{nuit}, and
set
\begin{equation}
H(q) = \f{1}{2} \Int{-q}{q} \dd \la \dd \mu
  \f{\nu'(\la)\, \nu(\mu) - \nu'(\mu)\, \nu(\la) }{\la-\mu}
+ \sul{\eps=\pm}{} \eps \nu_\eps\, \ln \kappa (\eps q; q ).
%
\end{equation}
Then,
\begin{equation}
 2\nu_+  \D{}{q} \pac{\ln\kappa\pa{q;q}}
  - 2\nu_-  \D{}{q} \pac{\ln\kappa\pa{-q;q}}
  - \f{\pa{\nu_+-\nu_-}^2}{q} = \D{H\pa{q}}{q}.
\label{egalite pour la cste du log det }
\end{equation}
\end{lemme}

\Proof
Using \eqref{nuit}, one can express the derivative of $H\pa{q}$ as
\begin{equation}
\D{H\pa{q}}{q} = \sul{\eps=\pm}{} \nu\pa{\eps q } \Int{-q}{q} \dd
\mu \f{\nu'\pa{\eps q} - \nu'\pa{\mu} }{\eps q -\mu}
+ \sul{\eps= \pm}{} \eps \nu\pa{\eps q} \D{}{q}\pac{\ln \kappa
\pa{\eps q; q }}.
\end{equation}
Thus, proving \eqref{egalite pour la cste du log det } amounts to
establishing the equality
\begin{equation}
\sul{\eps= \pm}{} \eps \nu\pa{\eps q} \D{}{q}\pac{\ln \kappa
\pa{\eps q; q }} - \f{\pa{\nu_+-\nu_-}^2}{q}
= \sul{\eps=\pm}{} \nu\pa{\eps q } \Int{-q}{q} \dd \mu
\f{\nu'\pa{\eps q} - \nu'\pa{\mu} }{\eps q -\mu} \;.
\end{equation}
The latter follows from an integration by parts:
\begin{align*}
\sul{\eps= \pm}{} \eps \nu\pa{\eps q}\,  & \D{}{q}\pac{\ln \kappa \pa{\eps q; q }}\\
 &=\sul{\eps= \pm}{} \eps\nu_{\eps} \paa{\nu'_{\eps}+\f{\nu_+-\nu_-}{2q}
   + \eps \Int{-q}{q} \dd \mu
         \f{\nu\pa{\mu}-\nu_{\eps}-\nu'_{\eps}\pa{\mu-\eps q}}
           { \pa{\mu-\eps q}^{2} } }
    \\
 &=\sul{\eps= \pm}{} \eps\nu_{\eps} \paa{\nu'_{\eps}+\f{\nu_+-\nu_-}{2q}
   + \f{\nu_{-\eps}-\nu_{\eps}+2\eps q\nu'_{\eps} }{ -2\eps q }
   + \eps \Int{-q}{ q} \dd \mu \f{\nu'\pa{\mu}-\nu'_{\eps}}{ \mu-\eps q } }
  \\
 &= \f{\pa{\nu_+ - \nu_-}^2}{q}
   +\sul{\eps=\pm}{} \nu_{\eps}
   \Int{-q}{ q} \dd \mu \f{\nu'\pa{\mu}-\nu'_{\eps}}{ \mu-\eps q } .
\end{align*}
This ends the proof.  \qed


\section{The density Theorem \label{DTh}}

\begin{theorem}
\label{theoreme de densite}
Let $U$, $W$ be two open neighbourhoods of $\intff{-q}{q}$,
and let $\mc{F}_n\pa{\ba{c}\paa{\la} \\ \paa{z} \ea }$ be a holomorphic function on $U^n\times W^n$, symmetric separately
in the $n$ variables $\la$ and in the $n$ variables $z$. Then, for any compact subsets $K$ (resp. $P$) of U (resp.W)
there exists a sequence $\pa{\varphi_p,\phi_p}_{p\in
\mathbb{N}}$ in $\mc{H}\pa{K}\times \mc{H}\pa{P}$ such that
\begin{equation}
\mc{F}_n\pa{ \ba{c} \paa{\la} \\ \paa{z} \ea }
=\sul{p=0}{+\infty}
\pl{i=1}{n} \varphi_{p}\pa{\la_i}\phi_p\pa{z_i} \qquad
\text{uniformly on }  K^n\times P^n\, .
\end{equation}
%
%
%
\end{theorem}

\Proof
Let $K$ and $P$ be as above.

Let $X=K^n\times P^n/\sim$, where the relation $\sim$ is defined as follows:
$\pa{\la,z}\sim\pa{\la',z'}$ if there exists a couple of permutations $\pa{\sg,\pi} \in
\mf{G}_n\times\mf{G}_n$ such that
$\pa{\la^{\sg},z^{\pi}}=\pa{\la',z'}$, where $\la^{\sg}$ stands for
$\pa{\la_{\sg\pa{1}},\dots, \la_{\sg\pa{n}}}$.
Since $\mf{G}_n\times\mf{G}_n$ is a discrete group, its action on
$K^n\times P^n$ is by definition proper, i.e. $\forall L \supset
K^n\times P^n$
\begin{equation}
\paa{\pa{\sg,\pi} \in \mf{G}_n\times\mf{G}_n : L^{\sg,\pi} \cap L =
\emptyset} \quad \e{is}\; \e{discrete.}
\end{equation}
This ensures that $X$ is a compact Hausdorff  topological space.
Moreover the space $\Cont{}{X,\Cx}$ of continuous functions on X is
canonically identified
with the space of continuous functions on $K^n\times P^n$ that are symmetric in the first or the last $n$ variables.

Define the subspace $S$ of $\Cont{}{X,\Cx}$ as the subset of functions $\mc{F}_n^{(\varphi,\phi)}$ of the form
\begin{equation}
\mc{F}_n^{(\varphi,\phi)}\pa{\ba{c}\paa{\la} \\ \paa{z} \ea }
=\pl{i=1}{n} \varphi\pa{\la_i}\phi\pa{z_i},
\end{equation}
where $\pa{\varphi,\phi} \in \mc{H}\pa{K}\times\mc{H}\pa{P}$,
and let $\mc{S}$ be the $C^*$-algebra generated by $S$. We
have that $S$ and hence $\mc{S}$ separates points in $X$. Indeed,
let   $\pa{\la,z}$ and $\pa{\mu,y}$ be any two representatives in
$K^n\times P^n$ of two distinct points in $X$. Thus
\begin{itemize}
\item
 there exists $\la_i \in K$ such that exactly $p$ of the $n$ coordinates of the $n$-tuple $\la$ are equal to $\la_i$, whereas exactly $q$ of the $n$ coordinates of the $n$-tuple $\mu$ are equal to $\la_i$, with $p\not=q$;
\item or there exists $z_i \in P$ such that exactly $p$ of the $n$ coordinates of the $n$-tuple $z$ are equal to $z_i$, whereas exactly $q$ of the $n$ coordinates of the $n$-tuple $y$ are equal to $z_i$, with $p\not=q$.
\end{itemize}
The situation is similar in the case of the first $n$ and last
$n$ variables, therefore we only treat the first case. By Lagrange
interpolation there exist a polynomial $Q$ such that, for any coordinate $\la_k$ of $\la$ and any coordinate $\mu_k$ of $\mu$ satisfying $\la_k \not= \la_i$ and $\mu_k \not=\la_i$,
\begin{equation}
%
Q\pa{\la_k}=Q\pa{\mu_k}=1 \quad \e{and} \quad Q\pa{\la_i}=2  .
\end{equation}
\noindent The function
\begin{equation}
\mc{F}_n^{(Q,1)}\pa{\ba{c}\paa{\la} \\ \paa{z} \ea } =\pl{p=1}{n} Q\pa{\la_p} \in
\mc{S}
\end{equation}
\noindent separates the projections of $\pa{\la,z}$ and $\pa{\mu,y}$
on $X$. Thus $\mc{S}$ is a $C^*$-subalgebra of $\Cont{}{X;\Cx}$ that
separates points. It then follows by the Stone-Weierstrass theorem
that $\mc{S}=\Cont{}{X;\Cx}$.

Let $\mc{F}_n$ be holomorphic on $U^n\times W^n$ and symmetric in the first and in the last $n$ variables.  There
exists compact sets $K_{\eps}\subset U$ and $P_{\eps}\subset W$ such that $K \subset
\overset{\circ}{K_{\eps}}$ and $P \subset
\overset{\circ}{P_{\eps}}$. Here, $\overset{\circ}{K_{\eps}}$ stands for the interior of $K_{\eps}$. Thus the restriction of $\mc{F}_n$ to $K_{\eps}^n\times P_{\eps}^n$ also belongs to
$\Cont{}{X_\eps;\Cx}$, with $X_\eps=K_{\eps}^n\times P_{\eps}^n / \sim$, and therefore there
exists $\big(\widetilde{\varphi}_p,\widetilde{\phi}_p\big)_{p\in\mathbb{N}}$ in
$\Cont{}{K_{\eps};\Cx}\times \Cont{}{P_{\eps};\Cx}$ such that
\begin{equation}
\mc{F}_n\pa{\ba{c}\paa{\la} \\ \paa{z} \ea }=\sul{p=0}{+\infty}
\pl{i=1}{n}\widetilde{\varphi}_p\pa{\la_i}
\widetilde{\phi}_p\pa{z_i}  \qquad \e{uniformly} \; \e{on} \;\;\;
K_{\eps}^n\times P_{\eps}^n.
\label{approx de F par des fontions continues}
\end{equation}
\noindent In particular the sequence converges uniformly to
$\mc{F}_n$ on $ \pa{\Dp{}{}K_{\eps} }^n\times \pa{\Dp{}{}P_{\eps}}^n
$, the latter set being compact. Therefore we have
\begin{multline}
\sul{p=0}{N}\Int{\Dp{}{}K_\eps}{} \f{\dd^n \mu}{\pa{2i\pi}^{n}}
\Int{\Dp{}{}P_\eps}{} \f{\dd^n y}{\pa{2i\pi}^n}
\pl{i=1}{n}\f{\widetilde{\varphi}_p\pa{\mu_i}
\widetilde{\phi}_p\pa{y_i}}{\pa{\mu_i-\la_i}\pa{y_i-z_i}}
= \sul{p=0}{N} \pl{i=1}{n}\varphi_p\pa{\la_i}\phi\pa{z_i}  \\
\limit{N}{+\infty} \Int{\Dp{}{}K_\eps}{} \f{\dd^n \mu}{\pa{2i\pi}^{n}}  \Int{\Dp{}{}P_\eps}{}
\f{\dd^n y}{\pa{2i\pi}^n}
\f{\mc{F}_n\pa{\paa{\mu}\mid
\paa{y}}}{\pl{i=1}{n}\pa{\mu_i-\la_i}\pa{y_i-z_i}} =
\mc{F}_n\pa{\paa{\la}\mid \paa{z}},
\end{multline}
uniformly in $\pa{\la,z} \in K^n\times P^n$. Moreover,
\begin{equation}
\varphi_p\pa{\la}=\Int{\Dp{}{}K_{\eps}}{} \f{\dd \mu}{2i\pi}
\f{\widetilde{\varphi}_p\pa{\mu}}{\mu-\la}
\quad \e{and} \quad \phi_p\pa{z}=\Int{\Dp{}{}P_{\eps}}{} \f{\dd
y}{2i\pi} \f{\widetilde{\phi}_p\pa{y}}{y-z}
\end{equation}
are holomorphic in $K$, resp. $P$.
\qed

\section{Form of the sub-leading terms in $I_n^{\e{sub}}$}
\label{append-subleading}

In this appendix, we focus on the general structure of the sub-leading asymptotics of cyclic integrals. We show that the $\tf{1}{x^N}$ term in the non-oscillating part can be obtained as an action of at most $N$ partial derivatives of the function $\mc{F}_n$ followed by an evaluation at $\pm q$ or by an integration over $[{-q};{q}]$.

In principle, the contour integrals defining \eqref{H_calligraphique} can be computed to the end.
However, the result is quite intricate, and we do not need, for the further applications, the formula in its whole generality. Indeed, we are interested in a particular sub-class of such integrals. More precisely we shall focus on the sub-class that is susceptible to produce the highest possible derivatives of the function $\mc{F}_n$. Here, by highest derivative we mean the total degree of all the partial derivatives that might act on the integrand. This subclass is identified in the upcoming lemma.

\begin{lemme}
\label{lemme des petales de Fleur}
Let $r,t\in\mathbb{N}$ with $r+t\geq 1$ label negations
$\sg_1,\ldots,\sg_{r+t}\in\{\pm\}$. Also introduce sufficiently small numbers $0<\de_1<\dots<\de_{r+t}<q$ as well as positive integers
$k_1,\ldots,k_{r+t}$. Finally, let $G\in \mc{H}\pa{ \ov{D}_{\sg_1 q,\de_1}\times\cdots\times \ov{D}_{\sg_{r+t} q,\de_{r+t}}}$ and
\begin{equation}
\mc{G}_{r,\, t}^{(\{\sg_i\},\{k_i\})}\pac{G}
=\Int{\Dp{}D_{\sg q,\de}^{\times \pa{r+t} }}{}
\hspace{-1mm}
 \f{\dd^{r+t}z}{\pa{2i\pi}^{r+t}}
 \pl{\ell=2}{r}\f{1}{z_{\ell-1}-z_{\ell}}\pl{\ell=r+2}{r+t}\f{1}{z_{\ell-1}-z_{\ell}}
\pl{\ell=1}{r+t}\f{1}{\pa{z_{\ell}-\sg_\ell q}^{k_{\ell}}} \;
G(\{z\})\, .
\end{equation}
with  $\Dp{}D_{\sg q,\de}^{\times \pa{r+t} }\equiv\Dp{}D_{\sg_1 q,\de_1}\times \dots \times \Dp{}D_{\sg_{r+t} q, \de_{r+t}}$.

Then, the integral $\mc{G}_{r,\,t}^{(\{\sg_i\},\{k_i\})}\pac{G}$ can be computed as some combinatorial  sum involving derivatives of $G$ at the points $\sg_i q$, the maximal order of such derivatives being equal to
\begin{equation}\label{ordre-der}
\sul{i=1}{r+t} k_i -n_r-n_t+\delta_{r,0}+\delta_{t,0}-2.
\end{equation}
Here $n_r$, resp. $n_t$, is the number of times the sequence $\pa{\sg_1,\dots,\sg_r}$, resp.
$\pa{\sg_{r+1},\dots, \sg_{r+t}}$, changes sign, and $\delta_{r,0},\,\delta_{t,0}$ denote the usual Kronecker symbols.
\end{lemme}

\Proof
Let us prove the claim by induction on $r+t$.

First, for $r+t=1$, \eqref{ordre-der} is obviously satisfied. Indeed,
\begin{equation}\label{G10}
 \mc{G}_{1,0}^{(\sg,k)}\pac{G}=\mc{G}_{0,1}^{(\sg,k)}\pac{G}
 =\frac{1}{(k-1)!}\pa{\Dp{z}^{k-1} G} (\sg q).
\end{equation}

Let us now assume that the result holds for any function $G$ up to some value of $r+t$. We will prove that it also holds for $r+t+1$.

Note first that $\mc{G}_{r,\,t}^{(\{\sg_i\},\{k_i\})}\big[G\big]
=\mc{G}_{t,\,r}^{(\{\tilde\sg_i\},\{\tilde{k}_i\})}\big[\tilde{G}\big]$, in which $\tilde{G}$, $\{\tilde\sg_i\}$, $\{\tilde{k}_i\}$ are obtained from $G$, $\{\sg_i\}$, $\{k_i\}$ by a reordering of the variables.
Hence, it is enough to prove the claim for $\mc{G}_{r+1,\,t}$.
We will have to distinguish two cases, depending on whether $r+1=1$ or $r+1>1$.

In the case $r+1=1$, it is easy to see that
\begin{equation}
\mc{G}_{1,\,t}^{(\{\sg_i\},\{k_i\})}\pac{G}
=\f{1}{\pa{k_1-1}!}\,
    \mc{G}_{0,\,t}^{(\sg_2,\ldots,\sg_{t+1},k_2,\ldots,k_{t+1})}
    \pac{\Dp{z_1}^{k_1-1}G\pa{\sg_1q,\paa{z_i}_{i=2}^{t+1}}}  \, ,
\end{equation}
which means that $\mc{G}_{1,\,t}^{(\{\sg_i\},\{k_i\})}\pac{G}$ can be expressed in terms of derivatives of $G$ of maximal order $\big(\sum_{i=2}^{t+1} k_i -1 -n_t +\delta_{t,0}\big) +(k_1-1)$, hence the result.

Let us now consider the case $r+1>1$. We have
\begin{align}
\mc{G}_{r+1,\,t}^{(\{\sg_i\},\{k_i\})}\pac{G}
&=\frac{1}{(k_1-1)!}\Int{\Dp{}D_{\sg q,\de}^{\times \pa{r+t} }}{}
 \pl{\ell=2}{r+t+1}\f{\dd z_{\ell}}{2i\pi} \;
 \pl{\ell=3}{r+1}\f{1}{z_{\ell-1}-z_{\ell}}
 \pl{\ell=r+3}{r+t+1}\f{1}{z_{\ell-1}-z_{\ell}}
 \pl{\ell=2}{r+t+1}\f{1}{\pa{z_{\ell}-\sg_\ell q}^{k_{\ell}}}  \nonumber\\
&\hspace{3cm}
\times
 \partial_{z_1}^{k_1-1}\pa{\frac{G(\{z\})}{z_1-z_2}}\pour{z_1=\sg_1 q}
              \nonumber\\
&= -
 \Int{\Dp{}D_{\sg q,\de}^{\times \pa{r+t} }}{}
 \pl{\ell=2}{r+t+1}\f{\dd z_{\ell}}{2i\pi} \;
 \pl{\ell=3}{r+1}\f{1}{z_{\ell-1}-z_{\ell}}
 \pl{\ell=r+3}{r+t+1}\f{1}{z_{\ell-1}-z_{\ell}}
 \pl{\ell=2}{r+t+1}\f{1}{\pa{z_{\ell}-\sg_\ell q}^{k_{\ell}}}  \nonumber\\
&\hspace{3cm}
\times
\sul{k=0}{k_1-1}\frac{1}{k!}
  \f{1}{\pa{z_2-\sg_1 q}^{k_1-k}}
  \pa{\Dp{z_1}^{k}G}\pa{\sg_1 q, \paa{z_i}_{i=2}^{r+t+1}}\, .
\label{rec-Grt}
\end{align}
At this point one should distinguish between the two possible cases: $\sg_1\sg_2=1$ or $\sg_1\sg_2=-1$. We first assume
$\sg_1\sg_2=1$ (i.e. that there is no change of sign between $\sg_1$ and $\sg_2$),
and set $\hat{G}^+_k\big(z_2,\ldots,z_{r+t+1}\big)
 =\Dp{z_1}^{k}G\big(z_1,\ldots,z_{r+t+1}\big)\mid_{z_1=\sg_1 q}$. Then
\begin{equation}
\mc{G}_{r+1,\,t}^{(\{\sg_i\},\{k_i\})}\pac{G}
=- \sul{k=0}{k_1-1} \f{1}{k!}\,
\mc{G}_{r,t}^{(\{\sg_i\}_2^{r+t+1},\{k_2+k_1-k\}\cup\{k_i\}_3^{r+t+1})}
\pac{\hat{G}^+_k}
 \, .
\label{G-r+1-t-egal}
\end{equation}
The latter can be expressed in terms of derivatives of $G$ of maximal order  $k+\big(k_2+k_1-k+\sum_{i=3}^{r+t+1}k_i-n_r-n_t+\delta_{t,0}-2\big)=
\sum_{i=1}^{r+t+1}k_i -2-n_{r+1}-n_t+\delta_{t,0}$.

We now assume that  $\sg_1\sg_2=-1$. This leads to
\begin{equation}
\mc{G}_{r+1,\,t}^{(\{\sg_i\},\{k_i\})}\pac{G}
= - \sul{k=0}{k_1-1} \f{1}{k!}\,
\mc{G}_{r,t}^{(\{\sg_i\}_2^{r+t+1},\{k_i\}_2^{r+t+1})}
\pac{\hat{G}^-_k}\, ,
\label{G-r+1-t-diff}
\end{equation}
where the function
\begin{equation}
 \hat{G}^-_k\big(z_2,\ldots,z_{r+t+1}\big)
 =\frac{\Dp{z_1}^{k}G\big(z_1,\ldots,z_{r+t+1}\big)\mid_{z_1=\sg_1 q} }{(z_2+\sg_2 q)^{k_1-k}  }
\end{equation}
is holomorphic inside the integration contour $\Dp{}D_{\sg_2 q,\de_2}\times \dots \times \Dp{}D_{\sg_{r+t} q, \de_{r+t}}$.
Once again, the result will be expressed in terms of derivatives of $G$ and the maximal order of these derivatives will be $k_1-1+\big(\sum_{i=2}^{r+t+1}k_i-n_r-n_t+\delta_{t,0}-2\big)
=\sum_{i=1}^{r+t+1} k_i -n_{r+1}-n_t
+\delta_{t,0}-2$, which ends the proof of Lemma~\ref{lemme des petales de Fleur}.
\qed

\begin{rem}
 The integral can be explicitly computed using the recurrence formulas \eqref{G-r+1-t-egal} and \eqref{G-r+1-t-diff}.
In particular, in the simplest case $\sg_1=\dots =\sg_{r}$ and $\sg_{r+1}=\dots=\sg_{r+t}$, we have
\begin{equation}
\mc{G}_{r,\, t}^{(\{\sg_i\},\{k_i\})}\pac{G}
= \pa{-1}^{r+t-\delta_{r,0}-\delta_{t,0}}
\sum_{\substack{u_1,\ldots,u_{r+t} \\ u_\ell\,\in\,\Gamma_\ell  } }
\pl{\ell=1}{r+t} \f{1}{u_\ell! }\;
 \Dp{z_{r+t}}^{u_{r+t}}\ldots\Dp{z_{1}}^{u_{1}}
 G\pa{\{z\}}\pour{z_i=\sg_i q} \,
,
\label{formule pour G caligraphique avec plusieurs derivees}
\end{equation}
in which the parameters $u_\ell$ are summed over sets $\Gamma_\ell$ defined as
\begin{xalignat}{3}
 &\Gamma_\ell=\Bigg\{0,\ldots,\sum_{j=1}^\ell k_j-\sum_{j=1}^{\ell-1}u_j -1\Bigg\}, &
 &\Gamma_{r+\ell}=\Bigg\{0,\ldots,\sum_{j=r+1}^{r+\ell} k_j
                       -\sum_{j=r+1}^{r+\ell-1}u_j -1\Bigg\}, &
 &(1 \le \ell < r),
 \label{def-Gammal-1}\\
 &\Gamma_r=\Bigg\{\sum_{j=1}^r k_j-\sum_{j=1}^{r-1}u_j -1 \Bigg\}, &
 &\Gamma_{r+t}=\Bigg\{\sum_{j=r+1}^{r+t} k_j-\sum_{j=r+1}^{r+t-1}u_j -1 \Bigg\}.
 & &
 \label{def-Gammal-2}
\end{xalignat}
\end{rem}

\begin{cor}\label{cor-ordre-der}
 The subleading terms of order $N$ in the asymptotic expansion \eqref{series-Hn-sub} for the cycle integral $\mc{I}_n[\mc{F}_n]$ are obtained in terms of derivatives of the function $\mc{F}_n$. More precisely, the non-oscillating term $I_n^{(N;\, \e{nosc})}[\mc{F}_n]$ involves derivatives of $\mc{F}_n$ of total order at most equal to $N$, whereas the oscillating one $I_n^{(N;\, \e{osc})}[\mc{F}_n]$ involves derivatives of $\mc{F}_n$ of total order at most equal to $N-2$.
\end{cor}

\Proof
In order to apply Lemma~\ref{lemme des petales de Fleur} to the integral over $\partial D_{\sg q,\delta}^{\times (r+t)}$ in \eqref{H_calligraphique}, let us set
\begin{multline}
G\pa{\paa{z}}
= \f{ \wt{\mf{D}}^{\pa{s}}_{r,\, t}\begin{bmatrix}
              \{p_{\ell j}\},\{\eps_i\} \\ \{\sg_i\}, \{k_i\}
              \end{bmatrix}
\pa{\paa{z_i} ; \ln x}  }
{ \pa{\la-z_1}^2 \pa{\la-z_{r+1}}}
\pl{\ell=1}{r+t} \pa{\f{z_{\ell}-\sg_\ell q}
                    {p(z_{\ell})-p_{\sg_{\ell}} } }^{k_{\ell}}  \\
\times
\pl{\ell=1}{r+t}  \pl{m=1}{ n } \pl{j=1}{p_{\ell m}}
\paa{
\Int{-q}{q} \f{\dd \mu_{\ell, m, j}  }{z_{\ell}-\mu_{\ell,m,j}}\; \eth^{\pa{m}}_{z_\ell}(\mu_{\ell,m,j})
}
\cdot
\mc{F}_n\pab{
\big\{ \paa{z_{\ell}}^{\bar{p}_\ell } \big\}_{1\le \ell \le r+t} }
{ \big\{ \paa{z_{\ell}}^{\bar{p}_\ell+\eps_{\ell}  }   \big\}_{1\le \ell \le r+t}  }\, .
\end{multline}
The poles at $z_1=\la$ and $z_{r+1}=\la$ being outside of the skeleton $\Dp{}D^{\times (r+t)}_{\sg q,\de}$, this function is indeed holomorphic in a vicinity of the polydisc $D^{\times (r+t)}_{\sg q,\de}$.

Applying the result of Lemma~\ref{lemme des petales de Fleur} to this function and using the fact that $\sum k_i=N+1$ in \eqref{H non oscillant}, it follows immediately that the expression of $I_n^{(N;\,\e{nosc})}[\mc{F}_n]$ cannot involve derivatives of the function $\mc{F}_n$ of order higher than $N$.  This maximal order of derivatives  corresponds to $t=0$ and  $\forall \, i \; \sg_i=\sg$ with $\sg=\pm$ in \eqref{H non oscillant}.

Similarly, as in \eqref{H oscillant} $\sum k_i\le N$, $I_n^{(N;\,\e{osc})}[\mc{F}_n]$ cannot involve derivatives of $\mc{F}_n$ of order higher than $N-1$. Moreover, due to the constraints $\sum\eps_\ell=0$ and $\sum \eps_\ell p_{\sg_\ell}\not= 0$, it follows that the variables $\sg_i$ have to take both values $+$ and $-$, which means that either $t\ge 1$ or $n_r\ge 1$ in \eqref{ordre-der} (we recall that $r\ge 1$). Hence $I_n^{(N;\,\e{osc})}[\mc{F}_n]$ cannot involve derivatives of $\mc{F}_n$ of order higher than $N-2$.
\qed



\begin{thebibliography}{10}

\bibitem{AchiezerKacFormulaforTruncatedWienerHopf}
N.~I. Akhiezer, \emph{{"The continuous analogues of some Theorems on Toeplitz
  matrices."}}, Ukrainian Math. J. \textbf{16} (1964), 445--462.

\bibitem{BarnesDoubleGaFctn2}
E.~W. Barnes, \emph{{"The theory of the double gamma function."}}, Philos.
  Trans. Roy. Soc. London, Ser. A \textbf{\bf196}.

\bibitem{BarnesDoubleGaFctn1}
E.~W. Barnes, \emph{{"Genesis of the double gamma function."}}, Proc. London Math.
  Soc. \textbf{\bf{31}} (1900), 358--381.

\bibitem{BasorTracyProblemsWithTauFunctionAndSineKernel}
E.~L. Basor and C.~A. Tracy, \emph{{"Some problems associated with the
  asymptotics of $\tau$ functions."}}, Suaikaguku $no$3 \textbf{\bf 30} (1992),
  71--76.

\bibitem{BogoluibovIzerginKorepin1986}
N.~M. Bogoliubov, A.~G. Izergin, and V.~E. Korepin, \emph{{"Critical exponents
  for integrable models"}}, Nucl. Phys. B. \textbf{275} (1986), 687.

\bibitem{BogoliubiovIzerginKorepinBookCorrFctAndABA}
N.~M. Bogoliubov, A.~G. Izergin, and V.~E. Korepin, \emph{{"Quantum Inverse
  Scattering Method, Correlation Functions and Algebraic Bethe Ansatz."}},
  Cambridge monograph on mathematical physics., 1993.

\bibitem{BudynBuslaevPureGammaSineKernelAsympt}
A.~M. Budylin and V.~S. Buslaev, \emph{{"Quasiclassical asymptotics of the
  resolvent of an integral convolution operator with a sine kernel on a finite
  interval."}}, Algebra i Analiz $no$ 6 \textbf{\bf{7}} (1995), 79--103.

\bibitem{CheianovZvonarevZeroTempforFreeFermAndPureSine}
V.V.~Cheianov and M.R.~Zvonarev, \emph{{"Zero temperature correlation functions for the impenetrable fermion gas."}},
J. Phys. A:Math. Gen.,{\textbf{37}} (2004), 2261-2297.

\bibitem{ColIKT92}
F.~Colomo, A.~G. Izergin, V.~E. Korepin, and V.~Tognetti, \emph{Correlators in
  the {H}eisenberg {XX0} chain as {F}redholm determinants}, Phys. Lett. A
  \textbf{169} (1992), 237--247.

\bibitem{ColIKT93}
F.~Colomo, A.~G. Izergin, V.~E. Korepin, and V.~Tognetti, \emph{Temperature correlation functions in the {XX0} {H}eisenberg
  chain}, Teor. Mat. Fiz. \textbf{94} (1993), 19--38.

\bibitem{DeiIZ93}
P.~A. Deift, A.~R. Its, and X.~Zhou, \emph{Long-time asymptotics for integrable
  nonlinear wave equations}, Important developments in soliton theory, Springer
  Ser. Nonlinear Dynam., Springer, Berlin, 1993, pp.~181--204.

\bibitem{DeiZ94}
P.~A. Deift and X.~Zhou, \emph{Long-time behaviour of the non-focusing nonlinear
  schr\:odinger equation - a case study}, Lectures in Mathematical Sciences,
  vol.~5, University of Tokyo, Tokyo, 1994.

\bibitem{DieftItsZhouSineKernelOnUnionOfIntervals}
P.~A. Deift, A.R. Its and X.~Zhou, \emph{{"A Riemann--Hilbert approach to
  asymptotics problems arising in the theory of random matrix models and also
  in the theory of integrable statistical mechanics."}}, Ann. Math.
  \textbf{\bf{146}} (1997), 149--235.
    

\bibitem{DieftZhouSteepestDescentForOscillatoryRHP}
P.~A. Deift and X.~Zhou, \emph{{"A steepest descent method for oscillatory
  Riemann--Hilbert problems."}}, Intl. Math. Res. \textbf{\bf{6}} (1997),
  285--299.
  
\bibitem{DeiftIKZ2007}
P.~A. Deift, A.~R. Its, I. Krasovsky and X.~Zhou, \emph{The Widom-Dyson constant for 
the gap probability in random matrix theory},  J. Comput. Appl. Math.  
202  (2007),  no. 1, 26--47.

\bibitem{DeiftIK2008}
P. Deift, A. R. Its and I. Krasovsky, \emph{Toeplitz and Hankel determinants with 
singularities: announcement of results}, arXiv:0809.2420.


\bibitem{DeiftIK2009}
P. Deift, A. R. Its and I. Krasovsky, \emph{Asymptotics of Toeplitz, Hankel, and Toeplitz+Hankel determinants with Fisher--Hartwig singularities}, arXiv:0905.0443.


\bibitem{DescloizeauxMethaSineKernelFirstAsympotics}
J.~des Cloizeaux and M.~L. Mehta, \emph{{"Asymptotic behaviour of spacing
  distributions for the eigenvalues of random matrices."}}, J. Math. Phys.
  \textbf{\bf{14}} (1973), 1648--1650.

\bibitem{DysonSineKernelInverseScatteringAsymptoticExpansions}
F.~Dyson, \emph{{"Fredholm determinants and inverse scattering problems."}},
  Comm. Math. Phys. \textbf{\bf{47}} (1976), 171--183.

\bibitem{Ehrhardt2006}
T.~Ehrhardt, \emph{"dyson's constant in the asymptotics of the fredholm
  determinant of the sine kernel."}, Comm. Math. Phys. \textbf{\bf{262}}
  (2006), 317--341.

\bibitem{GaudinGUELevelSpacingasSineKernel}
M.~Gaudin, \emph{{"Sur la loi limite de l'espacement des valeurs propres d'une
  matrice al\'eatoire."}}, Nucl. Phys. \textbf{\bf{25}} (1961), 447--458.

\bibitem{GaudinMehtaDensityOfEigenvaluesRandomMatrices}
M.~Gaudin and M.~L. Mehta, \emph{{"On the density of eigenvalues of a random
  matrix."}}, Nucl. Phys \textbf{\bf 18} (1960), 420--427.

\bibitem{Haldane1980}
F.~D.~M. Haldane, \emph{{"General relation of correlation exponents and
  spectral properties of one-dimensional Fermi systems: Application to the
  anisotropic s = 1/2 Heisenberg chain."}}, Phys. Rev. Lett. \textbf{45}
  (1980), 1358.

\bibitem{Haldane1981a}
F.~D.~M. Haldane, \emph{{"Demonstration of the ``Luttinger liquid`` character of
  Bethe-ansatz soluble models of 1-D quantum fluids"}}, Phys. Lett. A
  \textbf{81} (1981), 153.

\bibitem{Haldane1981b}
F.~D.~M. Haldane, \emph{{"Luttinger liquid theory of one-dimensional quantum fluids: I.
  Properties of the Luttinger model and their extension to the general 1D
  interacting spinless Fermi gas"}}, J. Phys. C: Solid State Phys. \textbf{14}
  (1981), 2585.

\bibitem{Its81}
A.~R. Its, \emph{Asymptotic behaviour of the solutions to the nonlinear
  {S}chr\"odinger equation, and isomonodromic deformations of systems of linear
  differential equations}, Dokl. Akad. Nauk SSSR \textbf{261} (1981), no.~1,
  14--18.

\bibitem{ItsIK90}
A.R. Its, A.G. Izergin, and V.E. Korepin, \emph{Long-distance asymptotics of
  temperature correlators of the impenetrable {B}ose gas}, Commun. Math. Phys.
  \textbf{130} (1990), 471--488.

\bibitem{ItsIK90a}
A.R. Its, A.G. Izergin, and V.E. Korepin, \emph{Temperature correlators of the impenetrable {B}ose gas as an
  integrable system}, Commun. Math. Phys. \textbf{129} (1990), 205--222.

\bibitem{ItsIzerginkorepinSlavnov1993}
A.R. Its, A.G. Izergin, V.E. Korepin, and N.~A. Slavnov, \emph{Temperature
  correlations of quantum spins}, Phys. Rev. Lett. \textbf{70} (1993), 1704.

\bibitem{ItsIzerginKorepinSlavnovDifferentialeqnsforCorrelationfunctions}
A.R. Its, A.G. Izergin, V.E. Korepin, and N.A. Slavonv, \emph{{"Differential
  equations for quantum correlation functions."}}, Int. J. Mod. Physics
  \textbf{\bf{B4}} (1990), 1003--1037.

\bibitem{ItsIzerginkorepinVarguzin1991}
A.R. Its, A.G. Izergin, V.E. Korepin, and G.~G. Varguzin, \emph{Large time and
  distance asymptotics of the correlator of the impenetrable bosons at finite
  temperature}, Physica D \textbf{54} (1991), 351.
  
\bibitem{ItsK2008}
A. R. Its, I. Krasovsky, \emph{Hankel determinant and orthogonal polynomials for
the Gaussian weight with a jump}, Contemporary Mathematics 
(AMS series) \textbf{458} (2008) 215--247.

\bibitem{JimMiwaMoriSatoSineKernelPVForBoseGAz}
M.~Jimbo, T.~Miwa, Y.~Mori, and M.~Sato, \emph{{"Density matrix of an
  impenetrable Bose gas and the fifth Painlevé transcendent."}}, Physica D
  \textbf{\bf{1}} (1980), 80--158.

\bibitem{KacAcheizerTruncatedWienerHopf}
M.~Kac, \emph{{"Toeplitz matrices, translation kernels and related problem in
  probability."}}, Duke Math. J. \textbf{21} (1954), 501--510.

\bibitem{KitKMST009}
N.~Kitanine, K.~K. Kozlowski, J.-M. Maillet, N.A. Slavnov, and V.~Terras,
  \emph{{"Algebraic Bethe ansatz approach to the asymptotic
behavior of correlation functions."}}, J. Stat. Mech. (2009) P04003.

\bibitem{KorS90}
V.~E. Korepin and N.~A. Slavnov, \emph{The time dependent correlation function
  of an impenetrable {B}ose gas as a {F}redholm minor. {I}}, Comm. Math. Phys.
  \textbf{129} (1990), no.~1, 103--113.

\bibitem{KozWienerHopfWithFischerHartwig}
K.~K. Kozlowski, \emph{{"Truncated Wiener-Hopf operators with Fischer-Hartwig
  singularities."}}, math.FA/ 0805.3902.

\bibitem{Krasovsky2004}
V.~I. Krasovsky, \emph{{"Gap probability in the spectrum of random matrices and
  asymptotics of polynomials orthogonal on an arc of the unit circle."}}, Int.
  Math. Res. Not. \textbf{\bf{2004}} (2004), 1249--1272.

\bibitem{KuilajaarsMVVUniformAsymptoticsForModifiedJacobiOrthogonalPolynomials}
A.B.J. Kuijlaars, K.T.-R. McLaughlin, W.~Van Assche, and M.~Vanlessen,
  \emph{{"The Riemann--Hilbert approach to strong asymptotics for orthogonal
  polynomials on [-1,1]."}}, Advances in Math. \textbf{188} (2004), 337--398.

\bibitem{Len64}
A.~Lenard, \emph{Momentum distribution in the ground state of the
  one-dimensional system of impenetrable bosons}, J. Math. Phys. \textbf{5}
  (1964), 930--943.

\bibitem{Len66}
A.~Lenard, \emph{One-dimensional impenetrable bosons in thermal equilibrium}, J.
  Math. Phys. \textbf{7} (1966), 1268--1272.

\bibitem{McCoyTangSineKernelSubleadingFromPainleveV}
B.~M. McCoy and S. Tang \emph{{"Connection formulae for Painlev\'e V functions II. The
  delta function Bose gas problem."}}, Physica D \textbf{20} (1986), 187--216.

\bibitem{McCoyPArkShrockSpinTimeAutoCorrAsModSineKernel}
B.~M. McCoy, J.~H.~H. Perk, and R.~E. Shrock, \emph{{"Time-dependent
  correlation functions of the transverse Ising chain at the critical magnetic
  field."}}, Nucl. Phys. B \textbf{\bf{220}} (1983), 35--47.

\bibitem{WidomSzegoLimitonCircularArcs}
H.~Widom, \emph{{"The strong Szegö limit theorem for circular arcs."}}, Indiana
  Univ. Math. J. \textbf{21} (1971), 277--283.

\bibitem{WidomSinekernelOnSingleIntervals}
H.~Widom, \emph{{"The asymptotics of a continuous analogue of orthogonal
  polynomials."}}, J. Approx. th. \textbf{\bf{77}} (1994), 51--64.

\bibitem{WidomSinekernelOnUnionOfIntervals}
H.~Widom, \emph{{"Asymptotics for the Fredholm determinant of the Sine Kernel on
  a Union of Intervals."}}, Comm. Math. Phys. \textbf{\bf{171}} (1995),
  159--180.

\end{thebibliography}
\end{document}